\documentclass[twocolumn,letterpaper,pre,showpacs]{revtex4}
\newlength{\twocolumnwidth}

\newlength{\crossitoutwidth} 
\newlength{\crossitoutheight} 
\newlength{\vAlength} 
\usepackage{graphicx}
\usepackage{bm}
\usepackage{amsmath}
\usepackage{ifthen}
\begin{document} 
\title{Quantum-field-theoretical approach to phase-space techniques:\\ Symmetric Wick theorem and multitime Wigner representation. }
\author{L.I.Plimak}
\thanks{{\em Present address:\/} Max Born Institute for Nonlinear Optics and Short Pulse Spectroscopy, Division A1, 12489 Berlin, Germany. {\em E-mail\/}: Lev.Plimak@mbi-berlin.de.}
\affiliation{Institut f\"ur Quantenphysik, Universit\"at Ulm, 89069 Ulm, Germany.}
\author{M.K.Olsen}
\affiliation{ARC Centre of Excellence for Quantum-Atom Optics, School of Physical Sciences, University of Queensland, Brisbane, Qld 4072, Australia.}
\date{\today} 
\begin{abstract} 
In this work we present the formal background used to develop the methods used in earlier works to extend the truncated Wigner representation of quantum and atom optics in order to address multi-time problems. 
The truncated Wigner representation has proven to be of great practical use, especially in the numerical study of the quantum dynamics of Bose condensed gases. In these cases, it allows for the simulation of effects which are missed entirely by other approximations, such as the Gross-Pitaevskii equation, but does not suffer from the severe instabilities of more exact methods. The numerical treatment of interacting many-body quantum systems is an extremely difficult task, and the ability to extend the truncated Wigner beyond single-time situations adds another powerful technique to the available toolbox. This article gives the formal mathematics behind the development of our ``time-Wigner ordering'' which allows for the calculation of the multi-time averages which are required for such quantities as the Glauber correlation functions which are applicable to bosonic fields. 
\end{abstract}
\maketitle
\section{Introduction}
This paper is a formal corollary to the recent Ref.\ \cite{Bettina}, where the truncated Wigner representation of quantum optics \cite{Wminus} was extended to multitime problems. 
(For a discussion of operator orderings, such as normal, time-normal and symmetric, we refer the reader to the classic treatise of Mandel and Wolf \cite{MandelWolf}.) The goal of Ref.\ \cite{Bettina} was to develop a practical computational tool, while formal analyses were reduced to the bare necessities. In this paper we give proper justification to the formal techniques underlying Ref.\ \cite{Bettina}. The functional techniques used here are to a large extent borrowed from Vasil'ev \cite{VasF}; they were first outlined in preprint \cite{PreprintGenPW}. However, in \cite{PreprintGenPW} a number of important results (notably, continuity of time-symmetrically ordered operator products) were overlooked. 

Important for the putting the results of this paper in perspective is the connection between phase-space techniques and the so-called {\em real-time quantum field theory\/} (for details and references see \cite{DirResp}). According to \cite{DirResp}, the {\em Keldysh rotation\/} underlying the latter is a generalisation of Weyl's ordering to Heisenberg operators. This generalisation is nothing but the aforementioned time-symmetric ordering. This paper could thus equally be titled, ``Phase space approach to real-time quantum field theory''. 

For physical motivation and literature we refer the reader to the introduction to \cite{Bettina}. Here we only briefly touch upon the more recent developments. The truncated Wigner representation has found its main utility for the investigation and numerical modelling of Bose-Einstein condensates (BEC), where the presence of a significant third-order nonlinearity makes the exact positive-P method unstable except for very short times \cite{WBEC1}, although it is still also used in quantum optics. Over the last few years it has been used for an increasing number of investigations, with the most relevant mentioned in what follows.

The ability to include the effects of initial quantum states other than coherent was first shown numerically for trapped BEC molecular photoassociation in three papers by Olsen and Plimak \cite{MeLev}, Olsen \cite{WigstateMe}, and Olsen, Bradley, and Cavalcanti \cite{UFAL}, using methods later published by Olsen and Bradley \cite{OCstates} to sample the required quantum states. 
Johnsson and Hope calculated the multimode quantum limits to the linewidth of an atom laser \cite{JohnssonHope}.
The quantum dynamics of superflows in a toroidal trapping geometry were treated by Jain {\em et al.\/} \cite{Pyush}. Ferris {\em et al.\/} \cite{AJFerris} investigated dynamical instabilities of BEC at the band edge in one-dimensional optical lattices, while Hoffmann, Corney, and Drummond \cite{seatlicker} made an attempt to combine the truncated Wigner and positive-P representations into a hybrid method for Bose fields. Corney {\em et al.\/} \cite{Joelfibre} used the truncated Wigner to simulate polarisation squeezing in optical fibres, a system which is mathematically analogous to BEC. The mode entanglement and Einstein-Podolsky paradox were numerically investigated in the process of degenerate four-waving mixing of BEC in an optical lattice by Olsen and Davis \cite{MeMavis} and Ferris, Olsen, and Davis \cite{AndyMeMavis}. Also analysing BEC in an optical lattice, Shrestha, Javanainen, and Rusteokoski \cite{Finns} looked at the quantum dynamics of instability-induced pulsations. Midgley {\em et al.\/} performed a comparison of the truncated Wigner with the exact positive-P representation and a Hartree-Fock-Bogoliubov approximate method for the simulation of molecular BEC dissociation \cite{BECSarah}, concluding that the truncated Wigner representation was the most useful in practical terms. This practical usefulness has been demonstrated in studies of BEC interferometry \cite{Hoffperv}, domain formation in inhomogeneous ferromagnetic dipolar condensates \cite{Sau}, vortex unbinding following a quantum quench \cite{unbind}, a reverse Kibble-Zurek mechanism in two-dimensional superfluids \cite{reverseKZ}, the quantum and thermal effects of dark solitons in one-dimensional Bose gases \cite{AndyMartin}, the quantum dynamics of multi-well Bose-Hubbard models \cite{CVC1,CVC2}, and analysis of a method to produce Einstein-Podolsky-Rosen states in two-well BEC \cite{2wellEPR}.
Along with its continuing use in quantum optics, the above examples demonstrate that the truncated Wigner representation is an extremely useful approximation method, allowing for the numerical simulation of a number of processes for which nothing else is known to be as effective.

The task in \cite{Bettina} was split naturally in two. Firstly, we constructed a path-integral approach in phase space, which we called {\em multitime Wigner representation\/}. Within this approach, truncated Wigner equations emerged as an approximation to exact quasi-stochastic equations for paths. Secondly, we developed a way of bringing time-symmetric products of {Heisenberg}\ operators \cite{Bettina,PreprintGenPW}, expressed by the path integral, to time-normal order \cite{MandelWolf}. The respective relations, called {\em generalised phase-space correspondences\/}, originate in Kubo's formula for the linear response function \cite{KuboIrrevI,KuboTodaHashitsumeII}. However, the properties of time-symmetric operator products \cite{Bettina,PreprintGenPW} were formulated without proof. Details of the limiting procedure defining the path integral were ignored as irrelevant within the truncated Wigner approximation. 

The logic of the present paper is best understood by drawing an analogy with our Ref.\ \cite{BWO}. In \cite{BWO}, our goal was to extend {\em normal-ordering-based\/} approaches of quantum stochastics beyond quartic Hamiltonians. This paper is an attempt to extend the techniques of Ref.\ \cite{BWO} further, this time to the {\em Weyl-ordering-based\/} method of Ref.\ \cite{Bettina}. Particulars aside, in \cite{BWO} we applied {\em causal\/} \cite{BWO}, or {\em response\/} \cite{API,APII,APIII}, transformation to Perel-Keldysh diargam series \cite{Perel,Keldysh} for the system in question. The result was a Wyld-type \cite{Wyld} diagram series, which we called {\em causal series\/}. These kind of diagram techniques are well known in the theory of classical stochastic processes \cite{ZinnJ,VasF,VasR}. It is therefore not surprising that we could reverse engineer the causal series resulting in a stochastic differential equation (SDE) for which this series was a formal solution. For quartic (collisional) interactions the result is the well-known positive-P representation of quantum optics. For other interactions, finding an SDE in the true meaning of the term is not always possible because of the Pawula theorem \cite{Pawula}. In such cases the causal series may be only approximated by a stochastic {\em difference\/} equation (S$\Delta $E) in discretised time. Attempts to simulate emerging S$\Delta $Es numerically have not been encouraging \cite{BWO,EPL01}. This is the primary reason why in Ref.\ \cite{Bettina} we resorted to approximate methods. 

The key point of \cite{BWO} is a formal link existing between conventional quantum field theory (Schwinger-Perel-Keldysh's closed-time-loop formalism) and conventional quantum optics (time-normal operator ordering and positive-P representation). The existence of such a link was first pointed out by one of the present authors in \cite{Corresp}. For the purposes of this paper, both backgrounds are inadequate or missing and have to be built from scratch. On the quantum-field-theoretical side, we develop a formal framework of ``symmetric Wick theorems'' expressing time-ordered operator products by symmetrically-ordered ones. Similar to Wick's theorem proper, this allows us to construct a ``symmetric'' variety of Perel-Keldysh diagram series with corresponding ``symmetric'' propagators. On the quantum-optical side, we introduce what we called a multitime Wigner representation generalising Weyl's ordering to multitime averages of Heisenberg operators. The necessary ``flavour'' of causal transformation can then be borrowed from \cite{PreprintGenPW}. The link to time-normal ordering is then based on a fundamental link between commutators of Heisenberg operators for different times and the response properties of quantum fields \cite{Corresp,BWO,API,APII,APIII}. All our results apply to nonlinear quantum systems with, to a large extent, arbitrary interaction Hamiltonians. 

The existence of symmetric Wick theorems for the change of time ordering to symmetric may be of independent interest for a wider audience than that for which this paper is primarily intended. It is for this wider audience that in the appendix we present the generalisation of symmetric Wick theorems to arbitrary quantised fields. 

The glue that holds the paper together is the concept of time-symmetric ordering of Heisenberg operators. Continuing the analogy with \cite{BWO}, this ordering replaces the time-normal operator ordering on which \cite{BWO} was built. On the quantum-field-theoretical side, we express ``symmetric'' propagators by the retarded Greens function. The induced restructuring of ``symmetric'' Perel-Keldysh series {\em automatically\/} turns them into perturbative series for quantum averages of time-symmetrically ordered products of Heisenberg operators. On the quantum-optical side, time-symmetric ordering appears as a natural generalisation of the conventional symmetric ordering to Heisenberg operators. Quantum field theory and quantum optics are not equal here. Time-symmetric ordering emerges in quantum field theory as a fully specified formal concept. At the same time, there does not seem to be any way of guessing, save deriving, this concept from within conventional phase-space techniques without a reference to Schwinger's closed-time-loop formalism \cite{SchwingerC}. 

The paper is organised as follows. In sections \ref{ch:Symm} and \ref{ch:Wick}, two formal concepts are prepared for later use. In section \ref{ch:Symm} we define time-symmetric ordering of Heisenberg operators and prove its most important properties: reality, continuity, and the fact that for free fields it reduces to conventional symmetric ordering. In section \ref{ch:Wick} we prove ``symmetric Wick theorems'' generalising Wick's theorem proper from normal to symmetric ordering. In section \ref{ch:Keldysh}, we introduce a functional framework and derive closed perturbative relations with symmetric ordering of operators. The multitime Wigner representation of an arbitrary bosonic system is formulated in section \ref{ch:CommResp}. In section \ref{ch:DynaPh} we construct a representation of time-symmetric operator averages by phase-space path integrals. Section \ref{ch:SecCaus} presents a discussion of causal regularisation needed to make our analyses mathematically defined. The problem of reordering Heisenberg operators is discussed in section \ref{ch:Order}. In the appendix, we extend symmetric Wick theorems to arbitrary quantised fields. 

\section{The time-symmetric ordering}\label{ch:Symm}
\subsection{The oscillator basics}\label{ch:Bas}
We introduce the common pair of bosonic creation and annihilation operators, 
\begin{gather} 
\begin{aligned}
\left[
\hat a ,\hat a ^{\dag}
\right]=1, 
\end{aligned}%
\label{eq:OscComm} 
\end{gather}%
and the usual free oscillator Hamiltonian, 
\begin{gather} 
\begin{aligned}
\hat{H}_0 = \omega_0\hat a ^{\dag}\hat a . 
\end{aligned}%
\label{eq:HOsc} 
\end{gather}%
We use units where $ \hbar =1$. The symmetric (Weyl) ordering of the creation and annihilation operators is conveniently defined in terms of the operator-valued characteristic function 
\begin{gather} 
\begin{aligned}
\chi (\eta,\eta^*) = \text{e}^{
{\eta^*\hat a + \eta\hat a^{\dag}}
} , 
\end{aligned}%
\label{eq:ChiDef} 
\end{gather}%
so that, by definition 
\begin{gather} 
\begin{aligned}
\frac{\partial^{m+n}\chi (\eta,\eta^*)}
{\partial \eta^n\partial \eta^{*m}} \Big |_{\eta=0} = 
W\hat a^m\hat a^{\dag n} . 
\end{aligned}%
\label{eq:WDef} 
\end{gather}%
For simplicity, we consider a single-mode case; multi-mode cases are recovered formally by attaching mode index to all quantities, cf.\ section \ref{ch:Mod} and the appendix. Operator orderings act mode-wise, so that extension to many modes is straightforward. 

The time-dependent creation and annihilation operators are defined as, 
\begin{align} 
\begin{aligned} 
\hat a(t) &=\hat a\text{e}^{-i\omega_0t} , &
\hat a^{\dag}(t) &=\hat a^{\dag}\text{e}^{i\omega_0t} . 
\end{aligned} 
\label{eq:at} 
\end{align}%
Formally, they are Heisenberg field operators with respect to the free Hamiltonian, e.g., 
\begin{align} 
\begin{aligned} 
i\partial_t\hat a(t) = \big [ 
\hat a(t),\hat H_0(t)
 \big ] , 
\end{aligned} 
\label{eq:60a} 
\end{align}%
where $\hat H_0(t)$ is $\hat H_0$ in the interaction picture, 
\begin{align} 
\begin{aligned} 
\hat H_0(t) = \omega_0\hat a^{\dag}(t)\hat a(t) . 
\end{aligned} 
\label{eq:62a} 
\end{align}%
We shall term $\hat a(t),\hat a^{\dag}(t)$ free-field, or interaction-picture, operators, because this is the role they play in real problems with interactions. 
Symmetric ordering is extended to free fields postulating that, 
\begin{multline} 
\hspace{0.4\columnwidth}\hspace{-0.4\twocolumnwidth}
W\hat a(t_1)\cdots\hat a(t_m)
\hat a^{\dag}(t_{m+1})\cdots\hat a^{\dag}(t_{m+n}) \hspace{0.075\columnwidth}
\\ = 
\text{e}^{-i\omega_0(t_1+\cdots+t_m-t_{m+1}-\cdots-t_{m+n})}\,
W\hat a^m\hat a^{\dag n} . 
\hspace{0.4\columnwidth}\hspace{-0.4\twocolumnwidth}%
\label{eq:53HQ} 
\end{multline}%
Furthermore, adding an arbitrary, and, in general, time-dependent, interaction term to $\hat{H}_0$, 
\begin{gather} 
\begin{aligned}
\hat{H} =\hat{H}_0 +\hat{H}_{\text{I}}, 
\end{aligned}%
\label{eq:H0I} 
\end{gather}%
allows us to introduce the pair of the Heisenberg fields operators proper,
\begin{gather} 
\begin{aligned}
{\hat{\mathcal A}}(t) &= {\hat{\mathcal U}}^{\dag}(t)\hat a {\hat{\mathcal U}}(t), & 
{\hat{\mathcal A}}^{\dag}(t) &= {\hat{\mathcal U}}^{\dag}(t)\hat a^{\dag}{\hat{\mathcal U}}(t)
, 
\end{aligned}%
\label{eq:Evol} 
\end{gather}%
where the evolution operator is defined through 
the Schr\"odinger equation, 
\begin{gather} 
\begin{aligned}
i \frac{d{\hat{\mathcal U}}(t)}{dt} & =\hat H{\hat{\mathcal U}}(t), &
\lim_{t\to -\infty}{\hat{\mathcal U}}(t) =\hat\openone . 
\end{aligned}%
\label{eq:HEq} 
\end{gather}%
Interaction picture is introduced in the usual way by splitting the evolution operator in two factors, 
\begin{align} 
\begin{aligned} 
{\hat{\mathcal U}}(t) = {\hat{\mathcal U}}_0(t){\hat{\mathcal U}}_{\text{I}}(t), 
\end{aligned} 
\label{eq:63a} 
\end{align}%
where ${\hat{\mathcal U}}_0(t)$ is the evolution operator with respect to $\hat H_0$, 
\begin{align} 
\begin{aligned} 
 &i \frac{d{\hat{\mathcal U}}_0(t)}{dt} & =\hat H_0{\hat{\mathcal U}}_0(t), & &
\lim_{t\to -\infty}{\hat{\mathcal U}}_0(t) =\hat\openone, 
\end{aligned} 
\label{eq:64a} 
\end{align}%
and ${\hat{\mathcal U}}_{\text{I}}(t)$ obeys the equation, 
\begin{align} 
\begin{aligned} 
 &i \frac{d{\hat{\mathcal U}}_{\text{I}}(t)}{dt} & =\hat H_{\text{I}}(t){\hat{\mathcal U}}_{\text{I}}(t), & &
\lim_{t\to -\infty}{\hat{\mathcal U}}_{\text{I}}(t) =\hat\openone, 
\end{aligned} 
\label{eq:65a} 
\end{align}%
with $\hat H_{\text{I}}(t)$ being $\hat H_{\text{I}}$ in the interaction picture, 
\begin{align} 
\begin{aligned} 
\hat H_{\text{I}}(t) = {\hat{\mathcal U}}_0^{\dag}(t)\hat H_{\text{I}}{\hat{\mathcal U}}_0(t) . 
\end{aligned} 
\label{eq:66a} 
\end{align}%

The reader should have noticed that we adhere to certain notational conventions. Calligraphic letters are reserved for evolution and Heisenberg operators. Plain letters denote {Schr\"odinger}\ and interaction-picture operators, which are in turn distinguished by the absence or presence of the time argument, respectively. The {Schr\"odinger}\ operator $\hat a$ becomes $\hat a(t)$ in the interaction picture and ${\hat{\mathcal A}}(t)$ in the Heisenberg picture, 
\begin{align} 
\begin{aligned} 
\hat a(t) &= {\hat{\mathcal U}}_0^{\dag}(t)\hat a{\hat{\mathcal U}}_0(t), \\ 
{\hat{\mathcal A}}(t) &= {\hat{\mathcal U}}^{\dag}(t)\hat a{\hat{\mathcal U}}(t) = {\hat{\mathcal U}}_{\text{I}}^{\dag}(t)\hat a(t){\hat{\mathcal U}}_{\text{I}}(t), 
\end{aligned} 
\label{eq:67a} 
\end{align}%
and similarly for other operators. 

We note that, with unspecified interaction, Heisenberg operators are in essence placeholders. By specifying interaction one {\em ipso facto\/} specifies all Heisenberg operators. 

\subsection{The time-symmetric operator products}\label{ch:TWDef}
Throughout the paper we make extensive use of the concept of {\em time-symmetric\/} product of the Heisenberg operators \cite{Bettina}. We define it by the following recursive procedure. For a single operator the ordering is irrelevant, 
\begin{gather} 
\begin{aligned}
{\cal T}^W\!{\hat{\mathcal A}}(t) &= {\hat{\mathcal A}}(t), & {\cal T}^W\!{\hat{\mathcal A}}^{\dag}(t) &= {\hat{\mathcal A}}^{\dag}(t). 
\end{aligned}%
\label{eq:TW1} 
\end{gather}%
Furthermore, if ${\hat{\mathcal P}}_{[>t]}$ is a time-symmetric product\ where all time arguments are larger than $t$, then 
\begin{gather} 
\begin{aligned}
{\cal T}^W\!{\hat{\mathcal P}}_{[>t]}\,{\hat{\mathcal A}}(t) 
&= 
\frac{1}{2} \big [ 
{\hat{\mathcal P}}_{[>t]},{\hat{\mathcal A}}(t)
 \big ] 
_+, \\ 
{\cal T}^W\!{\hat{\mathcal P}}_{[>t]}\,{\hat{\mathcal A}}^{\dag}(t) 
&= 
\frac{1}{2} \big [ 
{\hat{\mathcal P}}_{[>t]},{\hat{\mathcal A}}^{\dag}(t)
 \big ]_+ 
, 
\end{aligned}%
\label{eq:TWRec} 
\end{gather}%
where $ [ \cdots ]_+$ stands for the anticommutator, 
\begin{align} 
\begin{aligned} 
\big [ 
{\hat{\mathcal X}},{\hat{\mathcal Y}}
 \big ]_+ = {\hat{\mathcal X}}{\hat{\mathcal Y}}+ {\hat{\mathcal Y}}{\hat{\mathcal X}} . 
\end{aligned} 
\label{eq:3a} 
\end{align}%
This allows one to built a time-symmetric product of any number of factors, by applying (\ref{eq:TWRec}) in the order of {\em decreasing time arguments\/} ({\em increasing\/} in \cite{PreprintGenPW} is a typo). The result are nested anticommutators, (with ${\hat{\mathcal X}}_k = {\hat{\mathcal A}},{\hat{\mathcal A}}^{\dag}, \ k=1,\cdots,N$)
\begin{multline} 
\hspace{0.4\columnwidth}\hspace{-0.4\twocolumnwidth}
{\cal T}^W\!{\hat{\mathcal X}}_1(t_1){\hat{\mathcal X}}_2(t_2)\cdots{\hat{\mathcal X}}_n(t_n)
= \frac{1}{2^{N-1}} \\ \times
\big [ \cdots
\big [ 
\big [ 
{\hat{\mathcal X}}_1(t_1),{\hat{\mathcal X}}_2(t_2)
 \big ]_+ ,{\hat{\mathcal X}}_3(t_3)
 \big ]_+ ,\cdots,{\hat{\mathcal X}}_N(t_N)
 \big ]_+ , \\ 
t_1>t_2>\cdots>t_N . 
\hspace{0.4\columnwidth}\hspace{-0.4\twocolumnwidth}%
\label{eq:78JS} 
\end{multline}%
Equations (\ref{eq:TWRec}) and (\ref{eq:78JS}) imply that all time arguments in a {time-symmetric product} are different. Extension to coinciding time arguments may be given by continuity, see below.

For two factors we find plain symmetrised combinations, 
\begin{gather} 
\begin{aligned}
{\cal T}^W\!{\hat{\mathcal A}}(t_1){\hat{\mathcal A}}(t_2) & = \frac{1}{2}\big [ 
{\hat{\mathcal A}}(t_1){\hat{\mathcal A}}(t_2)+{\hat{\mathcal A}}(t_2){\hat{\mathcal A}}(t_1)
 \big ] 
, 
\\ 
{\cal T}^W\!{\hat{\mathcal A}}(t_1){\hat{\mathcal A}}^{\dag}(t_2) & = \frac{1}{2}\big [ 
{\hat{\mathcal A}}(t_1){\hat{\mathcal A}}^{\dag}(t_2)+{\hat{\mathcal A}}^{\dag}(t_2){\hat{\mathcal A}}(t_1)
 \big ] 
 , 
\\ 
{\cal T}^W\!{\hat{\mathcal A}}^{\dag}(t_1){\hat{\mathcal A}}^{\dag}(t_2) & = \frac{1}{2}\big [ 
{\hat{\mathcal A}}^{\dag}(t_1){\hat{\mathcal A}}^{\dag}(t_2)+{\hat{\mathcal A}}^{\dag}(t_2){\hat{\mathcal A}}^{\dag}(t_1)
 \big ] 
, 
\end{aligned}%
\label{eq:12} 
\end{gather}%
where the order of times does not matter. For three and more factors it already does, e.g., 
\begin{widetext} 
\begin{multline} 
{\cal T}^W\!{{\hat{\mathcal A}}}(t_1){{\hat{\mathcal A}}}(t_2){{\hat{\mathcal A}}}^{\dag}(t_3) 
= \frac{1}{4}\big [ 
\big [ 
{{\hat{\mathcal A}}}(t_1),{{\hat{\mathcal A}}}(t_2)
 \big ]_+, {{\hat{\mathcal A}}}^{\dag}(t_3)
 \big ]_+ 
\\ = \frac{1}{4} \big[
{{\hat{\mathcal A}}}(t_1){{\hat{\mathcal A}}}(t_2){{\hat{\mathcal A}}}^{\dag}(t_3) + 
\settowidth{\crossitoutwidth}{\ensuremath{{{\hat{\mathcal A}}}(t_1){{\hat{\mathcal A}}}^{\dag}(t_3){{\hat{\mathcal A}}}(t_2)}}%
\settoheight{\crossitoutheight}{\ensuremath{{{\hat{\mathcal A}}}(t_1){{\hat{\mathcal A}}}^{\dag}(t_3){{\hat{\mathcal A}}}(t_2)}}%
\ensuremath{{{\hat{\mathcal A}}}(t_1){{\hat{\mathcal A}}}^{\dag}(t_3){{\hat{\mathcal A}}}(t_2)}\hspace{-1\crossitoutwidth}%
\rule[0.253\crossitoutheight]{1\crossitoutwidth}{0.03em} + 
{{\hat{\mathcal A}}}^{\dag}(t_3){{\hat{\mathcal A}}}(t_1){{\hat{\mathcal A}}}(t_2)+ 
{{\hat{\mathcal A}}}(t_2){{\hat{\mathcal A}}}(t_1){{\hat{\mathcal A}}}^{\dag}(t_3)\\ +
\settowidth{\crossitoutwidth}{\ensuremath{{{\hat{\mathcal A}}}(t_2){{\hat{\mathcal A}}}^{\dag}(t_3){{\hat{\mathcal A}}}(t_1)}}%
\settoheight{\crossitoutheight}{\ensuremath{{{\hat{\mathcal A}}}(t_2){{\hat{\mathcal A}}}^{\dag}(t_3){{\hat{\mathcal A}}}(t_1)}}%
\ensuremath{{{\hat{\mathcal A}}}(t_2){{\hat{\mathcal A}}}^{\dag}(t_3){{\hat{\mathcal A}}}(t_1)}\hspace{-1\crossitoutwidth}%
\rule[0.253\crossitoutheight]{1\crossitoutwidth}{0.03em} + 
{{\hat{\mathcal A}}}^{\dag}(t_3){{\hat{\mathcal A}}}(t_2){{\hat{\mathcal A}}}(t_1) 
\big] 
, \ \ t_1>t_2>t_3, 
\label{eq:123} 
\end{multline}%
as opposed to 
\begin{multline} 
{\cal T}^W\!{{\hat{\mathcal A}}}(t_1){{\hat{\mathcal A}}}(t_2){{\hat{\mathcal A}}}^{\dag}(t_3) 
= \frac{1}{4}\big [ 
\big [ 
{{\hat{\mathcal A}}}(t_1),{{\hat{\mathcal A}}}^{\dag}(t_3)
 \big ]_+, {{\hat{\mathcal A}}}(t_2)
 \big ]_+ 
\\ = \frac{1}{4} \big[
\settowidth{\crossitoutwidth}{\ensuremath{{{\hat{\mathcal A}}}(t_1){{\hat{\mathcal A}}}(t_2){{\hat{\mathcal A}}}^{\dag}(t_3)}}%
\settoheight{\crossitoutheight}{\ensuremath{{{\hat{\mathcal A}}}(t_1){{\hat{\mathcal A}}}(t_2){{\hat{\mathcal A}}}^{\dag}(t_3)}}%
\ensuremath{{{\hat{\mathcal A}}}(t_1){{\hat{\mathcal A}}}(t_2){{\hat{\mathcal A}}}^{\dag}(t_3)}\hspace{-1\crossitoutwidth}%
\rule[0.253\crossitoutheight]{1\crossitoutwidth}{0.03em} + 
{{\hat{\mathcal A}}}(t_1){{\hat{\mathcal A}}}^{\dag}(t_3){{\hat{\mathcal A}}}(t_2) +
{{\hat{\mathcal A}}}^{\dag}(t_3){{\hat{\mathcal A}}}(t_1){{\hat{\mathcal A}}}(t_2) + 
{{\hat{\mathcal A}}}(t_2){{\hat{\mathcal A}}}(t_1){{\hat{\mathcal A}}}^{\dag}(t_3) \\ + 
{{\hat{\mathcal A}}}(t_2){{\hat{\mathcal A}}}^{\dag}(t_3){{\hat{\mathcal A}}}(t_1) + 
\settowidth{\crossitoutwidth}{\ensuremath{{{\hat{\mathcal A}}}^{\dag}(t_3){{\hat{\mathcal A}}}(t_2){{\hat{\mathcal A}}}(t_1)}}%
\settoheight{\crossitoutheight}{\ensuremath{{{\hat{\mathcal A}}}^{\dag}(t_3){{\hat{\mathcal A}}}(t_2){{\hat{\mathcal A}}}(t_1)}}%
\ensuremath{{{\hat{\mathcal A}}}^{\dag}(t_3){{\hat{\mathcal A}}}(t_2){{\hat{\mathcal A}}}(t_1)}\hspace{-1\crossitoutwidth}%
\rule[0.253\crossitoutheight]{1\crossitoutwidth}{0.03em}
\big] 
, \ \ t_1>t_3>t_2. 
\label{eq:132} 
\end{multline}%
\end{widetext}%
In these formulae, the crossed-out terms are those absent in time-symmetric operator products but occuring in symmetrised ones. 
\subsection{General properties of time-symmetric products}\label{ch:GenTW}
\begin{figure}[b]
\begin{center} 
\setlength{\unitlength}{0.49\columnwidth}
\vspace{0.05\unitlength}
\includegraphics[width=2\unitlength]
{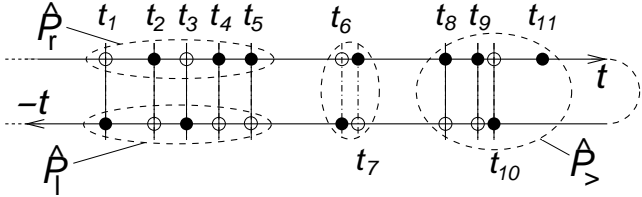}
\end{center} 
\caption{%
Graphical representation of Schwinger products produced by the recursion (\ref{eq:TWRec}). 
This particular example implies $N=11$ operators ${\hat{\mathcal X}}_k(t_k)$, $k=1,\cdots,N$, with $t_1<t_2<\cdots<t_{N}$. 
The operators are imagined as positioned on the C-contour travelling from $t=-\infty$ to $t=+\infty$ (the forward branch) and then back to $t=-\infty$ (reverse branch). 
Each operator may be either on the forward or on the reverse branch. The order of operators on the C-contour determines the order of factors in a particular Schwinger product (from right to left, to match Eq.\ (\ref{eq:TpTm})). 
So, the way of placing operators shown by dark circles corresponds to the product 
\mbox{$
{\hat{\mathcal X}}_{1}
{\hat{\mathcal X}}_{3}
{\hat{\mathcal X}}_{6}
{\hat{\mathcal X}}_{10}
{\hat{\mathcal X}}_{11}
{\hat{\mathcal X}}_{9}
{\hat{\mathcal X}}_{8}
{\hat{\mathcal X}}_{7}
{\hat{\mathcal X}}_{5}
{\hat{\mathcal X}}_{4}
{\hat{\mathcal X}}_{2}
$}, where time arguments are omitted for brevity. 
Placing of the latest time ($t_{11}$) does not affect the order of operators, so we put it arbitrarily on the forward branch. 
With all alternative placings of operators (light circles) we recover $2^{N-1}$ distinct Schwinger products. 
For the operator placing shown by dark circles, the quantities occuring in equation (\ref{eq:Quadr}) are ${{\hat{\mathcal P}}}_{\text{l}} = {\hat{\mathcal X}}_{1}(t_{1})
{\hat{\mathcal X}}_{3}(t_{3})
$, ${{\hat{\mathcal P}}}_{\text{r}} ={\hat{\mathcal X}}_{5}(t_{5})
{\hat{\mathcal X}}_{4}(t_{4})
{\hat{\mathcal X}}_{2}(t_{2})
$, ${{\hat{\mathcal P}}}_{>} = {\hat{\mathcal X}}_{10}(t_{10})
{\hat{\mathcal X}}_{11}(t_{11})
{\hat{\mathcal X}}_{9}(t_{9})
{\hat{\mathcal X}}_{8}(t_{8})
$, $t_6 = \min(t,t')$, and $t_7 = \max(t,t')$. 
}
\label{fig:TW}
\end{figure}

Equation (\ref{eq:78JS}) expresses a time-symmetrically ordered product of $N$ factors as a sum of $2^{N-1}$ products of field operators which differ in the order of factors. In all these products the time arguments first increase then decrease (whereas in crossed-out terms in (\ref{eq:123}) and (\ref{eq:132}) the earliest operator is in the middle). This kind of time-ordered structure is characteristic of Schwinger's closed-time-loop formalism (see Refs.\ \cite{SchwingerC,Keldysh} and sections \ref{ch:SWSchwinger} and \ref{ch:Keldysh} below), so we shall talk of {\em Schwinger (time) sequences\/} and of the {\em Schwinger order\/} of factors in {\em Schwinger products\/}. It is easy to see that all possible Schwinger products appear in the sum. Indeed, the earliest time in a Schwinger sequence can only occur either on the left or on the right. That recursions (\ref{eq:TWRec}) generate all Schwinger products can then be shown by induction, starting from pair products for which this is obvious. We may thus give an alternative definition of the time-symmetric product: 
\begin{center}
\parbox{0.9\columnwidth}{{\em The time-symmetric product of $N$ factors is $1/2^{N-1}$ times the sum of all possible Schwinger products of these factors.\/}}
\end{center}
This definition of the {time-symmetric product} is illustrated in Fig.\ \ref{fig:TW}, where the $2^{N-1}$ Schwinger products comprising it are visualised as $2^{N-1}$ distinct ways of placing the operators on the so-called C-contour \cite{SchwingerC,Perel,Keldysh,KamenevLevchenko}. The latter travels from $t=-\infty$ to $t=+\infty$ (the forward branch) and then back to $t=-\infty$ (reverse branch). 
The operators are imagined as positioned on the C-contour; 
each operator may be either on the forward or reverse branch, cf.\ Fig.\ \ref{fig:TW}. The order of operators on the C-contour determines the order of factors in a particular Schwinger product (from right to left, to match Eq.\ (\ref{eq:TpTm}) below). 

Definition of a {time-symmetric product} as a sum of all Schwinger products makes obvious the reality property, 
\begin{gather} 
\begin{aligned}
\left[
{\cal T}^W\!(\cdots)
\right] ^{\dag}= 
{\cal T}^W\!(\cdots)^{\dag}
. 
\end{aligned}%
\label{eq:TWConj} 
\end{gather}%
This follows from the trivial fact that if a particular sequence of times is in a Schwinger order, the reverse sequence is also in a Schwinger order. 
Furthermore, visualisation in Fig.\ \ref{fig:TW} helps us to prove the most important property of the time-symmetric products: their continuity at coinciding time arguments. Assume that a pair of times $t,t'$ can change their mutual order but both stay either earlier or later in respect of all other times, cf.\ Fig.\ \ref{fig:TW}. Assume also that placing of all times on the C-contour except $t,t'$ is fixed. If $t,t'$ are not the latest times, we are left with four ways of placing them on the C-contour resulting, up to the overall coefficient, in four terms (with ${\hat{\mathcal X}},{\hat{\mathcal Y}} = {\hat{\mathcal A}},{\hat{\mathcal A}}^{\dag}$)
\begin{gather} 
\begin{aligned}
&{\hat{\mathcal Q}}(t,t') = 
{{\hat{\mathcal P}}}_{\text{l}} {\hat{\mathcal X}}(t){{\hat{\mathcal P}}}_>{\hat{\mathcal Y}}(t') {{\hat{\mathcal P}}}_{\text{r}} + 
{{\hat{\mathcal P}}}_{\text{l}} {\hat{\mathcal Y}}(t){{\hat{\mathcal P}}}_>{\hat{\mathcal X}}(t') {{\hat{\mathcal P}}}_{\text{r}} 
\\ & \quad  + 
\begin{cases}
{{\hat{\mathcal P}}}_{\text{l}} {\hat{\mathcal X}}(t){\hat{\mathcal Y}}(t'){{\hat{\mathcal P}}}_> {{\hat{\mathcal P}}}_{\text{r}} + 
{{\hat{\mathcal P}}}_{\text{l}} {{\hat{\mathcal P}}}_> {\hat{\mathcal Y}}(t'){\hat{\mathcal X}}(t) {{\hat{\mathcal P}}}_{\text{r}} ,
 \ \ t<t' , \\ 
{{\hat{\mathcal P}}}_{\text{l}} {\hat{\mathcal Y}}(t'){\hat{\mathcal X}}(t){{\hat{\mathcal P}}}_> {{\hat{\mathcal P}}}_{\text{r}} + 
{{\hat{\mathcal P}}}_{\text{l}} {{\hat{\mathcal P}}}_> {\hat{\mathcal X}}(t){\hat{\mathcal Y}}(t') {{\hat{\mathcal P}}}_{\text{r}} ,
 \ \ t>t' 
. 
\end{cases}
\end{aligned}%
\label{eq:Quadr} 
\end{gather}%
Here, the operator product ${{\hat{\mathcal P}}}_>$ comprises all operator factors with time arguments larger than $t,t'$, ${{\hat{\mathcal P}}}_{\text{r}}$ comprises operators with time arguments less than $t,t'$ placed on the forward branch of the C-contour, and ${{\hat{\mathcal P}}}_{\text{l}}$ comprises operators with time arguments less than $t,t'$ placed on the reverse branch of the C-contour (see Fig.\ \ref{fig:TW}). Then, 
\begin{align}
&\lim_{t\nearrow t'}{\hat{\mathcal Q}}(t,t') - \lim_{t\searrow t'}{\hat{\mathcal Q}}(t,t') 
\nonumber \\ 
& \quad = 
{{\hat{\mathcal P}}}_{\text{l}}
\big [ 
{\hat{\mathcal X}}(t),{\hat{\mathcal Y}}(t)
 \big ] {{\hat{\mathcal P}}}_> {{\hat{\mathcal P}}}_{\text{r}} - 
{{\hat{\mathcal P}}}_{\text{l}} {{\hat{\mathcal P}}}_> 
\big [ 
{\hat{\mathcal X}}(t),{\hat{\mathcal Y}}(t)
 \big ] {{\hat{\mathcal P}}}_{\text{r}} = 0 , 
\end{align}%
because the commutator here may only be a c-number (or zero); this holds also in a multimode case. If $t,t'$ are the latest times, then ${{\hat{\mathcal P}}}_>=\hat\openone$ and the freedom of placing $t,t'$ reduces to two terms, 
\begin{gather} 
\begin{aligned}
{\hat{\mathcal Q}}(t,t') = 
{{\hat{\mathcal P}}}_{\text{l}} 
\big [ 
{\hat{\mathcal X}}(t){\hat{\mathcal Y}}(t') + {\hat{\mathcal Y}}(t){\hat{\mathcal X}}(t')
 \big ] {{\hat{\mathcal P}}}_{\text{r}} 
, 
\end{aligned}%
\end{gather}%
which is independent of the order of the times $t,t'$. Continuity of the time-symmetric products has thus been proven. Note that this spares us the necessity to specify their values at coinciding time arguments. 
\subsection{Equivalence of the time-symmetric and symmetric ordering of the free-field operators}\label{ch:EqTWtoW}
We now prove that, when applied to free fields, the time-symmetric ordering is just a fancy way of redefining conventional symmetric ordering defined by Eqs.\ (\ref{eq:WDef}) and (\ref{eq:53HQ}), 
\begin{multline} 
\hspace{0.4\columnwidth}\hspace{-0.4\twocolumnwidth}
{\cal T}^W\!\hat a(t_1)\cdots\hat a(t_m)
\hat a^{\dag}(t_{m+1})\cdots\hat a^{\dag}(t_{m+n}) 
\\ = 
W\hat a(t_1)\cdots\hat a(t_m)
\hat a^{\dag}(t_{m+1})\cdots\hat a^{\dag}(t_{m+n}) \hspace{0.075\columnwidth}
\\ = 
\text{e}^{-i\omega_0(t_1+\cdots+t_m-t_{m+1}-\cdots-t_{m+n})}\,
W\hat a^m\hat a^{\dag n}, 
\label{eq:TWAsW} 
\hspace{0.4\columnwidth}\hspace{-0.4\twocolumnwidth}
\end{multline}%
so that the order of times here in fact does not matter. 
At first glance, equations (\ref{eq:123}) and (\ref{eq:132}) serve as counter-examples. From Eq.\ (\ref{eq:123}) we get, 
\begin{multline} 
\hspace{0.4\columnwidth}\hspace{-0.4\twocolumnwidth}
{\cal T}^W\!\hat a(t_1)\hat a(t_2)\hat a^{\dag}(t_3) = 
\frac{\text{e}^{-i\omega_0(t_1+t_2-t_3)}}{2} \\ \times \left[
\hat a^2\hat a^{\dag}+\hat a^{\dag}\hat a^2 
\right], \ \ t_1>t_2>t_3 , 
\label{eq:aaad} 
\hspace{0.4\columnwidth}\hspace{-0.4\twocolumnwidth}
\end{multline}%
as contrasted by the formula obtained from Eq.\ (\ref{eq:132}), 
\begin{multline} 
\hspace{0.4\columnwidth}\hspace{-0.4\twocolumnwidth}
{\cal T}^W\!\hat a(t_1)\hat a(t_2)\hat a^{\dag}(t_3) = 
\frac{\text{e}^{-i\omega_0(t_1+t_2-t_3)}}{4}\\ \times \left[
\hat a^2\hat a^{\dag}+\hat a^{\dag}\hat a^2 + 2\hat a\hat a^{\dag}\hat a
\right],\ \ t_1>t_3>t_2 . 
\label{eq:aada} 
\hspace{0.4\columnwidth}\hspace{-0.4\twocolumnwidth}
\end{multline}%
Neither result coincides with the symmetrically-ordered product following from (\ref{eq:WDef}), 
\begin{multline} 
\hspace{0.4\columnwidth}\hspace{-0.4\twocolumnwidth}
W\hat a(t_1)\hat a(t_2)\hat a^{\dag}(t_3) = \frac{\text{e}^{-i\omega_0(t_1+t_2-t_3)}}{3}\\ \times \left[
\hat a^2\hat a^{\dag}+\hat a^{\dag}\hat a^2 +\hat a\hat a^{\dag}\hat a
\right]. 
\hspace{0.4\columnwidth}\hspace{-0.4\twocolumnwidth}
\end{multline}%
However by using the commutational relation (\ref{eq:OscComm}) it is easy to verify that 
\begin{multline} 
\hspace{0.4\columnwidth}\hspace{-0.4\twocolumnwidth}
\frac{1}{2} \left[
\hat a^2\hat a^{\dag}+\hat a^{\dag}\hat a^2 
\right] = \frac{1}{4} \left[
\hat a^2\hat a^{\dag}+\hat a^{\dag}\hat a^2 + 2\hat a\hat a^{\dag}\hat a
\right]\\ = \frac{1}{3} \left[
\hat a^2\hat a^{\dag}+\hat a^{\dag}\hat a^2 +\hat a\hat a^{\dag}\hat a
\right] = W\hat a^2\hat a^{\dag}. 
\hspace{0.4\columnwidth}\hspace{-0.4\twocolumnwidth}
\end{multline}%

To prove (\ref{eq:TWAsW}) in general we note that the recursion procedure defining the time-symmetric product may be started from 
the unity operator, 
\begin{gather} 
\begin{aligned}
{\cal T}^W\!\hat\openone & =\hat\openone . 
\end{aligned}%
\label{eq:TWOne} 
\end{gather}%
Then, 
\begin{gather} 
\begin{aligned}
{\cal T}^W\!{\hat{\mathcal A}}(t) 
&= 
\frac{1}{2} \big[
\hat\openone,{\hat{\mathcal A}}(t)
\big]_+ = {\hat{\mathcal A}}(t), \\ 
{\cal T}^W\!{\hat{\mathcal A}}^{\dag}(t) 
&= 
\frac{1}{2} \big[
\hat\openone,{\hat{\mathcal A}}^{\dag}(t)
\big]_+ = {\hat{\mathcal A}}^{\dag}(t)
,
\end{aligned}%
\label{eq:1a} 
\end{gather}%
in obvious agreement with (\ref{eq:TW1}). Furthermore, we can replace $\hat\openone$ 
in (\ref{eq:TWOne}) by $\chi (\eta,\eta^*)$ given by Eq.\ (\ref{eq:ChiDef}) and then apply the recursion (\ref{eq:TWRec}) starting from the latest time. 
With the end limit $\eta = 0$ this replacement does not affect the resulting time-symmetric products. So, assuming that $t'<t$, 
\begin{gather} 
\begin{gathered} 
{\cal T}^W\!\hat a(t) = \frac{1}{2}\big [ 
\chi (\eta,\eta^*),\hat a(t)
 \big ]_+|_{\eta =0} , \\
{\cal T}^W\!\hat a(t)\hat a^{\dag}(t') = \frac{1}{4}\big [ \big [ 
\chi (\eta,\eta^*),\hat a(t)
 \big ]_+,\hat a^{\dag}(t') \big ]_+|_{\eta =0} . 
\end{gathered} 
\label{eq:4a} 
\end{gather}%
Similar relations hold for larger number of factors. We now transform them using the standard phase-space correspondences. Recalling that 
\begin{gather} 
\begin{aligned}
\text{e}^{
{\eta^*\hat a + \eta\hat a^{\dag}}
} = \text{e}^{
{\eta\hat a^{\dag}}
}\text{e}^{
{\eta^*\hat a}
}\text{e}^{|\eta |^2/2} = 
\text{e}^{
{\eta^*\hat a}
}
\text{e}^{
{\eta\hat a^{\dag}}
}
\text{e}^{-|\eta |^2/2} 
\end{aligned}%
\end{gather}%
we find, 
\begin{gather} 
\begin{split}
\frac{\partial \chi (\eta,\eta^*)}{\partial \eta} & = 
\left(
\hat a ^{\dag}+ {\eta^*}/{2} 
\right)\chi (\eta,\eta^*) \\ & = \chi (\eta,\eta^*) \left(
\hat a ^{\dag}- {\eta^*}/{2} 
\right), \\ 
\frac{\partial \chi (\eta,\eta^*)}{\partial \eta^*} & = 
\left(
\hat a - {\eta}/{2} 
\right)\chi (\eta,\eta^*) \\ & = \chi (\eta,\eta^*) \left(
\hat a + {\eta}/{2} 
\right)
. 
\end{split}%
\end{gather}%
Pairwise combining these relations and multiplying them by suitable time exponents we obtain 
\begin{gather} 
\begin{gathered}
\frac{1}{2}\left[
\chi (\eta,\eta^*),\hat a(t)
\right]_+ = \text{e}^{-i\omega_0t}\frac{\partial \chi (\eta,\eta^*)}{\partial \eta^*}, \\ 
\frac{1}{2}\left[
\chi (\eta,\eta^*),\hat a^{\dag}(t)
\right]_+ = \text{e}^{i\omega_0t}\frac{\partial \chi (\eta,\eta^*)}{\partial \eta} 
. 
\end{gathered}%
\end{gather}%
This allows us to rewrite Eqs.\ (\ref{eq:4a}) as 
\begin{gather} 
\begin{gathered} 
{\cal T}^W\!\hat a(t) = \text{e}^{-i\omega_0 t}\, 
\frac{\partial\chi (\eta,\eta^*)}{\partial \eta ^*} \Big|_{\eta =0} , \\
{\cal T}^W\!\hat a(t)\hat a^{\dag}(t') = \text{e}^{-i\omega_0 (t-t')}\, 
\frac{\partial^2\chi (\eta,\eta^*)}
{\partial \eta \partial \eta ^*} \Big|_{\eta =0} . 
\end{gathered} 
\label{eq:5a} 
\end{gather}%
For an arbitrary number of factors we have, 
\begin{multline} 
\hspace{0.4\columnwidth}\hspace{-0.4\twocolumnwidth}
{\cal T}^W\!\hat a(t_1)\cdots\hat a(t_m)
\hat a^{\dag}(t_{m+1})\cdots\hat a^{\dag}(t_{m+n}) \\ = 
\text{e}^{-i\omega_0(t_1+\cdots+t_m-t_{m+1}-\cdots-t_{m+n})}\, 
\frac{\partial^{m+n}\chi (\eta,\eta^*)}
{\partial \eta^n\partial \eta^{*m}} \Big |_{\eta=0}
. 
\hspace{0.4\columnwidth}\hspace{-0.4\twocolumnwidth}%
\label{eq:TWDif} 
\end{multline}%
By virtue of (\ref{eq:WDef}) this is another form of (\ref{eq:TWAsW}). 

To conclude this paragraph we note that Eq.\ (\ref{eq:TWAsW}) is not as straightforward as the corresponding relation for the normal ordering: 
\begin{multline} 
\hspace{0.4\columnwidth}\hspace{-0.4\twocolumnwidth}
{\mbox{\rm\boldmath$:$}}\hat a^{\dag}(t_{m+1})\cdots\hat a^{\dag}(t_{m+n}) 
\hat a(t_1)\cdots\hat a(t_m){\mbox{\rm\boldmath$:$}}\ 
\\ = 
\text{e}^{-i\omega_0(t_1+\cdots+t_m-t_{m+1}-\cdots-t_{m+n})}\,{\mbox{\rm\boldmath$:$}}\hat a^{\dag n}\hat a^m{\mbox{\rm\boldmath$:$}}. 
\label{eq:TNAsN} 
\hspace{0.4\columnwidth}\hspace{-0.4\twocolumnwidth}
\end{multline}%
Here setting the free-field operators in the normal order {\em ipso facto\/} sets the creation and annihilation operators in the normal order, while in Eq.\ (\ref{eq:TWAsW}) the symmetric order is only recoved on rearranging the operators using (\ref{eq:OscComm}). Since the commutator contains dynamical information one may say that relation (\ref{eq:TNAsN}) is purely kinematical while equation (\ref{eq:TWAsW}) is dynamical. 
\subsection{The time-symmetric averages}
The time-symmetric ordering of free-field operators delivers us ``the best of both worlds.'' On the one hand, it is just symmetric ordering in disguise. On the other hand, the Schwinger products of which it is build are consistent with such ``big guns'' as Schwinger's closed-time-loop formalism. The time-symmetric ordering of the free-field operators correctly ``guesses'' certain fundamental structures underlying the interacting quantum field theory, making it an irreplaceable bridging concept when deriving perturbative relations with the symmetric ordering. 

In practice, it is convenient to eliminate symmetric ordering altogether by reexpressing all information about the initial state of the system directly in terms of time-symmetric averages. We introduce the Wigner distribution in the standard way by 
\begin{gather} 
\begin{aligned}
\left\langle \chi \left(
\eta ,\eta ^*
\right)\right\rangle = \int \frac{d^2\alpha}{\pi }\,W(\alpha,\alpha^*) 
\,\text{e}^{\eta ^* \alpha + \eta \alpha ^*}
 , 
\end{aligned}%
\end{gather}%
where the quantum averaging is over the initial state of the system. The symmetric averages characterising the inital state may then be written as 
\begin{gather} 
\begin{aligned}
\left\langle W\hat a^m\hat a^{\dag n}\right\rangle = \int \frac{d^2\alpha}{\pi }\,W(\alpha,\alpha^*) 
\,\alpha^n\alpha^{*n}. 
\end{aligned}%
\end{gather}%
By making use of this relation, the result of averaging Eq.\ (\ref{eq:TWAsW}) takes the form
\begin{multline} 
\hspace{0.4\columnwidth}\hspace{-0.4\twocolumnwidth}
\left\langle {\cal T}^W\!\hat a(t_1)\cdots\hat a(t_m)
\hat a^{\dag}(t_{m+1})\cdots\hat a^{\dag}(t_{m+n})\right\rangle \\ = \int \frac{d^2\alpha}{\pi } W(\alpha,\alpha^*)\alpha(t_1)\cdots\alpha(t_m)
\\ \times 
\alpha^*(t_{m+1})\cdots\alpha^*(t_{m+n}), 
\label{eq:WMom} 
\hspace{0.4\columnwidth}\hspace{-0.4\twocolumnwidth}
\end{multline}%
where 
\begin{align} 
\begin{aligned} 
 &\alpha(t) = \alpha\text{e}^{-i\omega_0t}, &
 &\alpha^*(t) = \alpha^*\text{e}^{i\omega_0t}. 
\end{aligned} 
\label{eq:alt} 
\end{align}%
If $W(\alpha,\alpha^*)>0$, this is nothing but stochastic moments of the random c-number field $\alpha(t)$, which is specified by its value at $t=0$ (initial condition) being distributed in accordance with the probability distribution $W(\alpha,\alpha^*)$. 
For nonpositive $W(\alpha,\alpha^*)$ this interpretation holds by replacing probability by quasi-probability. 
\section{Symmetric Wick theorems}\label{ch:Wick}
\subsection{Time ordering of operators}\label{ch:Tp}
The time ordering places operators from right to left in the order of increasing time arguments, e.g., 
\begin{multline} 
\hspace{0.4\columnwidth}\hspace{-0.4\twocolumnwidth}
T_+^W {{\hat{\mathcal A}}}(t) {{\hat{\mathcal A}}}^{\dag}(t') = 
\theta(t-t'){{\hat{\mathcal A}}}(t) {{\hat{\mathcal A}}}^{\dag}(t')\\ + 
\theta(t'-t){{\hat{\mathcal A}}}^{\dag}(t'){{\hat{\mathcal A}}}(t) , 
\hspace{0.4\columnwidth}\hspace{-0.4\twocolumnwidth}
\label{eq:Tp} 
\end{multline}%
and similarly for a larger number of factors. By definition, bosonic operators under the time ordering commute. 
The notation $T_+^W$ as distinct from $T_+$ used in \cite{BWO} emphasises that the $T_+^W$-ordering implies a different specification for coinciding time arguments. 
In \cite{BWO} this specification was by normal ordering, while, not quite unexpectedly, in this paper we imply symmetric ordering. 
In \cite{BWO}, such specification was enforced by the {\em causal regularisation\/} which was part of the dynamical approach. A similar (but different) regularisation scheme is part of the dynamical approach in this paper, cf.\ section \ref{ch:SecCaus}. 
Till that section we suppress all specifications related to operator orderings for coinciding time arguments. The notation $T_+^W$ should thus be regarded a placeholder for the concept full meaning of which will be made clear in section \ref{ch:SecCaus}. 
\subsection{Symmetric Wick theorem for the time ordering}\label{ch:SWDyson}
We now prove a generalisation of Wick's theorem to the symmetric ordering. As a major simplification, results of the previous section allow us to formulate and prove the {\em symmetric Wick theorem\/} as a relation between the time-ordered and 
time-symmetrically ordered operators products: 
\begin{center}
\parbox{0.9\columnwidth}{{\em A time-ordered product of interaction-picture operators equals a sum of time-symmetrically ordered products with all possible {\em symmetric contractions\/}, including the term without contractions.\/}}
\end{center}
The symmetric contraction is defined, in obvious analogy to Wick's theorem proper, as 
\begin{align} 
\begin{aligned} 
-i G^W_{++}(t-t') 
= T^W_+\hat a(t)\hat a^{\dag}(t') - {\cal T}^W\!\hat a(t)\hat a^{\dag}(t') 
. 
\end{aligned} 
\label{eq:GWpp} 
\end{align}%
One can use here the symmetric ordering in place of the time-symmetric one, but we follow our general principle of eliminating the $W$-ordering from considerations. The cumbersome notation we use for the symmetric contraction will be explained in section \ref{ch:SWSchwinger} below. 
For the oscillator, 
\begin{gather} 
\begin{aligned}
-i G^W_{++}(t) = \varepsilon (t)\left[
\hat a(t),\hat a^{\dag}(0)
\right] = \varepsilon (t) \text{e}^{-i\omega_0t}, 
\end{aligned}%
\label{eq:SCont} 
\end{gather}%
where $\varepsilon (t)$ is the odd stepfunction, 
\begin{gather} 
\begin{aligned}
\varepsilon (t) = \frac{1}{2}\left[
\theta(t) - \theta(-t)
\right] . 
\end{aligned}%
\end{gather}%

The symmetric Wick theorem obviously holds for one and two operators; in the latter case it coincides with Eq.\ (\ref{eq:GWpp}). A general proof follows by induction. Assume the symmetric Wick theorem has been proven up to a certain number of factors and consider a time-ordered product with one ``spare'' factor. We choose this spare factor as the earliest one, which is thus on the right of the time ordered product. Let it be $\hat a(t)$. The whole time-ordered product may then be written as 
\begin{gather} 
\begin{aligned}
T^W_+\hat a^{\dag}(t_1)\cdots\hat a^{\dag}(t_m)\hat a(t_{m+1}) \cdots\hat a(t_{m+n})\hat a(t)
 = {\hat{\mathcal P}}\,\hat a(t), 
\end{aligned}%
\end{gather}%
where we have introduced a notation for the product without the ``spare'' factor 
\begin{gather} 
\begin{aligned}
{\hat{\mathcal P}} = T^W_+\hat a^{\dag}(t_1)\cdots\hat a^{\dag}(t_m)\hat a(t_{m+1}) \cdots\hat a(t_{m+n}) . 
\end{aligned}%
\end{gather}%
We wish to have the spare factor symmetrically on either side of ${\hat{\mathcal P}}$ so we write 
\begin{gather} 
\begin{aligned}
{\hat{\mathcal P}}\,\hat a(t) = \frac{1}{2} \big[
{\hat{\mathcal P}},\hat a(t)
\big]_+ + \frac{1}{2} \big[
{\hat{\mathcal P}},\hat a(t)
\big]. 
\end{aligned}%
\label{eq:CA} 
\end{gather}%
The commutator is easily calculated 
\begin{gather} 
\begin{aligned}
\frac{1}{2} \big[
{\hat{\mathcal P}},\hat a(t)
\big] = \frac{1}{2} \sum_{k=1}^m{\hat{\mathcal P}}'_k \big[
\hat a^{\dag}(t_k),\hat a(t)
\big] , 
\end{aligned}%
\label{eq:Comm} 
\end{gather}%
where ${\hat{\mathcal P}}'_k$ is ${\hat{\mathcal P}}$ without the factor $\hat a^{\dag}(t_k)$; note that ${\hat{\mathcal P}}'_k$ remain time-ordered. For $t<t_k$, 
\begin{gather} 
\begin{aligned}
\frac{1}{2}\left[
\hat a^{\dag}(t_k),\hat a(t)
\right] = -i G^W_{++}(t-t_k), 
\end{aligned}%
\label{eq:Contr} 
\end{gather}%
and we find 
\begin{multline} 
\hspace{0.4\columnwidth}\hspace{-0.4\twocolumnwidth}
T^W_+\hat a^{\dag}(t_1)\cdots\hat a^{\dag}(t_m)\hat a(t_{m+1}) \cdots\hat a(t_{m+n})\hat a(t) \\ = 
\frac{1}{2} \big[
{\hat{\mathcal P}},\hat a(t)
\big]_+ -i \sum_{k=1}^m{\hat{\mathcal P}}'_k G^W_{++}(t-t_k)
. 
\label{eq:IndStep} 
\hspace{0.4\columnwidth}\hspace{-0.4\twocolumnwidth}
\end{multline}%
We now use the induction assumption and expand ${\hat{\mathcal P}}$ and all ${\hat{\mathcal P}}'_k$ according to the symmetric Wick theorem. By virtue of (\ref{eq:TWRec}) the anticommutator in (\ref{eq:IndStep}) is then a sum of time-symmetric products. It gives us the sum of terms where contractions exclude $\hat a(t)$. The sum in (\ref{eq:IndStep}) deliveres the terms where $\hat a(t)$ is involved in a contraction. It is straightforward to verify that all terms reguired by the symmetric Wick theorem are recovered this way, each occuring only once. 

The case when the earliest term is $\hat a^{\dag}(t)$ is treated by simply swapping in the above $\hat a \leftrightarrow\hat a^{\dag}$. Equation (\ref{eq:Contr}) is replaced by 
\begin{gather} 
\begin{aligned}
\frac{1}{2}\left[
\hat a(t_k),\hat a^{\dag}(t)
\right] & = -i G^W_{++}(t_k-t), & t<t_k
\end{aligned}%
\end{gather}%
which indeed holds, cf.\ Eq.\ (\ref{eq:SCont}). The symmetric Wick theorem has thus been proven. We note that the regularisation we mentioned after Eq.\ (\ref{eq:Tp}) results, in particular, in $G^W_{++}(0) = 0$ without mathematical ambiguity. Hence the caveat of Wick's theorem, that ``no contractions should occur between operators with equal time arguments,'' also applies to the symmetric Wick theorem. 
\subsection{Symmetric Wick theorem for the closed-time-loop ordering%
}\label{ch:SWSchwinger}
\begin{figure}[t] 
\begin{center} 
\includegraphics[width=0.75\columnwidth]{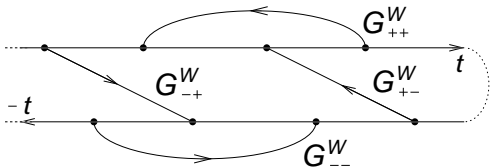}
\end{center} 
\caption{The C-contour and the contractions. The contractions are directed from creation to annihilation, cf. Eq.\ (\ref{eq:GWpm}).}
\label{fig:C}
\end{figure} 
If the initial state of the system is not vacuum, closed equations of motion cannot be written in terms of averages of the time-ordered operator products only, cf., e.g., Ref.\ \cite{KamenevLevchenko}. The necessary type of ordering was introduced, rather indirectly, in Schwinger's seminal paper \cite{SchwingerC}, and later applied to developing diagram techniques for nonequilibrium nonrelativistic quantum problems by Konstantinov and Perel \cite{Perel}. Extension to relativistic problems was given by Keldysh \cite{Keldysh} under whose name the approach became known. We note that the concept of the {\em closed time loop\/} and the related operator ordering is much more general than the Keldysh diagram techniques commonly associated with it. As was shown in paper \cite{BWO}, it underlies such a phase-space concept as the time-normal ordering of Glauber and Kelly and Kleiner. In this paper we investigate the relation between the closed-time-loop and the symmetric orderings. 

The closed-time-loop operator ordering is commonly defined as an ordering on the so-called C-contour which can be seen in Figs.\ \ref{fig:TW} and \ref{fig:C}. The C-contour travels from $t=-\infty$ to $t=+\infty$ (forward branch) and then back to $t=-\infty$ (reverse branch). The operators are formally assigned an additional binary argument which distinguishes operators on the forward branch from those on the reverse branch. ``Earlier'' and ``later'' are then generalised to match the travelling rule along the C-contour. All operators on the forward branch are by definition ``earlier'' than those on the reverse branch and go to the right. Among themselves, the operators on the forward branch are set from right to left in the order of increasing time arguments ($T^W_+$-ordered), while those on the reverse branch are set from right to left in the order of decreasing time arguments ($T^W_-$-ordered). Using that Hermitian conjugation inverts the time order of factors, the $T^W_-$-ordering may be defined as 
\begin{gather} 
\begin{aligned}
T^W_- (\cdots) = \big[T^W_+ (\cdots)^{\dag}\big]^{\dag}. 
\end{aligned}%
\label{eq:TpConj} 
\end{gather}%
The closed-time-loop-ordered operator product may thus be alternatively defined as a {\em double-time-ordered\/} product, 
\begin{gather} 
\begin{aligned}
T^W_- {\hat{\mathcal P}}_-\,T^W_+ {\hat{\mathcal P}}_+, 
\end{aligned}%
\label{eq:TpTm} 
\end{gather}%
where ${\hat{\mathcal P}}_{\pm}$ are two independent operator products. In terms of the C-contour, the operators in ${\hat{\mathcal P}}_+$ are visualised as positioned on the forward branch of the C-contour, and those in ${\hat{\mathcal P}}_-$---on the reverse branch. 
Factors in a double-time-ordered product are always arranged in a Schwinger sequence, cf.\ the opening paragraph of section \ref{ch:GenTW}; for specifications pertaining to coinciding time arguments we refer the reader to section \ref{ch:SecCaus}. 
In this paper we employ Eq.\ (\ref{eq:TpTm}) as a primary definition and use the concept of closed time loop only for illustration purposes. 

Extension of the symmetric Wick theorem to the double time ordering employs the linear order of the C-contour. A formal generalisation of the symmetric Wick theorem to an arbitrary linearly ordered set may be found in the appendix. Hori's form of the symmetric Wick theorem given by Eq.\ (\ref{eq:DWC1}) below, of which the derivation is our goal, is in fact a particular case of the general structural relation (\ref{eq:52HP}). Nonetheless we believe that the direct proof we present here will benefit the reader. 

The symmetric Wick theorem is generalised to the double-time ordering by making the symmetric contraction dependent on the positioning of the contracted pair in the double-time-ordered product (or, which is the same, on the C-contour, cf.\ Fig.\ \ref{fig:C}). As a result we recover four contractions: 
\begin{gather} 
\begin{aligned}
-i G^W_{++}(t-t') &= T^W_+\hat a(t)\,\hat a^{\dag}(t') 
- {\cal T}^W\!\hat a(t)\hat a^{\dag}(t'), 
\\
-i G^W_{--}(t-t') &= T^W_-\hat a(t)\,\hat a^{\dag}(t') 
- {\cal T}^W\!\hat a(t)\hat a^{\dag}(t'), 
\\
-i G^W_{-+}(t-t') &= T^W_-\hat a(t)\, T^W_+\hat a^{\dag}(t') 
- {\cal T}^W\!\hat a(t)\hat a^{\dag}(t'), 
\\
-i G^W_{+-}(t-t') &= T^W_-\hat a^{\dag}(t')\, T^W_+\hat a(t) 
- {\cal T}^W\!\hat a(t)\hat a^{\dag}(t') 
. 
\end{aligned}%
\label{eq:GWpm} 
\end{gather}%
Unlike in Ref.\ \cite{BWO}, there are four nonzero symmetric contractions, not three. To make the genesis of the contractions clearer we retained the orderings also where they are redundant, e.g., 
$T^W_-\hat a(t)\, T^W_+\hat a^{\dag}(t') =\hat a(t)\hat a^{\dag}(t')$. The contraction $-i G^W_{++}(t)$ is exactly that defined by (\ref{eq:GWpp}) explaining the notation. As c-number kernels, $-i G^W_{++}(t)$ and $-i G^W_{--}(t)$ are coupled by Hermitian conjugation, 
while $-i G^W_{+-}(t)$ and $-i G^W_{-+}(t)$ are Hermitian: 
\begin{gather} 
\begin{aligned}
-i G^W_{--}(t-t') &= \big[-i G^{W}_{++}(t'-t)\big]^*, \\ 
-i G^W_{+-}(t-t') & = \big[-i G^{W}_{+-}(t'-t)\big]^*, 
\\ 
-i G^W_{-+}(t-t') & = \big[-i G^{W}_{-+}(t'-t)\big]^* 
. 
\end{aligned}%
\label{eq:GpmConj} 
\end{gather}%
In obtaining this we used (\ref{eq:TWConj}) as well as the similar relation for the double-time ordering, 
\begin{gather} 
\begin{aligned}
\big[T^W_- {\hat{\mathcal P}}_-\,T^W_+ {\hat{\mathcal P}}_+\big]^{\dag}= 
T^W_- {\hat{\mathcal P}}_+^{\dag}\,T^W_+ {\hat{\mathcal P}}_-^{\dag}
, 
\end{aligned}%
\label{eq:TpTmConj} 
\end{gather}%
cf.\ Eq.\ (\ref{eq:TpConj}). For the oscillator, 
\begin{gather} 
\begin{aligned}
-i G^W_{++}(t) &= i G^W_{--}(t) 
= \varepsilon (t) \text{e}^{-i\omega_0t}, 
\\
-i G^W_{-+}(t) &=i G^W_{+-}(t) 
= \frac{1}{2} \text{e}^{-i\omega_0t}, 
\end{aligned}%
\label{eq:GWpmExp} 
\end{gather}%
cf.\ Eq.\ (\ref{eq:SCont}). 

With the contractions defined by (\ref{eq:GWpm}) the symmetric Wick theorem reads: 
\begin{center}
\parbox{0.9\columnwidth}{{\em A double-time-ordered product of interaction-picture operators equals a sum of time-symmetrically ordered products with all possible {\em symmetric contractions\/}, including the term without contractions.\/}}
\end{center}
This includes the time-ordered products as a special case. Importantly, the generalisation to the double-time-ordering makes the symmetric Wick theorem invariant under Hermitian conjugation. This follows from Eqs.\ (\ref{eq:TWConj}) and (\ref{eq:TpTmConj}), supplemented by the observation that Hermitian conjugation turns every Schwinger time sequence into a reversed sequence, so that arguments of all contractions must change sign. Equations (\ref{eq:GpmConj}) show that this is exactly the effect the compex conjugation has on the contractions, including the replacement $G^W_{++}\leftrightarrow G^W_{--}$. It is therefore only necessary to generalise the inductive step in the above proof to the case when the ``spare'' operator is the earliest one under the $T^W_+$ ordering. Again, let us firstly assume that this operator is $\hat a(t)$. Equation (\ref{eq:CA}) holds with ${\hat{\mathcal P}}$ now being a double-time-ordered product, 
\begin{gather} 
\begin{aligned}
{\hat{\mathcal P}} = T^W_- {\hat{\mathcal P}}_- \,T^W_+ {\hat{\mathcal P}}_+. 
\end{aligned}%
\end{gather}%
Equation (\ref{eq:Comm}) applies as well, by assuming that ${\hat{\mathcal P}}'_k$ ``knows'' whether $\hat a ^{\dag}(t_k)$ originates from ${\hat{\mathcal P}}_-$ or from ${\hat{\mathcal P}}_+$. The critical observation is that the contractions depend in fact only on the visual order of times, so that, remembering that $t<t_k$, 
\begin{gather} 
\begin{aligned}
\frac{1}{2}\left[
\hat a^{\dag}(t_k),\hat a(t)
\right] = -i G^W_{++}(t-t_k) = -i G^W_{+-}(t-t_k). 
\end{aligned}%
\label{eq:2Contr} 
\end{gather}%
With this observation equation (\ref{eq:IndStep}) is replaced by 
\begin{multline} 
\hspace{0.4\columnwidth}\hspace{-0.4\twocolumnwidth}
T^W_- {\hat{\mathcal P}}_-\, T^W_+ {\hat{\mathcal P}}_+\hat a(t) = 
\frac{1}{2} \big[
{\hat{\mathcal P}},\hat a(t)
\big]_+ 
\\ 
-i \sum_{t_k\in{\hat{\mathcal P}}_-} {\hat{\mathcal P}}'_k G^W_{+-}(t-t_k) 
-i \sum_{t_k\in{\hat{\mathcal P}}_+} {\hat{\mathcal P}}'_k G^W_{++}(t-t_k) 
, 
\label{eq:IndStepD} 
\hspace{0.4\columnwidth}\hspace{-0.4\twocolumnwidth}
\end{multline}%
where $\sum_{t_k\in{\hat{\mathcal P}}_{+}}$ means summation over all $k$ such that $\hat a^{\dag}(t_k)$ originates from ${\hat{\mathcal P}}_{+}$, and similarlly for ${\hat{\mathcal P}}_{-}$. The case when the earliest operator under the $T^W_+$-ordering is $\hat a^{\dag}(t)$ is treated similarly, the ``critical observation'' in this case being that, for $t_k>t$,
\begin{gather} 
\begin{aligned}
\frac{1}{2}\left[
\hat a(t_k),\hat a^{\dag}(t)
\right] = -i G^W_{++}(t_k-t) = -i G^W_{-+}(t_k-t). 
\end{aligned}%
\label{eq:2ContrX} 
\end{gather}%
This completes the inductive step of the proof. The symmetric Wick theorem has thus also been proven for the double time ordering. 
\subsection{Functional form of the symmetric Wick theorem}\label{ch:SWHori}
One technical tool in Ref.\ \cite{BWO} was Hori's form of Wick's theorem expressing it as an application of a functional differential operator, cf.\ Eq.\ (18) in \cite{BWO}. 
Except for the redefinition of contractions, the symmetric Wick theorem coincides with Wick's theorem proper. 
This makes the functional form of Wick's theorem equally applicable to the symmetric Wick theorem. 
All we need is to redefine the differential operator $\Delta_C$ given by Eq.\ (19) in \cite{BWO} as 
\begin{multline} 
\hspace{0.4\columnwidth}\hspace{-0.4\twocolumnwidth}
\Delta ^W_C\bigg [ 
\frac{\delta}{\delta a_+},
\frac{\delta}{\delta \bar a_+},
\frac{\delta}{\delta a_-},
\frac{\delta}{\delta \bar a_-}
 \bigg ] \\ = -i \sum_{c,c'=\pm} \int dt dt'
G^W_{cc'}(t-t')
\frac{\delta^2}{\delta a_{c}(t)\delta {\bar a}_{c'}(t')} , 
\hspace{0.4\columnwidth}\hspace{-0.4\twocolumnwidth}%
\label{eq:DWC1} 
\end{multline}%
where $a_{\pm}(t),\bar a_{\pm}(t)$ are four independent c-number fields. Square brackets signify functionals. The symmetric contractions were defined by Eqs.\ (\ref{eq:GWpm}). With these replacements we find Hori's form of the symmetric Wick theorem, 
\begin{multline} 
\hspace{0.4\columnwidth}\hspace{-0.4\twocolumnwidth}
T_- P_-\big [ 
\hat a,\hat a^{\dag}
 \big ] \,T_+P_+\big [ 
\hat a,\hat a^{\dag}
 \big ] \\ = {\cal T}^W\!\bigg \{ 
\exp 
\Delta ^W_C\bigg [ 
\frac{\delta}{\delta a_+},
\frac{\delta}{\delta \bar a_+},
\frac{\delta}{\delta a_-},
\frac{\delta}{\delta \bar a_-}
 \bigg ]
\\ \times 
P_-\big [ 
a_-, \bar a_-
 \big ]P_+\big [ 
a_+, \bar a_+
 \big ] 
\Big|_{ a \to {\hat a}} 
 \bigg \}\ . 
\hspace{0.4\columnwidth}\hspace{-0.4\twocolumnwidth}%
\label{eq:HoriW} 
\end{multline}%
Here, $P_{\pm}[\cdot,\cdot]$ are two independent functionals of time-dependent fields, and $ a \to {\hat a}$ stands for the replacement of the c-numbers by operators, 
\begin{gather} 
\begin{aligned}
a_{\pm}(t) &\to {\hat a}(t) 
, &
{\bar a}_{\pm}(t) &\to { {\hat a}}^{\dag}(t)
. 
\end{aligned}%
\label{eq:C2A} 
\end{gather}%
Once again, exact meaning to Eq.\ (\ref{eq:HoriW}) will be assigned by regularisations in section \ref{ch:SecCaus}. 

\section{The closed-time-loop formalism}\label{ch:Keldysh}
\subsection{The model}\label{ch:Mod}
All definitions and majority of formulae in this and the following sections do not depend on the system Hamiltonian and can be applied to any bosonic system. To be specific, and to maintain connections with our previous paper \cite{Bettina} we demostrate our techniques for a one-dimensional Bose-Hubbard (1D BH) chain. This system combines simplicity of the formal model with practical relevance: starting from the seminal paper by Jaksch {\em et al.\/} \cite{Jaksch} the BH model evolved into a standard approach to cold atoms trapped in optical lattices. The fact that a particular relation below is specific to this model will always be stated explicitly. The simplest version of the 1D BH model describes a chain of $N$ anharmonic oscillators with Josephson's tunneling (``hopping'') between the nearest neighbours, 
\begin{multline} 
\hspace{0.4\columnwidth}\hspace{-0.4\twocolumnwidth}
\hat H = \sum_{k=1}^N \big[\omega_0\hat a^{\dag}_k\hat a_k + 
\frac{\kappa }{2} \hat a^{\dag 2}_k\hat a^2_k 
\\ - J(\hat a_{k}^{\dag}\hat a_{k+1} + \hat a_{k+1}^{\dag}\hat a_{k})\big] 
,
\label{eq:BHH} 
\hspace{0.4\columnwidth}\hspace{-0.4\twocolumnwidth}
\end{multline}%
where $\hat n_k = \hat a^{\dag}_k\hat a _k$, and $\hat a_k^{\dag},\hat a_k$ is the standard creation-annihilation pair for the $k$-th site, $\big[\hat a_k,\hat a^{\dag}_{k'}\big] = \delta_{kk'}$. The indices are understood modulo $N$, $\hat a_{N+1} = \hat a_1$; in other words, we consider a ring rather than a chain. As usual we break Hamiltonian (\ref{eq:BHH}) into the ``free'' and ``interaction'' Hamiltonians, 
\begin{gather} 
{\hat{\mathcal H}} = {\hat{\mathcal H}}_0 + {\hat{\mathcal H}}_{\text{I}}, \ \ 
{\hat{\mathcal H}}_{0} = \sum_{k=1}^N{\hat{\mathcal H}}_{0k} = 
\sum_{k=1}^N \omega_0\hat n{_k}.
\label{eq:H0kI} 
\end{gather}%
The free-field and the Heisenberg operators, $\hat a_k(t),\hat a_k^{\dag}(t)$ and ${\hat{\mathcal A}}_k(t),{\hat{\mathcal A}}_k^{\dag}(t)$ are defined in detailed analogy to section \ref{ch:Bas}.
We exclude hopping from the free Hamiltonian making our reasoning most easlily adaptable to arbitrary multi-mode bosonic systems. 
In particular this allows the analyses in the previous two sections to be generalised simply by applying them mode-wise. 
\subsection{Perel-Keldysh Green functions}
We now introduce the formal framework that will serve us throughout the rest of the paper. A brief historical introduction into the Schwinger-Perel-Keldysh closed-time-loop formalism was given at the beginning of section \ref{ch:SWSchwinger}. The basic quantity in this formalism is a double-time-ordered Green function 
\begin{align} 
&\Big \langle 
T^W_- 
{\hat{\mathcal A}}_{k_1}(t_1)
\cdots 
{\hat{\mathcal A}}_{k_{m}}(t_m)
{\hat{\mathcal A}}_{k_{m+1}}^{\dag}(t_{m+1})
\cdots 
{\hat{\mathcal A}}_{k_{m+\bar m}}^{\dag}(t_{m+\bar m}) 
\nonumber \\ & \times 
T^W_+ 
{\hat{\mathcal A}}_{k'_1}(t'_1)
\cdots 
{\hat{\mathcal A}}_{k'_{n}}(t'_n)
{\hat{\mathcal A}}_{k'_{n+1}}^{\dag}(t'_{n+1})
\cdots 
{\hat{\mathcal A}}_{k'_{n+\bar n}}^{\dag}(t'_{n+\bar n}) 
 \Big \rangle 
,
\label{eq:GF} 
\end{align}%
where $m,\bar m, n, \bar n\geq 0$ are arbitrary integers. Specifications of Eq.\ (\ref{eq:GF}) for equal time arguments are postponed till section \ref{ch:SecCaus}, cf.\ the remarks at the end of section \ref{ch:SWDyson}. A convenient interface to the whole assemblage of functions (\ref{eq:GF}) is their generating, or characteristic, functional 
\begin{multline} 
\hspace{0.4\columnwidth}\hspace{-0.4\twocolumnwidth}
\Xi^W 
\big [ 
{\mbox{\rm\boldmath$\zeta$}}_-,\bar{{\mbox{\rm\boldmath$\zeta$}}}_-,{\mbox{\rm\boldmath$\zeta$}}_+,\bar{{\mbox{\rm\boldmath$\zeta$}}}_+
 \big ] = 
\Big \langle 
T^W_-\exp
\big(
-i\bar{{\mbox{\rm\boldmath$\zeta$}}}_- 
{\mbox{\rm\boldmath$\mathcal{A}$}}\hspace{-1\vAlength}
\hspace{0.27\vAlength}\hat{\phantom{{\mbox{\rm\boldmath$\mathcal{A}$}}}}
\hspace{-0.27\vAlength}-
i{\mbox{\rm\boldmath$\zeta$}}_- 
{\mbox{\rm\boldmath$\mathcal{A}$}}\hspace{-1\vAlength}
\hspace{0.27\vAlength}\hat{\phantom{{\mbox{\rm\boldmath$\mathcal{A}$}}}}
\hspace{-0.27\vAlength}^{\dag}
 \big) \\ \times 
T^W_+\exp
\big(
i\bar{{\mbox{\rm\boldmath$\zeta$}}}_+ 
{\mbox{\rm\boldmath$\mathcal{A}$}}\hspace{-1\vAlength}
\hspace{0.27\vAlength}\hat{\phantom{{\mbox{\rm\boldmath$\mathcal{A}$}}}}
\hspace{-0.27\vAlength}
+i{\mbox{\rm\boldmath$\zeta$}}_+ 
{\mbox{\rm\boldmath$\mathcal{A}$}}\hspace{-1\vAlength}
\hspace{0.27\vAlength}\hat{\phantom{{\mbox{\rm\boldmath$\mathcal{A}$}}}}
\hspace{-0.27\vAlength}^{\dag}
 \big) 
 \Big \rangle 
, 
\hspace{0.4\columnwidth}\hspace{-0.4\twocolumnwidth}%
\label{eq:XiW} 
\end{multline}%
where $\zeta_{k\pm}(t),\zeta_{k\pm}^{\dag}(t)$ are four arbitrary c-number functions per mode. Notationally we treat them, as well as the mode operators, as vectors. We follow the sign conventions of Ref.\ \cite{APIII}. 
To emphasise the structural side of our analyses we employ a condensed notation, 
\begin{align} 
\begin{aligned} 
{\mbox{\rm\boldmath$x$}}{\mbox{\rm\boldmath$y$}} &= \sum_{k=1}^N \int dt\, {x}_k(t)
{y}_k(t), \\
{\mbox{\rm\boldmath$x$}} X{\mbox{\rm\boldmath$y$}} &= \sum_{k,k'=1}^N \int dtdt'\, {x}_k(t)
X_{kk'}(t-t')
{y}_{k'}(t') 
. 
\end{aligned} 
\label{eq:Cond} 
\end{align}%
Here, ${x}_k(t)$ and ${y}_k(t)$ are arbitrary functions and \mbox{$X_{kk'}(t-t')$} is an arbitrary kernel; for q-numbers the order of factors matters. 
\subsection{Closed perturbartive relations}\label{ch:ClPert}
The main advantage of the functional framework is that the {\em universal structural\/} part of the perturbative calculations can be expressed as a small set of {\em closed perturbative relations\/}. This was demonstrated in \cite{BWO} for the normal ordering, and will be demonstrated in this paper for the symmetric ordering. 
We wish to construct a perturbative formula the generating functional (\ref{eq:XiW}). 
By the same means as equation (24) was found in Ref.\ \cite{BWO} we rewrite Eq.\ (\ref{eq:XiW}) in the interaction picture as 
\begin{multline} 
\hspace{0.4\columnwidth}\hspace{-0.4\twocolumnwidth}
\Xi^W \big[{\mbox{\rm\boldmath$\zeta$}}_-,\bar{{\mbox{\rm\boldmath$\zeta$}}}_-,{\mbox{\rm\boldmath$\zeta$}}_+,{\mbox{\rm\boldmath$\zeta$}}_+^{\dag}\big] 
\\ = 
\Big\langle
T_-^W \exp\left(
-i\bar{{\mbox{\rm\boldmath$\zeta$}}}_- \hat{{\mbox{\rm\boldmath$a$}}}
-i{\mbox{\rm\boldmath$\zeta$}}_- \hat{{\mbox{\rm\boldmath$a$}}}^{\dag}
+ i L_{\text{I}}^{W} 
\big [ 
\hat{{\mbox{\rm\boldmath$a$}}},\hat{{\mbox{\rm\boldmath$a$}}}^{\dag}
 \big ] 
\right) \\ \times 
T_+^W \exp\left(
i\bar{{\mbox{\rm\boldmath$\zeta$}}}_+ \hat{{\mbox{\rm\boldmath$a$}}}+
i{\mbox{\rm\boldmath$\zeta$}}_+ \hat{{\mbox{\rm\boldmath$a$}}}^{\dag}
- i 
L_{\text{I}}^W
\big [ 
\hat{{\mbox{\rm\boldmath$a$}}},\hat{{\mbox{\rm\boldmath$a$}}}^{\dag}
 \big ] 
\right) 
\Big\rangle. 
\hspace{0.4\columnwidth}\hspace{-0.4\twocolumnwidth}%
\label{eq:XiPert} 
\end{multline}%
Here, $L_{\text{I}}^{W}$ is a functional of two arguments, 
\begin{gather} 
\begin{aligned}
L_{\text{I}}^W
\big [ 
\hat{{\mbox{\rm\boldmath$a$}}},\hat{{\mbox{\rm\boldmath$a$}}}^{\dag}
 \big ] 
= \int dt\, 
H_{\text{I}}^W 
\big(
\hat{{\mbox{\rm\boldmath$a$}}}(t),\hat{{\mbox{\rm\boldmath$a$}}}^{\dag}(t) 
 \big) 
, 
\end{aligned}%
\label{eq:SM} 
\end{gather}%
where the c-number {\em function\/} $H_{\text{I}}^W (
\cdot,\cdot 
 ) 
$ is the {\em symmetric form\/} of the interaction Hamiltonian, 
\begin{gather} 
\begin{aligned}
\hat H_{\text{I}} = \text{W}H_{\text{I}}^W
\big(
\hat{{\mbox{\rm\boldmath$a$}}},\hat{{\mbox{\rm\boldmath$a$}}}^{\dag}
 \big) 
. 
\end{aligned}%
\label{eq:HWI} 
\end{gather}%
The last equation is written in the {Schr\"odinger}\ picture. Note that 
\begin{align} 
\begin{aligned} 
{\hat{\mathcal U}}_{\text{I}} = T_+^W \exp
\big(
-iL_{\text{I}}^W
\big [ 
\hat{{\mbox{\rm\boldmath$a$}}},\hat{{\mbox{\rm\boldmath$a$}}}^{\dag}
 \big ]
 \big) 
\end{aligned} 
\label{eq:33a} 
\end{align}%
is the interaction-picture S-matrix. 
Equation (\ref{eq:HWI}) is general and holds for any interaction. 
For the Bose-Hubbard chain, 
\begin{multline} 
\hspace{0.4\columnwidth}\hspace{-0.4\twocolumnwidth}
\hat{H}_{\text{I}} 
= \sum_{k=1}^N \bigg[
\frac{\kappa }{2}\hat a^{\dag 2}_k\hat a_k^2
- 
J\Big(\hat a_{k}^{\dag}\hat a_{k+1} +\hat a_{k+1}^{\dag}\hat a_{k}\Big)\bigg] 
\\ = 
W\sum_{k=1}^N \bigg [ 
\frac{\kappa }{2 }\hat a^{\dag 2}_k\hat a_k^2 - 
\kappa\hat a^{\dag}_k\hat a_k + \frac{ \kappa }{4}
\\ - 
J\Big(
\hat a_{k}^{\dag}\hat a_{k+1} +\hat a_{k+1}^{\dag}\hat a_{k} 
 \Big)
 \bigg ] \equiv \hat H^W_{\text{I}}, 
\hspace{0.4\columnwidth}\hspace{-0.4\twocolumnwidth}%
\label{eq:HWDef} 
\end{multline}%
cf.\ Eq.\ (\ref{eq:BHH}). 
In deriving this we have used that, for the harmonic oscillator, 
\begin{align} 
\begin{aligned} 
\text{W} 
\left \{ 
{\hat{a}^{\dagger}} {\hat{a}} 
\right \}
 &=\frac{1}{2}\left ( 
{\hat{a}^{\dagger}} {\hat{a}} +\hat{a} 
\hat{a}^{\dagger}\right ) , 
\\ 
\text{W} 
\left \{ 
{\hat{a}^{\dagger 2}} {\hat{a}}^{2} 
\right \} 
 &= 
\frac{1}{6}\big ( 
{\hat{a}^{\dagger 2}} {\hat{a}}^{2} +\hat{a}^{\dagger}\hat{a} 
\hat{a}^{\dagger}\hat{a} + 
\hat{a}^{\dagger}{\hat{a}}^{2}\hat{a}^{\dagger} 
\\ &\quad+\hat{a} 
{\hat{a}^{\dagger 2}}\hat{a} +\hat{a}\hat{a}^{\dagger} 
\hat{a}\hat{a}^{\dagger} +{\hat{a}}^{2} {\hat{a}^{\dagger 2}} 
\big ). 
\end{aligned} 
\label{eq:9a} 
\end{align}%
The reason why the symmetric form of the interaction must be used is exactly that why in \cite{BWO} we had to use the normal form. The key property of the $T_+$-ordering in Ref.\ \cite{BWO} 
is that it does not affect the single-time normally ordered operator products, so that, in particular, 
\begin{align} 
T_+ 
\big\{ {\mbox{\rm\boldmath$:$}}
\hat H _{\text{I}} (t) {\mbox{\rm\boldmath$:$}}
\big\} = \ {\mbox{\rm\boldmath$:$}}
\hat H _{\text{I}} (t){\mbox{\rm\boldmath$:$}} . 
\end{align}%
We have the same consistency between 
$T^W_+$ and $\hat H^W_{\text{I}} (t)$: 
\begin{align} 
T_+^W 
\big\{ 
\hat{H}^W_{\text{I}} (t) 
\big\} = 
\hat{H}^W_{\text{I}} (t). 
\end{align}%
Since the order of operators under $T_+^W$ is {\em fully\/} decided by this ordering, the only way to prevent it from redefining the interaction is to put the latter into symmetric form. 

Operators may be eliminated from Eq.\ (\ref{eq:XiPert}) altogether by, firstly, applying the symmetric Wick theorem (\ref{eq:HoriW}) so as to bring the operator construct under the quantum averaging to a (time-)symmetrically ordered form, and, secondly, using a multimode generalisation of Eq.\ (\ref{eq:WMom}) to express the average. The result of this transformation reads 
\begin{widetext} 
\begin{multline} 
\Xi^W 
\big [ 
{\mbox{\rm\boldmath$\zeta$}}_-,\bar{{\mbox{\rm\boldmath$\zeta$}}}_-,{\mbox{\rm\boldmath$\zeta$}}_+,\bar{{\mbox{\rm\boldmath$\zeta$}}}_+
 \big ] = 
\int 
\frac{d^2 \alpha_{1} }{\pi } 
\cdots 
\frac{d^2 \alpha_{N} }{\pi } 
W\big(
{\mbox{\rm\boldmath$\alpha$}} , 
{\mbox{\rm\boldmath$\alpha$}}^* 
 \big) 
\Bigg \{ 
\exp
\Delta ^W_C\bigg [ 
\frac{\delta}{\delta a_+},
\frac{\delta}{\delta \bar a_+},
\frac{\delta}{\delta a_-},
\frac{\delta}{\delta \bar a_-}
 \bigg ] \\ \times 
\exp\Big(
-i\bar{{\mbox{\rm\boldmath$\zeta$}}}_- {{\mbox{\rm\boldmath$a$}}}_- 
-i{\mbox{\rm\boldmath$\zeta$}}_- \bar{{\mbox{\rm\boldmath$a$}}}_- 
+ i L_{\text{I}}^{W} 
\big [ 
{{\mbox{\rm\boldmath$a$}}}_-,\bar{{\mbox{\rm\boldmath$a$}}}_- 
 \big ] 
+i\bar{{\mbox{\rm\boldmath$\zeta$}}}_+ {{\mbox{\rm\boldmath$a$}}}_+ +
i{\mbox{\rm\boldmath$\zeta$}}_+ \bar{{\mbox{\rm\boldmath$a$}}}_+ 
- i 
L_{\text{I}}^W
\big [ 
{{\mbox{\rm\boldmath$a$}}}_+,\bar{{\mbox{\rm\boldmath$a$}}}_+ 
 \big ] 
 \Big)
 \Bigg \} 
\Big|_{a\to\alpha} 
. 
\label{eq:32a} 
\end{multline}%
\end{widetext}%
In the above, 
\begin{align} 
\begin{aligned} 
 &{\mbox{\rm\boldmath$\alpha$}}(t) = {\mbox{\rm\boldmath$\alpha$}}\text{e}^{-i\omega_0 t} , &
 &{\mbox{\rm\boldmath$\alpha$}}^*(t) = {\mbox{\rm\boldmath$\alpha$}}^*\text{e}^{i\omega_0 t} , 
\end{aligned} 
\label{eq:31a} 
\end{align}%
$W({\mbox{\rm\boldmath$\alpha $}},{\mbox{\rm\boldmath$\alpha $}}^*)$ is the multimode Wigner function (cf.\ Eq.\ (\ref{eq:WMom})), and $a\to\alpha $ stands for the replacement, 
\begin{align} 
\begin{aligned} 
 &{\mbox{\rm\boldmath$a$}}_{\pm}(t)\to{\mbox{\rm\boldmath$\alpha$}}(t), &
 &\bar{{\mbox{\rm\boldmath$a$}}}_{\pm}(t)\to{\mbox{\rm\boldmath$\alpha$}}^*(t) . 
\end{aligned} 
\label{eq:85JZ} 
\end{align}%

Not to be confused by Eq.\ (\ref{eq:32a}), note 
that the functional differential operation within the curly brackets is applied under the condition that ${\mbox{\rm\boldmath$a$}}_{\pm}(t),\bar{{\mbox{\rm\boldmath$a$}}}_{\pm}(t)$ are four arbitrary c-number vector functions. These functions are then replaced pairwise by ${\mbox{\rm\boldmath$\alpha $}}(t),{\mbox{\rm\boldmath$\alpha $}}^*(t)$, which depend only on the initial condition ${\mbox{\rm\boldmath$\alpha $}}$. 
This replacement turns the {\em functional expression\/} in curly brackets into a {\em function\/} of the {\em initial condition\/} ${\mbox{\rm\boldmath$\alpha $}}$, making averaging over the Wigner {\em function\/} $W({\mbox{\rm\boldmath$\alpha $}},{\mbox{\rm\boldmath$\alpha $}}^*)$ meaningful. 

\subsection{Formal solution to the quantum-statistical response problem}\label{ch:QResp} 
Following Kubo, we add a source term to the Hamiltonian (\ref{eq:BHH}), 
\begin{gather} 
\begin{aligned}
\hat {H}' =\hat {H} - \sum_{k=1}^N\big [ 
\hat a_k^{\dag}s_k(t) +\hat a_k s_k^*(t)
 \big ] . 
\end{aligned}%
\label{eq:Hpr} 
\end{gather}%
The Heisenberg operators corresponding to $\hat H '$ will be denoted as ${{\hat{\mathcal A}}'}_k(t)$. 
Similar to (\ref{eq:XiW}), we introduce a characteristic functional for the double-time-ordered averages of ${{\hat{\mathcal A}}}'_k(t),{{\hat{\mathcal A}}}^{\prime\dag}(t)$: 
\begin{multline} 
\hspace{0.4\columnwidth}\hspace{-0.4\twocolumnwidth}
\Xi ^W
{\big[ 
{\mbox{\rm\boldmath$\zeta$}}_-,\bar{{\mbox{\rm\boldmath$\zeta$}}}_-,{\mbox{\rm\boldmath$\zeta$}}_+,\bar{{\mbox{\rm\boldmath$\zeta$}}}_+ \big| {\mbox{\rm\boldmath$s$}},{\mbox{\rm\boldmath$s$}}^*
 \big]} \\ = 
\Big \langle 
T_-^W\exp
\big(
-i\bar{{\mbox{\rm\boldmath$\zeta$}}}_- 
{\mbox{\rm\boldmath$\mathcal{A}$}}\hspace{-1\vAlength}
\hspace{0.27\vAlength}\hat{\phantom{{\mbox{\rm\boldmath$\mathcal{A}$}}}}
\hspace{-0.27\vAlength}' -i
{\mbox{\rm\boldmath$\zeta$}}_- 
{\mbox{\rm\boldmath$\mathcal{A}$}}\hspace{-1\vAlength}
\hspace{0.27\vAlength}\hat{\phantom{{\mbox{\rm\boldmath$\mathcal{A}$}}}}
\hspace{-0.27\vAlength}^{\prime\dag} 
 \big) \\ \times 
T_+^W\exp
\big(
i\bar{{\mbox{\rm\boldmath$\zeta$}}}_+ 
{\mbox{\rm\boldmath$\mathcal{A}$}}\hspace{-1\vAlength}
\hspace{0.27\vAlength}\hat{\phantom{{\mbox{\rm\boldmath$\mathcal{A}$}}}}
\hspace{-0.27\vAlength}' +
i{\mbox{\rm\boldmath$\zeta$}}_+ 
{\mbox{\rm\boldmath$\mathcal{A}$}}\hspace{-1\vAlength}
\hspace{0.27\vAlength}\hat{\phantom{{\mbox{\rm\boldmath$\mathcal{A}$}}}}
\hspace{-0.27\vAlength}^{\prime\dag} 
 \big) 
 \Big \rangle 
. 
\hspace{0.4\columnwidth}\hspace{-0.4\twocolumnwidth}%
\label{eq:XiWS} 
\end{multline}%
The condensed notation we use here was defined by (\ref{eq:Cond}). 

With a taste for the paradoxical, the message of this section may be formulated as, {\em the quantum response problem does not exist, because the information on the response properties of the system is already present in the commutators of the field operator\/} \cite{API,APII,APIII}. Formally, this is expressed by the following relation between the characteristic functionals, 
\begin{multline} 
\hspace{0.4\columnwidth}\hspace{-0.4\twocolumnwidth}
\Xi^W \big[{\mbox{\rm\boldmath$\zeta$}}_-,\bar{{\mbox{\rm\boldmath$\zeta$}}}_-,{\mbox{\rm\boldmath$\zeta$}}_+,{\mbox{\rm\boldmath$\zeta$}}_+^{\dag}
\big| {\mbox{\rm\boldmath$s$}}, {\mbox{\rm\boldmath$s$}}^*\big] 
\\ = 
\Xi^W \big [ 
{\mbox{\rm\boldmath$\zeta$}}_- + {\mbox{\rm\boldmath$s$}} ,
\bar{{\mbox{\rm\boldmath$\zeta$}}}_- + {\mbox{\rm\boldmath$s$}}^*,
{\mbox{\rm\boldmath$\zeta$}}_+ + {\mbox{\rm\boldmath$s$}}, 
{\mbox{\rm\boldmath$\zeta$}}_+^{\dag}+ {\mbox{\rm\boldmath$s$}}^* 
 \big ] 
. 
\hspace{0.4\columnwidth}\hspace{-0.4\twocolumnwidth}%
\label{eq:XiSXi} 
\end{multline}%
This formula is a trivial consequence of the closed perturbative relation (\ref{eq:XiPert}); it suffices to note that adding the source term to the Hamiltonian results in the following replacement in Eq.\ (\ref{eq:32a}), 
\begin{align} 
\begin{aligned} 
L_{\text{I}}^W
\big [ 
{{\mbox{\rm\boldmath$a$}}},\bar{{\mbox{\rm\boldmath$a$}}} 
 \big ] 
 \to 
L_{\text{I}}^W
\big [ 
{{\mbox{\rm\boldmath$a$}}},\bar{{\mbox{\rm\boldmath$a$}}} 
 \big ] 
- \bar{{\mbox{\rm\boldmath$a$}}}{\mbox{\rm\boldmath$s$}} 
- {{\mbox{\rm\boldmath$a$}}}{\mbox{\rm\boldmath$s$}}^* 
. 
\end{aligned} 
\label{eq:28a} 
\end{align}%
Equation (\ref{eq:XiSXi}) shows that all information one needs in order to predict response of a quantum system to an external source is already present in the Heisenberg field operators defined without the source. A closer inspection of Eq.\ (\ref{eq:XiSXi}) reveals that this information is ``stored'' in the commutators of the Heisenberg operators. For details see Refs.\ \cite{API,APII,APIII}. 
\section{The Wigner representation}\label{ch:CommResp}
\subsection{Symmetric Wick theorem and causality in phase space}%
\label{ch:Gen}
Similar to \cite{BWO}, the starting point of the ensuing derivation is understanding the causal structure of the symmetric contractions (\ref{eq:GWpm}). We begin with the observation that
\begin{gather} 
\begin{aligned}
\bigg(
\omega_0 -i \frac{d}{dt}
 \bigg) 
G_{++}^W(t) = \delta(t) . 
\end{aligned}%
\label{eq:GppAsFG} 
\end{gather}%
This characterises the symmetric contraction $G_{++}^W(t)$ as a Green function of the c-number equation 
\begin{gather} 
\begin{aligned}
\bigg(
\omega_0 -i \frac{d}{dt}
 \bigg) 
a_k(t) = s_k(t) + \bar s_k(t) . 
\end{aligned}%
\label{eq:EqA} 
\end{gather}%
To spare us future redefinitions we wrote (\ref{eq:EqA}) as an equation for the phase-space trajectories $a_k(t)$. We also broke the total source into the given {\em external\/} source $s_k(t)$ and the {\em self-action\/} source $\bar s_k(t)$ accounting for the interactions. 
In general, $\bar s_k(t)$ is quasi-stochastic and field-dependent. 
Note that here we work with a quasi-stochastic equation in the classical phase space, unlike in Ref.\ \cite{BWO} where stochastic equations in the phase space of doubled dimension were considered. As a result, here the reader encounters complex-conjugate field pairs instead of independent field pairs as in ref.\ \cite{BWO}. 

While $G_{++}^W(t)$ is obviously one of Green functions of Eq.\ (\ref{eq:EqA}), it is wrong as far as natural causality is concerned. 
In phase space as well as in classical mechanics physics are associated with {\em causal response\/} defined through 
the {\em retarded\/} Green function, 
\begin{gather} 
\begin{gathered} 
G_{\text{R}}(t) = i \theta (t) \text{e}^{-i\omega_0t}. 
\end{gathered} 
\label{eq:24a} 
\end{gather}%
This quantity also obeys equation (\ref{eq:GppAsFG}) but, unlike $G_{++}^W(t)$, is retarded. Causal solutions of Eq.\ (\ref{eq:EqA}) are specified by replacing it by the integral equation
\begin{gather} 
\begin{aligned}
a_k(t) = \alpha _k(t) + \int dt' G_{\text{R}}(t-t') 
\big [ 
s_k(t') + \bar s_k (t')
 \big ] , 
\end{aligned}%
\label{eq:EqI} 
\end{gather}%
with the condition $a_k(t)\to \alpha _k(t)$ as $t\to -\infty$. 
By analogy with Ref.\ \cite{BWO} we expect the in-field ${\mbox{\rm\boldmath$\alpha $}}(t)$ to be defined by the multimode generalisation of Eq.\ (\ref{eq:WMom}). That is, the initial condition for Eqs.\ (\ref{eq:EqA}) is defined by the initial state of the system. Consistency of these assumptions and the way Eq.\ (\ref{eq:EqA}) is linked to quantum averages remain subject to verification. 
\subsection{The causal transformation}
As in Ref.\ \cite{BWO} we proceed by noticing that all contractions may be expressed by $G_{\text{R}}(t)$. 
Simply by trial and error it is easy to get 
\begin{gather} 
\begin{aligned}
G^W_{++}(t) &= - G^W_{--}(t) = \frac{1}{2} \big[
G_{\text{R}}(t) +G_{\text{R}}^*(-t)
\big] , \\ 
G^W_{-+}(t) &= - G^W_{+-}(t) = \frac{1}{2} \big[G_{\text{R}}(t) - G_{\text{R}}^*(-t)\big] 
. 
\end{aligned}%
\label{eq:GWbyG} 
\end{gather}%
Similar to \cite{BWO}, we are looking for variables which would bring $\Delta ^W_C$ given by (\ref{eq:DWC1}) to the form 
\begin{gather} 
\Delta_C^W = 
\frac{\delta }{\delta {\mbox{\rm\boldmath$a$}}}%
G_{\text{R}}%
\frac{\delta }{\delta 
{\mbox{\rm\boldmath$\xi$}} 
^{ *}%
} 
+ 
\frac{\delta }{\delta {\mbox{\rm\boldmath$a$}}%
^{ *}%
}%
G_{\text{R}}^*%
\frac{\delta }{\delta {\mbox{\rm\boldmath$\xi$}} } 
. 
\label{eq:CausDWC} 
\end{gather}%
We use here condensed notation defined by Eq.\ (\ref{eq:Cond}). 
Equation (\ref{eq:CausDWC}) takes us to the substitution 
\begin{gather} 
\begin{aligned}
{\mbox{\rm\boldmath$a$}}_+(t) &= {\mbox{\rm\boldmath$a$}}(t) - \frac{i{\mbox{\rm\boldmath$\xi$}} (t)}{2} , & 
\bar{{\mbox{\rm\boldmath$a$}}}_+(t) &= {\mbox{\rm\boldmath$a$}}^{ *}(t) -\frac{i{\mbox{\rm\boldmath$\xi$}}^{ *}(t)}{2} 
 , \\ 
{\mbox{\rm\boldmath$a$}}_-(t) &= {\mbox{\rm\boldmath$a$}}(t) + \frac{i{\mbox{\rm\boldmath$\xi$}} (t)}{2} , & 
\bar{{\mbox{\rm\boldmath$a$}}}_-(t) &= {\mbox{\rm\boldmath$a$}}%
^{ *}(t) + \frac{i{\mbox{\rm\boldmath$\xi$}}^{ *}(t)}{2} 
. 
\end{aligned}%
\label{eq:CausAW} 
\end{gather}%
These relations imply that 
\begin{align} 
\begin{aligned} 
 &{\mbox{\rm\boldmath$a$}}_+^*(t) = \bar{{\mbox{\rm\boldmath$a$}}}_-(t), &
 &{\mbox{\rm\boldmath$a$}}_-^*(t) = \bar{{\mbox{\rm\boldmath$a$}}}_+(t). 
\end{aligned} 
\label{eq:86KA} 
\end{align}%
Under this condition the symmetric Wick theorem (\ref{eq:HoriW}) remains valid. It is also consistent with the substitution (\ref{eq:85JZ}), which in variables ${\mbox{\rm\boldmath$a$}}(t),{\mbox{\rm\boldmath$\xi $}}(t)$ becomes, 
\begin{align} 
\begin{aligned} 
 &{\mbox{\rm\boldmath$a$}}(t) \to {\mbox{\rm\boldmath$\alpha $}}(t), &
 &{\mbox{\rm\boldmath$a$}}^*(t) \to {\mbox{\rm\boldmath$\alpha $}}^*(t), &
 &{\mbox{\rm\boldmath$\xi $}}(t)\to 0. 
\end{aligned} 
\label{eq:88KC} 
\end{align}%
In the below we assume that Eq.\ (\ref{eq:86KA}) holds. 
\subsection{Response in the Wigner representation}
Continuing the analogy with Ref.\ \cite{BWO}, we impose the condition, 
\begin{multline} 
\hspace{0.4\columnwidth}\hspace{-0.4\twocolumnwidth}
-i\bar{{\mbox{\rm\boldmath$\zeta$}}}_-(t) {{\mbox{\rm\boldmath$a$}}}_-(t) 
-i{\mbox{\rm\boldmath$\zeta$}}_-(t) \bar{{\mbox{\rm\boldmath$a$}}}_-(t) 
+i\bar{{\mbox{\rm\boldmath$\zeta$}}}_+(t) {{\mbox{\rm\boldmath$a$}}}_+(t) \\ +
i{\mbox{\rm\boldmath$\zeta$}}_+(t) \bar{{\mbox{\rm\boldmath$a$}}}_+(t) = 
{\mbox{\rm\boldmath$\zeta$}}(t){\mbox{\rm\boldmath$a$}}^*(t) 
+ {\mbox{\rm\boldmath$\zeta$}}^* (t)
{\mbox{\rm\boldmath$a$}}(t) 
\\ + {\mbox{\rm\boldmath$\xi$}}(t) 
{\mbox{\rm\boldmath$\sigma$}}^*(t) 
+ {\mbox{\rm\boldmath$\xi$}} ^*(t) 
{\mbox{\rm\boldmath$\sigma$}}(t) ,
\hspace{0.4\columnwidth}\hspace{-0.4\twocolumnwidth}%
\label{eq:87KB} 
\end{multline}%
on the linear form in Eq.\ (\ref{eq:32a}). This results in another substitution, this time in functionals (\ref{eq:XiW}) and (\ref{eq:XiWS}): 
\begin{align} 
\begin{aligned} 
{\mbox{\rm\boldmath$\zeta$}}_+(t) &
= {\mbox{\rm\boldmath$\sigma$}}(t) - \frac{i{\mbox{\rm\boldmath$\zeta$}}(t)}{2}
, &
\bar{{\mbox{\rm\boldmath$\zeta$}}}_+(t) &
= {\mbox{\rm\boldmath$\sigma$}}^*(t) - \frac{i{\mbox{\rm\boldmath$\zeta$}}^*(t)}{2} , \\ 
{\mbox{\rm\boldmath$\zeta$}}_-(t) &
= {\mbox{\rm\boldmath$\sigma$}}(t) + \frac{i{\mbox{\rm\boldmath$\zeta$}}(t)}{2} 
, &
\bar{{\mbox{\rm\boldmath$\zeta$}}}_-(t) &
= {\mbox{\rm\boldmath$\sigma$}}^*(t)+ \frac{i{\mbox{\rm\boldmath$\zeta$}}^*(t)}{2} 
. 
\end{aligned}
\label{eq:SubA} 
\end{align}%
The inverse substitution reads, 
\begin{align} 
\begin{aligned} 
{\mbox{\rm\boldmath$\zeta$}} (t) &= i\big [ 
{\mbox{\rm\boldmath$\zeta$}}_+(t) - {\mbox{\rm\boldmath$\zeta$}}_-(t)
 \big ] , 
 &
{{\mbox{\rm\boldmath$\zeta$}}}^* (t) &= i\big [ 
\bar{{\mbox{\rm\boldmath$\zeta$}}}_+(t) - \bar{{\mbox{\rm\boldmath$\zeta$}}}_-(t)
 \big ], 
\\ 
{{\mbox{\rm\boldmath$s$}}} (t) &= \frac{1}{2} \big [ 
{{\mbox{\rm\boldmath$\zeta$}}}_+(t) + {{\mbox{\rm\boldmath$\zeta$}}}_-(t)
 \big ], 
 &
{{\mbox{\rm\boldmath$s$}}}^* (t) &= \frac{1}{2} \big [ 
\bar{{\mbox{\rm\boldmath$\zeta$}}}^*_+(t) + \bar{{\mbox{\rm\boldmath$\zeta$}}}^*_-(t)
 \big ] . 
\end{aligned}
\label{eq:CausZW} 
\end{align}%
showing that (\ref{eq:SubA}) is a genuine change of functional variables. 
Similar to Eqs.\ (\ref{eq:CausAW}) and (\ref{eq:86KA}), these relations impose conditions on the functional arguments, 
\begin{align} 
\begin{aligned} 
 &{\mbox{\rm\boldmath$\zeta$}}_+^*(t) = \bar{{\mbox{\rm\boldmath$\zeta$}}}_-(t), &
 &{\mbox{\rm\boldmath$\zeta$}}_-^*(t) = \bar{{\mbox{\rm\boldmath$\zeta$}}}_+(t). 
\end{aligned} 
\label{eq:89KD} 
\end{align}%
These conditions do not interfere with (\ref{eq:CausAW}) and (\ref{eq:86KA}) serving as characteristic functionals for the corresponding Green function, nor with (\ref{eq:SubA}) being a one-to-one substitution. 

By definition, the {\em generalised multitime Wigner representation\/} emerges by applying substitution (\ref{eq:SubA}) to functionals (\ref{eq:CausAW}) and (\ref{eq:86KA}). To start with, we note that the replacement, 
\begin{align} 
\begin{aligned} 
 &{\mbox{\rm\boldmath$\zeta$}}_{\pm}(t)\to{\mbox{\rm\boldmath$\zeta$}}_{\pm}(t)+{\mbox{\rm\boldmath$s$}}(t), &
 &\bar{{\mbox{\rm\boldmath$\zeta$}}}_{\pm}(t)\to\bar{{\mbox{\rm\boldmath$\zeta$}}}_{\pm}(t)+{\mbox{\rm\boldmath$s$}}(t), 
\end{aligned} 
\label{eq:91KF} 
\end{align}%
cf.\ Eq.\ (\ref{eq:XiSXi}), in variables ${\mbox{\rm\boldmath$\zeta $}}(t),{\mbox{\rm\boldmath$\sigma $}}(t)$ becomes simply, 
\begin{align} 
\begin{aligned} 
{\mbox{\rm\boldmath$\sigma $}}(t)\to{\mbox{\rm\boldmath$\sigma $}}(t)+{\mbox{\rm\boldmath$s$}}(t) , 
\end{aligned} 
\label{eq:92KH} 
\end{align}%
with variable ${\mbox{\rm\boldmath$\zeta $}}(t)$ unaffected. In variables ${\mbox{\rm\boldmath$\zeta $}}(t),{\mbox{\rm\boldmath$\sigma $}}(t)$ functionals ``with and without the sources'' are naturally expressed by a single functional $\Phi ^W$, 
\begin{widetext} 
\begin{align} 
\begin{aligned} 
\Phi^W 
{\big[ 
{\mbox{\rm\boldmath$\zeta$}} ,{\mbox{\rm\boldmath$\zeta$}} ^* \big| {\mbox{\rm\boldmath$\sigma$}},{\mbox{\rm\boldmath$\sigma$}}^*
 \big]} &= \Xi^W\bigg [ 
{\mbox{\rm\boldmath$\sigma$}} + \frac{i{\mbox{\rm\boldmath$\zeta$}}}{2} , 
{\mbox{\rm\boldmath$\sigma$}}^*+ \frac{i{\mbox{\rm\boldmath$\zeta$}}^*}{2} , 
{\mbox{\rm\boldmath$\sigma$}} - \frac{i{\mbox{\rm\boldmath$\zeta$}}}{2} , 
{\mbox{\rm\boldmath$\sigma$}}^* - \frac{i{\mbox{\rm\boldmath$\zeta$}}^*}{2} 
 \bigg ], \\ 
\Phi^W 
{\big[ 
{\mbox{\rm\boldmath$\zeta$}} ,{\mbox{\rm\boldmath$\zeta$}} ^* \big| {\mbox{\rm\boldmath$\sigma$}} +{\mbox{\rm\boldmath$s$}},{\mbox{\rm\boldmath$\sigma$}}^*+{\mbox{\rm\boldmath$s$}}^*
 \big]} &= \Xi^W{\bigg[ 
{\mbox{\rm\boldmath$\sigma$}} + \frac{i{\mbox{\rm\boldmath$\zeta$}}}{2} , 
{\mbox{\rm\boldmath$\sigma$}}^*+ \frac{i{\mbox{\rm\boldmath$\zeta$}}^*}{2} , 
{\mbox{\rm\boldmath$\sigma$}} - \frac{i{\mbox{\rm\boldmath$\zeta$}}}{2} , 
{\mbox{\rm\boldmath$\sigma$}}^* - \frac{i{\mbox{\rm\boldmath$\zeta$}}^*}{2} 
 \bigg| 
{\mbox{\rm\boldmath$s$}},{\mbox{\rm\boldmath$s$}}^* 
 \bigg]} . 
\end{aligned} 
\label{eq:90KE} 
\end{align}%
We see that the functional variable ${\mbox{\rm\boldmath$\sigma$}}(t)$ corresponds to the {\em formal input\/} of the system, which in turn is specified by the {\em formal c-number source\/} added to the Hamiltonian. We emphasise formality of both concepts. Under macroscopic conditions, an external source may become a good approximation for a laser (say). Importantly, even in this case, it remains a phenomenological model for a complex quantum device. 
\subsection{Time-symmetric operator ordering}%
\label{ch:CharTS}
If ${\mbox{\rm\boldmath$\sigma$}}(t)$ defines an input of the system, what does the output defined by variable ${\mbox{\rm\boldmath$\zeta$}}(t)$ stand for? Following \cite{PreprintGenPW}, we {\em postulate\/} that, in the Wigner representation, the formal output of a system is expressed by time-symmetric averages of the Heisenberg field operators, 
\begin{align} 
\begin{aligned} 
\Big \langle 
{\cal T}^W\!\exp
\Big(
{\mbox{\rm\boldmath$\zeta$}} ^* 
{\mbox{\rm\boldmath$\mathcal{A}$}}\hspace{-1\vAlength}
\hspace{0.27\vAlength}\hat{\phantom{{\mbox{\rm\boldmath$\mathcal{A}$}}}}
\hspace{-0.27\vAlength}^{\prime} + 
{\mbox{\rm\boldmath$\zeta$}} 
{\mbox{\rm\boldmath$\mathcal{A}$}}\hspace{-1\vAlength}
\hspace{0.27\vAlength}\hat{\phantom{{\mbox{\rm\boldmath$\mathcal{A}$}}}}
\hspace{-0.27\vAlength}^{\prime \dag}
 \Big) 
 \Big \rangle \equiv 
\Phi^W 
{\big[ 
{\mbox{\rm\boldmath$\zeta$}} ,{\mbox{\rm\boldmath$\zeta$}} ^* \big| {\mbox{\rm\boldmath$s$}},{\mbox{\rm\boldmath$s$}}^*
 \big]} = 
\Xi^W 
{\bigg[ 
\frac{i{\mbox{\rm\boldmath$\zeta$}}}{2} , 
\frac{i{\mbox{\rm\boldmath$\zeta$}}^*}{2} , 
\frac{-i{\mbox{\rm\boldmath$\zeta$}}}{2} , 
\frac{-i{\mbox{\rm\boldmath$\zeta$}} ^*}{2} 
 \bigg| 
{\mbox{\rm\boldmath$s$}},{\mbox{\rm\boldmath$s$}}^*
 \bigg]} . 
\end{aligned} 
\label{eq:XiSW} 
\end{align}%
For the operator without the source, 
\begin{align} 
\begin{aligned} 
\Big \langle 
{\cal T}^W\!\exp
\Big(
{\mbox{\rm\boldmath$\zeta$}} ^* 
{\mbox{\rm\boldmath$\mathcal{A}$}}\hspace{-1\vAlength}
\hspace{0.27\vAlength}\hat{\phantom{{\mbox{\rm\boldmath$\mathcal{A}$}}}}
\hspace{-0.27\vAlength}+ 
{\mbox{\rm\boldmath$\zeta$}} 
{\mbox{\rm\boldmath$\mathcal{A}$}}\hspace{-1\vAlength}
\hspace{0.27\vAlength}\hat{\phantom{{\mbox{\rm\boldmath$\mathcal{A}$}}}}
\hspace{-0.27\vAlength}^{\dag}
 \Big) 
 \Big \rangle \equiv 
\Phi^W 
{\big[ 
{\mbox{\rm\boldmath$\zeta$}} ,{\mbox{\rm\boldmath$\zeta$}} ^* \big| 0,0
 \big]} = 
\Xi^W 
\bigg [ 
\frac{i{\mbox{\rm\boldmath$\zeta$}}}{2} , 
\frac{i{\mbox{\rm\boldmath$\zeta$}}^*}{2} , 
\frac{-i{\mbox{\rm\boldmath$\zeta$}}}{2} , 
\frac{-i{\mbox{\rm\boldmath$\zeta$}} ^*}{2}
 \bigg ] . 
\end{aligned} 
\label{eq:94KK} 
\end{align}%
Then, 
\begin{multline} 
\Big \langle 
{\cal T}^W\!
{\hat{\mathcal A}}_{k_1}(t_1)
\cdots 
{\hat{\mathcal A}}_{k_{m}}(t_m)
\,
{\hat{\mathcal A}}_{k_{m+1}}^{\dag}(t_{m+1})
\cdots 
{\hat{\mathcal A}}_{k_{m+\bar m}}^{\dag}(t_{m+\bar m}) 
 \Big \rangle \\ = 
\frac{
\delta ^{m+\bar m}\Xi ^W 
\bigg [ \displaystyle
\frac{i{\mbox{\rm\boldmath$\zeta$}}}{2} ,\frac{i{\mbox{\rm\boldmath$\zeta$}}^*}{2},
\frac{-i{\mbox{\rm\boldmath$\zeta$}}}{2} ,\frac{-i{\mbox{\rm\boldmath$\zeta$}} ^*}{2}
 \bigg ] 
}{
\delta\zeta^*_{k_1}(t_1)
\cdots 
\delta\zeta^*_{k_{m}}(t_m)
\delta\zeta_{k_{m+1}}(t_{m+1})
\cdots 
\delta\zeta_{k_{m+\bar m}}(t_{m+\bar m}) 
} \bigg |_ {{\mbox{\rm\boldmath$\zeta$}} =0}\ , 
\label{eq:FWDiff} 
\end{multline}%
and similarly for the primed operator. 

It is easy to prove that Eq.\ (\ref{eq:FWDiff}) agrees with the recursive definition of section \ref{ch:TWDef}. 
Explicitly, 
\begin{gather} 
\begin{gathered} 
\Xi ^W 
\bigg [ 
\frac{i{\mbox{\rm\boldmath$\zeta$}}}{2} ,\frac{i{\mbox{\rm\boldmath$\zeta$}}^*}{2},
\frac{-i{\mbox{\rm\boldmath$\zeta$}}}{2} ,\frac{-i{\mbox{\rm\boldmath$\zeta$}} ^*}{2}
 \bigg ] = 
\bigg \langle 
T^W_- \exp\bigg(\frac{1}{2} {\mbox{\rm\boldmath$\zeta$}}
{\mbox{\rm\boldmath$\mathcal{A}$}}\hspace{-1\vAlength}
\hspace{0.27\vAlength}\hat{\phantom{{\mbox{\rm\boldmath$\mathcal{A}$}}}}
\hspace{-0.27\vAlength}^{\dag}+ \frac{1}{2} {\mbox{\rm\boldmath$\zeta$}}^*
{\mbox{\rm\boldmath$\mathcal{A}$}}\hspace{-1\vAlength}
\hspace{0.27\vAlength}\hat{\phantom{{\mbox{\rm\boldmath$\mathcal{A}$}}}}
\hspace{-0.27\vAlength}\bigg)
\,
T^W_+ \exp
\bigg(\frac{1}{2} {\mbox{\rm\boldmath$\zeta$}}
{\mbox{\rm\boldmath$\mathcal{A}$}}\hspace{-1\vAlength}
\hspace{0.27\vAlength}\hat{\phantom{{\mbox{\rm\boldmath$\mathcal{A}$}}}}
\hspace{-0.27\vAlength}^{\dag}+ \frac{1}{2} {\mbox{\rm\boldmath$\zeta$}}^*
{\mbox{\rm\boldmath$\mathcal{A}$}}\hspace{-1\vAlength}
\hspace{0.27\vAlength}\hat{\phantom{{\mbox{\rm\boldmath$\mathcal{A}$}}}}
\hspace{-0.27\vAlength}\bigg) 
 \bigg \rangle , 
\end{gathered}
\label{eq:FWDef} 
\end{gather}%
For simplicity we assume that all times on the LHS of (\ref{eq:FWDiff}) are different. We can then also assume that 
\begin{align} 
\begin{aligned} 
\zeta_k(t) = \sum_{l=1}^{m+\bar m}\zeta_{l,k}(t) = 
\zeta_{1,k}(t)+\zeta '_{k}(t), 
\end{aligned} 
\end{align}%
where $\zeta_{l,k}(t)$ is nonzero only in close vicinity of $t_l$, and that different $\zeta_{l,k}(t)$ do not overlap.
Furthermore, if $t_1$ is the smallest time in (\ref{eq:FWDiff}), isolating in (\ref{eq:FWDef}) the contribution linear in ${\mbox{\rm\boldmath$\zeta$}}^*_1$ we can write 
\begin{multline} 
\Xi ^W 
\bigg [ 
\frac{i{\mbox{\rm\boldmath$\zeta$}}}{2} ,\frac{i{\mbox{\rm\boldmath$\zeta$}}^*}{2},
\frac{-i{\mbox{\rm\boldmath$\zeta$}}}{2} ,\frac{-i{\mbox{\rm\boldmath$\zeta$}} ^*}{2}
 \bigg ]\Big |_{\text{linear in ${\mbox{\rm\boldmath$\zeta$}}^*_1$}} \\ = 
\bigg \langle 
\bigg(\frac{1}{2} {\mbox{\rm\boldmath$\zeta$}}^*_1
{\mbox{\rm\boldmath$\mathcal{A}$}}\hspace{-1\vAlength}
\hspace{0.27\vAlength}\hat{\phantom{{\mbox{\rm\boldmath$\mathcal{A}$}}}}
\hspace{-0.27\vAlength}\bigg)
\bigg [ 
T^W_- 
\exp\bigg(\frac{1}{2} {\mbox{\rm\boldmath$\zeta$}}'
{\mbox{\rm\boldmath$\mathcal{A}$}}\hspace{-1\vAlength}
\hspace{0.27\vAlength}\hat{\phantom{{\mbox{\rm\boldmath$\mathcal{A}$}}}}
\hspace{-0.27\vAlength}^{\dag}+ \frac{1}{2} {\mbox{\rm\boldmath$\zeta$}}^{\prime *}
{\mbox{\rm\boldmath$\mathcal{A}$}}\hspace{-1\vAlength}
\hspace{0.27\vAlength}\hat{\phantom{{\mbox{\rm\boldmath$\mathcal{A}$}}}}
\hspace{-0.27\vAlength}\bigg)
\,
T^W_+ 
\exp
\bigg(\frac{1}{2} {\mbox{\rm\boldmath$\zeta$}}'
{\mbox{\rm\boldmath$\mathcal{A}$}}\hspace{-1\vAlength}
\hspace{0.27\vAlength}\hat{\phantom{{\mbox{\rm\boldmath$\mathcal{A}$}}}}
\hspace{-0.27\vAlength}^{\dag}+ \frac{1}{2} {\mbox{\rm\boldmath$\zeta$}}^{\prime *}
{\mbox{\rm\boldmath$\mathcal{A}$}}\hspace{-1\vAlength}
\hspace{0.27\vAlength}\hat{\phantom{{\mbox{\rm\boldmath$\mathcal{A}$}}}}
\hspace{-0.27\vAlength}\bigg) 
 \bigg ] 
 \bigg \rangle
\\ + 
\bigg \langle 
\bigg [ 
T^W_- 
\exp\bigg(\frac{1}{2} {\mbox{\rm\boldmath$\zeta$}}'
{\mbox{\rm\boldmath$\mathcal{A}$}}\hspace{-1\vAlength}
\hspace{0.27\vAlength}\hat{\phantom{{\mbox{\rm\boldmath$\mathcal{A}$}}}}
\hspace{-0.27\vAlength}^{\dag}+ \frac{1}{2} {\mbox{\rm\boldmath$\zeta$}}^{\prime *}
{\mbox{\rm\boldmath$\mathcal{A}$}}\hspace{-1\vAlength}
\hspace{0.27\vAlength}\hat{\phantom{{\mbox{\rm\boldmath$\mathcal{A}$}}}}
\hspace{-0.27\vAlength}\bigg)
\,
T^W_+ 
\exp
\bigg(\frac{1}{2} {\mbox{\rm\boldmath$\zeta$}}'
{\mbox{\rm\boldmath$\mathcal{A}$}}\hspace{-1\vAlength}
\hspace{0.27\vAlength}\hat{\phantom{{\mbox{\rm\boldmath$\mathcal{A}$}}}}
\hspace{-0.27\vAlength}^{\dag}+ \frac{1}{2} {\mbox{\rm\boldmath$\zeta$}}^{\prime *}
{\mbox{\rm\boldmath$\mathcal{A}$}}\hspace{-1\vAlength}
\hspace{0.27\vAlength}\hat{\phantom{{\mbox{\rm\boldmath$\mathcal{A}$}}}}
\hspace{-0.27\vAlength}\bigg) 
 \bigg ] 
\bigg(\frac{1}{2} {\mbox{\rm\boldmath$\zeta$}}^*_1
{\mbox{\rm\boldmath$\mathcal{A}$}}\hspace{-1\vAlength}
\hspace{0.27\vAlength}\hat{\phantom{{\mbox{\rm\boldmath$\mathcal{A}$}}}}
\hspace{-0.27\vAlength}\bigg)
 \bigg \rangle . 
\label{eq:TWRecF} 
\end{multline}%
\end{widetext}%
The square brackets of the RHS of this formula delineate the range to which the remaining time orderings are applied. Clearly equation (\ref{eq:TWRecF}) is nothing but an elaborate form of the recursive definition (\ref{eq:TWRec}). The case when the ``earliest'' operator is one of the ${\hat{\mathcal A}}^{\dag}$s follows by complex conjugation of (\ref{eq:TWRecF}). 

\section{Dynamics in phase-space}
\subsection{Phase-space path integral}\label{ch:DynaPh}
Physically, the most natural way of looking at the system is through time-symmetric averages of the field operator in the presence of the source. Their characteristic functional $\Phi^W 
{\big[ 
{\mbox{\rm\boldmath$\zeta$}} ,{\mbox{\rm\boldmath$\zeta$}} ^* \big| {\mbox{\rm\boldmath$s$}},{\mbox{\rm\boldmath$s$}}^*
 \big]}$ is given by Eq.\ (\ref{eq:XiSW}). On the one hand, it allows one full access to quantum properties of the system expressed by Perel-Keldysh Green functions (\ref{eq:GF}). On the other hand, it has an obvious phase-space interpretation found by generalising Eq.\ (\ref{eq:WMom}) to Heisenberg operators. This equation may be written as, 
\begin{align} 
\begin{aligned} 
\big \langle 
{\cal T}^W\!\exp\big(
{\mbox{\rm\boldmath$\zeta$}} ^* \hat{{\mbox{\rm\boldmath$a$}}}+ 
{\mbox{\rm\boldmath$\zeta$}} \hat{{\mbox{\rm\boldmath$a$}}}^{\dag}
 \big) 
 \big \rangle = \overline{\hspace{0.1ex}\exp\big(
{\mbox{\rm\boldmath$\zeta$}} ^*{\mbox{\rm\boldmath$\alpha $}} + 
{\mbox{\rm\boldmath$\zeta$}}{\mbox{\rm\boldmath$\alpha $}} ^* 
 \big)\hspace{0.1ex}} , 
\end{aligned} 
\label{eq:95KL} 
\end{align}%
where, we remind, 
\begin{align} 
\begin{aligned} 
 &{\mbox{\rm\boldmath$\zeta$}} ^* \hat{{\mbox{\rm\boldmath$a$}}}= \sum_{k=1}^N\int dt \zeta_k^*(t) \hat a_k(t), &
 &{\mbox{\rm\boldmath$\zeta$}} ^* {\mbox{\rm\boldmath$\alpha $}} = \sum_{k=1}^N\int dt \zeta_k^*(t) \alpha_k (t), 
\end{aligned} 
\label{eq:96KM} 
\end{align}%
and similarly for other ``products.'' Functions $\alpha _k(t) = \alpha_k \text{e}^{-i\omega _0 t}$ depend on the initial condition ${\mbox{\rm\boldmath$\alpha$}} $. The bar in (\ref{eq:95KL}) symbolises quasi-averaging over the Wigner function, cf. Eq.\ (\ref{eq:WMom}). Following the pattern of Eq.\ (\ref{eq:95KL}) we postulate the {\em path-integral\/} representation of the functional (\ref{eq:XiSW}), 
\begin{align} 
\begin{aligned} 
\big \langle 
{\cal T}^W\!\exp
\big(
{\mbox{\rm\boldmath$\zeta$}} ^* 
{\mbox{\rm\boldmath$\mathcal{A}$}}\hspace{-1\vAlength}
\hspace{0.27\vAlength}\hat{\phantom{{\mbox{\rm\boldmath$\mathcal{A}$}}}}
\hspace{-0.27\vAlength}^{\prime} + 
{\mbox{\rm\boldmath$\zeta$}} 
{\mbox{\rm\boldmath$\mathcal{A}$}}\hspace{-1\vAlength}
\hspace{0.27\vAlength}\hat{\phantom{{\mbox{\rm\boldmath$\mathcal{A}$}}}}
\hspace{-0.27\vAlength}^{\prime \dag}
 \big) 
 \big \rangle = \overline{\hspace{0.1ex}\exp\big(
{\mbox{\rm\boldmath$\zeta$}}{\mbox{\rm\boldmath$a$}}^* + 
{{\mbox{\rm\boldmath$\zeta$}}}^* {\mbox{\rm\boldmath$a$}} \big)
\hspace{0.1ex}} \Big |_{{{\mbox{\rm\boldmath$s$}}},{{\mbox{\rm\boldmath$s$}}}^*}\ .
\end{aligned}
\label{eq:PI} 
\end{align}%
In this relation, ${\mbox{\rm\boldmath$a$}}(t)$ is the c-number phase-space trajectory, and the bar denotes a path integral over these trajectories. We assume that trajectories obey the generic equation (\ref{eq:EqI}), making them dependent (conditional) on the external source ${\mbox{\rm\boldmath$s$}}(t)$ and on the initial condition ${\mbox{\rm\boldmath$\alpha $}}$. The bar includes quasi-averaging over the initial condition and the own quasi-stochasticity of trajectories coming from the self-action source $\bar s_k(t)$ in (\ref{eq:EqI}). For comparison, in (\ref{eq:95KL}) the quasi-averaging is over the initial condition while the way trajectories evolve in time is nonstochastic. 
\subsection{Closed perturbative relations in the Wigner representation}
Rewriting Eq.\ (\ref{eq:32a}) in variables ${\mbox{\rm\boldmath$a$}}(t)$, ${\mbox{\rm\boldmath$\sigma $}}(t)$, ${\mbox{\rm\boldmath$\zeta$}} (t)$, and ${\mbox{\rm\boldmath$\xi$}} (t)$ defined by Eqs.\ (\ref{eq:CausZW}) and (\ref{eq:CausAW}), and replacing ${\mbox{\rm\boldmath$\sigma $}}(t)$ by ${\mbox{\rm\boldmath$s$}}(t)$, we obtain 
\begin{widetext} 
\begin{multline} 
\big \langle 
{\cal T}^W\!\exp
\big(
{\mbox{\rm\boldmath$\zeta$}} ^* 
{\mbox{\rm\boldmath$\mathcal{A}$}}\hspace{-1\vAlength}
\hspace{0.27\vAlength}\hat{\phantom{{\mbox{\rm\boldmath$\mathcal{A}$}}}}
\hspace{-0.27\vAlength}^{\prime} + 
{\mbox{\rm\boldmath$\zeta$}} 
{\mbox{\rm\boldmath$\mathcal{A}$}}\hspace{-1\vAlength}
\hspace{0.27\vAlength}\hat{\phantom{{\mbox{\rm\boldmath$\mathcal{A}$}}}}
\hspace{-0.27\vAlength}^{\prime \dag}
 \big) 
 \big \rangle = 
\int 
\frac{d^2 \alpha_1 }{\pi } 
\cdots 
\frac{d^2 \alpha _N }{\pi } 
W\big(
{\mbox{\rm\boldmath$\alpha $}},{\mbox{\rm\boldmath$\alpha $}}^* 
 \big) 
\bigg \{ 
\exp \bigg(
\frac{\delta }{\delta {\mbox{\rm\boldmath$a$}}} 
G_{\text{R}}%
\frac{\delta }{\delta 
{\mbox{\rm\boldmath$\xi$}} 
^*%
} 
+ 
\frac{\delta }{\delta{\mbox{\rm\boldmath$a$}}^* 
} 
G^*_{\text{R}}%
\frac{\delta }{\delta{\mbox{\rm\boldmath$\xi$}} } 
 \bigg) \\ \times 
\exp\Big( 
{\mbox{\rm\boldmath$\zeta$}}{\mbox{\rm\boldmath$a$}}^* 
+ {\mbox{\rm\boldmath$\zeta$}} 
^* 
{\mbox{\rm\boldmath$a$}} 
+ {\mbox{\rm\boldmath$\xi$}} 
{\mbox{\rm\boldmath$s$}} 
^* 
+ {\mbox{\rm\boldmath$\xi$}} 
^* 
{\mbox{\rm\boldmath$s$}} 
+ S^W {\big[ 
{\mbox{\rm\boldmath$\xi$}} ,{\mbox{\rm\boldmath$\xi$}} 
^* 
\big| {\mbox{\rm\boldmath$a$}} ,{\mbox{\rm\boldmath$a$}} 
^* 
 \big]} 
\Big) 
 \bigg \} 
\bigg|_{{\mbox{\rm\boldmath$a$}}(t)\to{\mbox{\rm\boldmath$\alpha $}}(t) ,\, {\mbox{\rm\boldmath$\xi$}}(t)\to 0}, 
\label{eq:XiW2} 
\end{multline}%
where 
\begin{multline} 
S^W {\big[ 
{\mbox{\rm\boldmath$\xi$}} ,{\mbox{\rm\boldmath$\xi$}} ^* \big| {\mbox{\rm\boldmath$a$}} ,{\mbox{\rm\boldmath$a$}} ^*
 \big]} 
= i\int dt 
\bigg [ 
 H ^W_{\text{I}}
\bigg(
{\mbox{\rm\boldmath$a$}}(t) + \frac{i{\mbox{\rm\boldmath$\xi$}}(t)}{2} , 
{\mbox{\rm\boldmath$a$}}^*(t) + \frac{i{\mbox{\rm\boldmath$\xi$}}^*(t)}{2} 
 \bigg) \\ - 
 H ^W_{\text{I}}
\bigg(
{\mbox{\rm\boldmath$a$}}(t) - \frac{i{\mbox{\rm\boldmath$\xi$}}(t)}{2} , 
{\mbox{\rm\boldmath$a$}}^*(t) - \frac{i{\mbox{\rm\boldmath$\xi$}}^*(t)}{2} 
 \bigg) 
 \bigg ] 
. 
\label{eq:SigW} 
\end{multline}%
For better understanding of Eq.\ (\ref{eq:XiW2}) the reader should reread the comments following Eq.\ (\ref{eq:32a}). 
\subsection{Multitime-Wigner equations for the Bose-Hubbard model}
Equations (\ref{eq:XiW2}), (\ref{eq:SigW}) are general and hold for any interaction specified by its symmetrically ordered form 
$ H ^W_{\text{I}}$. 
For the Bose-Hubbard chain we find (omitting time arguments in the integrand) 
\begin{multline} 
\label{eq:SW} 
S^W \big[
{\mbox{\rm\boldmath$\xi$}} ,{\mbox{\rm\boldmath$\xi$}} ^*\big|{\mbox{\rm\boldmath$a$}} ,{\mbox{\rm\boldmath$a$}} ^* 
\big] = - \sum_{k=1}^N \int dt \bigg\{
\kappa
\Big [ 
\Big(
\xi_k a^*_k 
+\xi_k ^* 
a_k 
 \Big) 
\Big(
a_k^* 
a_k -1 
 \Big) \Big ] - 
\frac{\kappa }{4} 
\Big(
\xi_k^2\xi%
_k^* 
a_k^* 
+\xi_k^{* 2} 
\xi_k a_k 
 \Big) \\ - J\Big(\xi_k^* a_{k+1} +\xi_{k+1} ^* a_{k} +\xi_{k} a_{k+1}^* +\xi_{k+1} a_{k}^*\Big)\bigg\}. 
\end{multline}%
\end{widetext}%
As was shown in \cite{BWO}, Eq.\ (\ref{eq:XiW2}) may be interpreted as a solution to the generic stochastic equation in phase space formulated in section \ref{ch:Gen}. The random source in Eq.\ (\ref{eq:EqA}) is characterised by its averages conditioned on the ``local field'' $a_k(t)$, 
\begin{multline} 
\hspace{0.4\columnwidth}\hspace{-0.4\twocolumnwidth}
\overline{\hspace{0.1ex}\exp\sum_{k=1}^N\int dt \left[
\xi ^*_k(t) \bar s_k(t) +\xi_k(t) \bar s ^*_k(t)
\right] \hspace{0.1ex}}\bigg |_{{\mbox{\rm\boldmath$a$}}} \\ = \exp S^W {\big[ 
{\mbox{\rm\boldmath$\xi$}} ,{\mbox{\rm\boldmath$\xi$}} ^* \big| {\mbox{\rm\boldmath$a$}} ,{\mbox{\rm\boldmath$a$}} ^*
 \big]}, 
\label{eq:SAv} 
\hspace{0.4\columnwidth}\hspace{-0.4\twocolumnwidth}
\end{multline}%
so that $S^W {\big[ 
{\mbox{\rm\boldmath$\xi$}} ,{\mbox{\rm\boldmath$\xi$}} ^* \big| {\mbox{\rm\boldmath$a$}} ,{\mbox{\rm\boldmath$a$}} ^*
 \big]}$ is a characteristic functional for the {\em cumulants\/} of the random source. 
For the Bose-Hubbard model we have, 
\begin{multline} 
\hspace{0.4\columnwidth}\hspace{-0.4\twocolumnwidth}
\bar s_k(t) = - \kappa a_k(t) \big[
\left | 
a_k(t)
\right |^2 - 1 
\big]\\ + J\big[
a_{k+1}(t) + a_{k-1}(t)
\big] + \bar s^{(3)}_k(t), 
\label{eq:SBar} 
\hspace{0.4\columnwidth}\hspace{-0.4\twocolumnwidth}
\end{multline}%
where the actual random contribution comes from the {\em third-order noise\/} $\bar s^{(3)}_k(t)$. It is specified by the conjugate pair of nonzero cumulants, 
\begin{gather} 
\begin{aligned}
& \overline{\hspace{0.1ex}\bar s^{(3)}_k(t)\bar s^{(3)}_{k'}(t')
\bar s^{(3) *}_{k''}(t'')\hspace{0.1ex}}\big |_{{\mbox{\rm\boldmath$a$}}} \\ & \ \ \ \ \ = 
\delta_{kk'}\delta_{kk''}\delta(t-t')\delta(t-t'')\frac{\kappa a_k(t)}{2} , \\ 
& \overline{\hspace{0.1ex}\bar s^{(3)}_k(t)\bar s^{(3) *}_{k'}(t')
\bar s^{(3) *}_{k''}(t'')\hspace{0.1ex}}\big |_{{\mbox{\rm\boldmath$a$}}} \\ & \ \ \ \ \ = 
\delta_{kk'}\delta_{kk''}\delta(t-t')\delta(t-t'')\frac{\kappa a^*_k(t)}{2} , 
\end{aligned}%
\label{eq:MainS3} 
\end{gather}%
while all other cumulants of $\bar s^{(3)}_k(t)$ are zero (this allowed us to specify the average in place of the cumulant). 
While the Wigner function responsible for the initial condition in (\ref{eq:EqA}) may be positive thus affording a statistical interpretation, the cubic noise is a purely pseudo-stochastic object. 
\subsection{The truncated Wigner representation}
Dropping the third-order noise turns the full Wigner representation into the so-called {\em truncated Wigner representation\/}. In more traditional phase-space techniques, the truncated Wigner is found by dropping the third-order derivatives in the generalised Fokker-Planck equation for the single-time Wigner distribution. In the absence of losses, the corresponding Langevin equation is non-stochastic. For the Hamiltonian (\ref{eq:BHH}), it coincides with Eq.\ (\ref{eq:EqA}), where the source is given by (\ref{eq:SBar}) without 
$\bar s^{(3)}_k$. However simple and straightforward, the conventional way of deriving the truncated Wigner leaves it unclear if it can be applied to any {\em multitime\/} quantum averages. In our approach here as well as in Ref.\ \cite{Bettina} the truncated-Wigner equations emerge as an approximation within rigorous techniques intended for calculation of the time-symmetric averages of the Heisenberg operators. The generalisation associated with extending the truncated-Wigner equations to multitime averages is thus highly nontrivial and requires a new concept: the time-symmetric ordering of the Heisenberg operators. There does not seem to be a way of guessing this concept from within the conventional phase-space techniques. 
\section{The causal regularisation}\label{ch:SecCaus}
Strictly speaking, all relations derived so far are leading considerations that require specifications. Two things need to be warranted: that the functional form of the symmetric Wick theorem conforms with the specification of $T^W_+$ as symmetric ordering for coinciding time arguments, and that the stochastic integral equation (\ref{eq:EqI}) is defined mathematically. In Ref.\ \cite{BWO}, both problems were taken care of ``in a single blow'' by the causal regularisation \cite{BWO,EPL98,ProcPathIntVI,EPL01} of the retarded Green function $G_{\text{R}}(t)$, cf.\ Eq.\ (\ref{eq:24a}).
Namely, $G_{\text{R}}(t)$ should be replaced by a sufficiently smooth function while preserving its causal nature, $G_{\text{R}}(t)\sim \theta (t)$. 
One may assume, for instance, that 
\begin{gather} 
\begin{aligned}
G_{\text{R}}(t) = i \theta (t) \big(1-\text{e}^{-\Gamma t}\big)^m \text{e}^{-i\omega_0t} , 
\end{aligned}%
\label{eq:Gr} 
\end{gather}%
where the limit $\Gamma \to \infty$ is implied. The regularised Green function is zero at $t=0$ and has $m-1$ zero derivatives. Equation (\ref{eq:Gr}) is a toy version of the Pauli-Villars regularisation used in the quantum field theory \cite{Schweber,Bogol} as part of the common renormalisation procedure (cf.\ also \cite{DirResp}). 

In this paper as well as in \cite{BWO} the causal regularisation of $G_{\text{R}}(t)$ has a two-fold effect. Firstly, it assigns mathematical sense to equations (\ref{eq:EqA}), (\ref{eq:EqI}), defining them as Ito equations. In \cite{BWO}, this was in agreement with the more traditional approach based on pseudo-distributions and generalised Fokker-Planck equations \cite{GardinerQN}. Furthermore, with $G_{\text{R}}(t)$ regularised, equation (\ref{eq:GWbyG}) assures that the kernels $G_{cc'}^W(t)$ are smooth functions and 
\begin{gather} 
\begin{aligned}
G_{cc'}^W(0) = 0 
\end{aligned}%
\end{gather}%
without any mathematical ambiguity. As a result, the symmetric Wick theorem (\ref{eq:HoriW}) leaves alone any product of operators with equal time arguments. Since the final expression in (\ref{eq:HoriW}) is ordered symmetrically, the symmetric ordering is also enforced for any same-time product on the LHS of (\ref{eq:HoriW}). Regularisation (\ref{eq:Gr}) applied to the symmetric Wick theorem (\ref{eq:HoriW}) thus indeed specifies the double-time-ordered product (\ref{eq:TpTm}) so that operators with equal time arguments are ordered symmetrically. 

An unwanted effect of regularisation is that symmetric ordering is also enforced for 
same-time groups of operators split between the $T^W_+$ and $T^W_-$ orderings. For example, with regularisation, 
\begin{multline} 
\hspace{0.4\columnwidth}\hspace{-0.4\twocolumnwidth}
T_-^W \hat a_{k'}^{\dag}(t')\,T_+^W \hat a_{k}(t) = 
\text{W}\hat a_{k}(t) \hat a_{k'}^{\dag}(t')\\ - \frac{\delta_{kk'}}{2} \big(1-\text{e}^{-\Gamma |t-t'|}\big)^m 
\text{e}^{-i\omega_0|t-t'|} . 
\hspace{0.4\columnwidth}\hspace{-0.4\twocolumnwidth}
\end{multline}%
This problem was also encountered in \cite{BWO}, and the solution to it remains the same: the limit $\Gamma \to \infty$ should always preceed $t'\to t$. Then, 
\begin{multline} 
\hspace{0.4\columnwidth}\hspace{-0.4\twocolumnwidth}
T_-^W \hat a_{k'}^{\dag}(t)\,T_+^W \hat a_{k}(t) \\ = 
\lim_{t'\to t}\lim_{\Gamma \to \infty} 
T_-^W \hat a_{k'}^{\dag}(t')\,T_-^W \hat a_{k}(t) \\ = 
\text{W}\hat a_{k}(t) \hat a_{k'}^{\dag}(t)-\frac{\delta_{kk'}}{2} = 
\hat a_{k'}^{\dag}(t)\hat a_{k}(t) 
\hspace{0.4\columnwidth}\hspace{-0.4\twocolumnwidth}
\end{multline}%
as expected. The general recipe is to keep time arguments of operators under the $T_+^W$ and $T_-^W$ orderings slightly different, which is equivalent to ignoring regularisation of $G^W_{+-}$ and $G^W_{-+}$. This recipe implies that the double-time-ordered averages in the limit $\Gamma \to \infty$ are continuous functions of all time differences $t_+-t_-$, where $t_+$ and $t_-$ are time arguments of an operator pair split between the $T_+^W$ and $T_-^W$ orderings. This is obviously consistent with continuity properties of the $G^W_{+-}$ and $G^W_{-+}$ kernels, 
\begin{gather} 
\begin{aligned}
\lim_{t\searrow 0}\lim_{\Gamma \to \infty}G^W_{-+}(t) & = 
\lim_{t\nearrow 0}\lim_{\Gamma \to \infty}G^W_{-+}(t) = 
\bar G^W_{-+}(0), \\ 
\lim_{t\searrow 0}\lim_{\Gamma \to \infty}G^W_{+-}(t) & = 
\lim_{t\nearrow 0}\lim_{\Gamma \to \infty}G^W_{+-}(t) = 
\bar G^W_{+-}(0), 
\end{aligned}%
\end{gather}%
where $\bar G^W_{-+}$ and $\bar G^W_{+-}$ are unregularised kernels. With this amendment all specifications enforced by regularisation only apply to operators under the time orderings. In fact, we only need to specify the $T^W_+$-ordering, defining it as time ordering for different times and symmetric ordering for equal times. The $T^W_-$-ordering remains defined by (\ref{eq:TpConj}). We return to the problem of ``unwanted effects'' of regularisation in the next section. 
\section{The reordering problem}\label{ch:Order}
Assume we know a way of calculating time-symmetric averages of Heisenberg operators while the physics demand time-normal ones, or vice versa. For two-time averages this problem is solved by Kubo's formula for the linear response function \cite{KuboIrrevI,KuboTodaHashitsumeII}, cf.\ Refs.\ \cite{QNDBWO,Bettina}. Here we consider a general approach to reordering Heisenberg operators. Among other results we show that this always reduces to considering a response problem of sorts. 

It is instructive to compare the formulae relating the time-symmetric and time-normal averages to the Perel-Keldysh Green functions (\ref{eq:GF}): 
\begin{widetext} 
\begin{align} 
\begin{aligned} 
\end{aligned} 
\Phi^W 
{\big[ 
{\mbox{\rm\boldmath$\zeta$}} ,{\mbox{\rm\boldmath$\zeta$}} ^* \big| {\mbox{\rm\boldmath$s$}},{\mbox{\rm\boldmath$s$}}^*
 \big]} &= 
\Xi^W \bigg[
\frac{i{\mbox{\rm\boldmath$\zeta$}}}{2} + {{\mbox{\rm\boldmath$s$}}} ,
\frac{i{\mbox{\rm\boldmath$\zeta$}}^*}{2} + {{\mbox{\rm\boldmath$s$}}^*}, 
\frac{-i{\mbox{\rm\boldmath$\zeta$}}}{2} + {{\mbox{\rm\boldmath$s$}}}, 
\frac{-i{\mbox{\rm\boldmath$\zeta$}}^*}{2} + {{\mbox{\rm\boldmath$s$}}^*}
\bigg] 
, 
\label{eq:43a} 
\\ 
\Phi^N 
{\big[ 
{\mbox{\rm\boldmath$\zeta$}} ,{\mbox{\rm\boldmath$\zeta$}} ^* \big| {\mbox{\rm\boldmath$s$}},{\mbox{\rm\boldmath$s$}}^*
 \big]} &= 
\Xi^N \big[
{i{\mbox{\rm\boldmath$\zeta$}}} + {{\mbox{\rm\boldmath$s$}}} ,
{{\mbox{\rm\boldmath$s$}}^*}, 
{{\mbox{\rm\boldmath$s$}}}, 
{-i{\mbox{\rm\boldmath$\zeta$}}^*} + {{\mbox{\rm\boldmath$s$}}^*}
\big] 
. 
\label{eq:44a} 
\end{align}%
\end{widetext}%
cf.\ section \ref{ch:CommResp}. The functionals $\Xi^W$ and $\Xi^N$ are both characteristic ones for Green functions (\ref{eq:GF}) but imply different specifications for coinciding time arguments: (\ref{eq:43a}) implies symmetric ordering while (\ref{eq:44a}) --- normal ordering. 
Such specifications only become of importance if Green functions (or their linear combinations) are considered for coinciding time arguments, and otherwise can be disregarded. 

Mathematically, $\Xi^W$ and $\Xi^N$ coincide up to a {\em singular part\/}. Hence, 
up to singular parts, the functionals $\Phi^W$ and $\Phi^N$ differ only in a functional substitution applied to Eq.\ (\ref{eq:XiW}). Time-symmetric averages appear with 
\begin{align} 
\begin{aligned} 
\eta_- &= \frac{i{\mbox{\rm\boldmath$\zeta$}}}{2} + {\mbox{\rm\boldmath$s$}} , &
\eta^{\dag}_- &= \frac{i{\mbox{\rm\boldmath$\zeta$}}^*}{2} + {\mbox{\rm\boldmath$s$}}^*, 
\\ 
\eta_+ &= \frac{-i{\mbox{\rm\boldmath$\zeta$}}}{2} + {\mbox{\rm\boldmath$s$}}, &
\eta^{\dag}_+ &= \frac{-i{\mbox{\rm\boldmath$\zeta$}}^*}{2} + {\mbox{\rm\boldmath$s$}}^*, 
\end{aligned} 
\label{eq:45a} 
\end{align}%
while the time-normal ones require 
\begin{align} 
\begin{aligned} 
\eta_- &= {i{\mbox{\rm\boldmath$\zeta$}}} + {\mbox{\rm\boldmath$s$}} , &
\eta^{\dag}_- &= {\mbox{\rm\boldmath$s$}}^*, 
\\ 
\eta_+ &= {\mbox{\rm\boldmath$s$}}, &
\eta^{\dag}_+ &= {-i{\mbox{\rm\boldmath$\zeta$}}^*} + {\mbox{\rm\boldmath$s$}}^*. 
\end{aligned} 
\label{eq:46a} 
\end{align}%
Both Eqs.\ (\ref{eq:45a}) and (\ref{eq:46a}) may be inverted, and it is straightforward to show that 
\begin{align} 
\begin{aligned} 
\Phi ^W {\big[ 
{\mbox{\rm\boldmath$\zeta$}} , {\mbox{\rm\boldmath$\zeta$}} ^* \big| {\mbox{\rm\boldmath$s$}}, {\mbox{\rm\boldmath$s$}}^*
 \big]} = 
\Phi ^N {\bigg[ 
{\mbox{\rm\boldmath$\zeta$}} , {\mbox{\rm\boldmath$\zeta$}} ^* \bigg| {\mbox{\rm\boldmath$s$}} - \frac{i {\mbox{\rm\boldmath$\zeta$}} }{2} , {\mbox{\rm\boldmath$s$}}^* + \frac{i {\mbox{\rm\boldmath$\zeta$}}^* }{2} 
 \bigg]}, 
\\ 
\Phi ^N {\big[ 
{\mbox{\rm\boldmath$\zeta$}} , {\mbox{\rm\boldmath$\zeta$}} ^* \big| {\mbox{\rm\boldmath$s$}}, {\mbox{\rm\boldmath$s$}}^*
 \big]} = 
\Phi ^W {\bigg[ 
{\mbox{\rm\boldmath$\zeta$}} , {\mbox{\rm\boldmath$\zeta$}} ^* \bigg| {\mbox{\rm\boldmath$s$}} + \frac{i {\mbox{\rm\boldmath$\zeta$}} }{2} , {\mbox{\rm\boldmath$s$}}^* - \frac{i {\mbox{\rm\boldmath$\zeta$}}^* }{2}
 \bigg]} . 
\end{aligned} 
\label{eq:47a} 
\end{align}%
In particular, for the time-normal and time-symmetric averages defined without the source we have 
\begin{align} 
\begin{aligned} 
\Phi ^W {\big[ 
{\mbox{\rm\boldmath$\zeta$}} , {\mbox{\rm\boldmath$\zeta$}} ^* \big| 0, 0
 \big]} = 
\Phi ^N {\bigg[ 
{\mbox{\rm\boldmath$\zeta$}} , {\mbox{\rm\boldmath$\zeta$}} ^* \bigg| - \frac{i {\mbox{\rm\boldmath$\zeta$}} }{2} , \frac{i {\mbox{\rm\boldmath$\zeta$}}^* }{2} 
 \bigg]}, 
\\ 
\Phi ^N {\big[ 
{\mbox{\rm\boldmath$\zeta$}} , {\mbox{\rm\boldmath$\zeta$}} ^* \big| 0, 0
 \big]} = 
\Phi ^W {\bigg[ 
{\mbox{\rm\boldmath$\zeta$}} , {\mbox{\rm\boldmath$\zeta$}} ^* \bigg| \frac{i {\mbox{\rm\boldmath$\zeta$}} }{2} , - \frac{i {\mbox{\rm\boldmath$\zeta$}}^* }{2}
 \bigg]} . 
\end{aligned} 
\label{eq:48a} 
\end{align}%
These relations make it evident that the reordering problem for the Heisenberg operators is indeed equivalent to the response problem. 

We remind the reader that Eqs.\ (\ref{eq:47a}) and (\ref{eq:48a}) hold only up to a singular part. All formulae for quantum averages following from them are only valid for different time arguments and otherwise require specifications. 

We now consider an example which also illustrates how to approach coinciding time arguments. 
We wish to express the time-normal average 
\begin{align} 
\begin{aligned} 
\big \langle 
{\hat{\mathcal A}}_{k}^{\dag}(t){\hat{\mathcal A}}_{k'}(t') 
 \big \rangle = 
\frac{\delta ^2\Phi ^N {\big[ 
{\mbox{\rm\boldmath$\zeta$}} , {\mbox{\rm\boldmath$\zeta$}} ^* \big| 0, 0
 \big]} 
}{ 
\delta\zeta_{k}(t)\delta\zeta_{k'}^*(t')
} \Big|_{\zeta =0}
\end{aligned} 
\label{eq:49a} 
\end{align}%
by the time-symmetric ones. Combining (\ref{eq:49a}) with the second of Eqs.\ (\ref{eq:48a}) we have 
\begin{multline} 
\hspace{0.4\columnwidth}\hspace{-0.4\twocolumnwidth}
\big \langle 
{\hat{\mathcal A}}_{k}^{\dag}(t){\hat{\mathcal A}}_{k'}(t') 
 \big \rangle = \big \langle 
{\cal T}^W\!{\hat{\mathcal A}}_{k}^{\dag}(t){\hat{\mathcal A}}_{k'}(t')
 \big \rangle \\ + \frac{i }{2} \bigg [ 
\frac{\delta \big \langle 
{\hat{\mathcal A}}_{k'}'(t')
 \big \rangle 
}{\delta s_{k}(t)} - 
\frac{\delta \big \langle 
{\hat{\mathcal A}}_{k}^{\prime\dag}(t)
 \big \rangle 
}{\delta s_{k'}^*(t')} 
 \bigg ] \bigg|_{s=0} . 
\hspace{0.4\columnwidth}\hspace{-0.4\twocolumnwidth}%
\label{eq:50a} 
\end{multline}%
In deriving this we used that 
\begin{align} 
\begin{aligned} 
\Phi ^{W,N}{\big[ 
0,0 \big| {\mbox{\rm\boldmath$s$}},{\mbox{\rm\boldmath$s$}}^*
 \big]} = 1 , 
\end{aligned} 
\label{eq:51a} 
\end{align}%
so that the derivatives purely by input arguments always disappear. In \cite{Bettina}, Eq.\ (\ref{eq:50a}) was derived by an {\em ad hoc\/} method; here we show that the same relation follows from the general equations (\ref{eq:48a}). 
For details on numerical implementation of Eq.\ (\ref{eq:50a}) see \cite{Bettina}.

Equation (\ref{eq:50a}) is well behaved in the limit $t'\to t$. For instance, let $t'>t$. Then, by causality \cite{APII}, 
\begin{align} 
\begin{aligned} 
\frac{\delta \big \langle 
{\hat{\mathcal A}}_{k}^{\dag}(t)
 \big \rangle 
}{\delta s_{k'}^*(t')} = 0 . 
\end{aligned} 
\label{eq:52a} 
\end{align}%
Using (\ref{eq:12}) we rewrite (\ref{eq:50a}) as 
\begin{align} 
\begin{aligned} 
\frac{\delta \big \langle 
{\hat{\mathcal A}}'_{k'}(t')
 \big \rangle 
}{\delta s(t)}\Big|_{s=0} &= i\Big \langle 
\big [ 
{\hat{\mathcal A}}_{k'}(t'),{\hat{\mathcal A}}^{\dag}_{k}(t) 
 \big ] 
 \Big \rangle , &t'>t.
\end{aligned} 
\label{eq:53a} 
\end{align}%
We have thus rederived Kubo's famous relation for the linear response function. Both sides here have a well-defined limit $i\delta_{kk'} $ for $t\nearrow t'$. Similarly, 
\begin{align} 
\begin{aligned} 
\frac{\delta \big \langle 
{\hat{\mathcal A}}^{\prime\dag}_{k}(t)
 \big \rangle 
}{\delta s^*_{k'}(t')}\Big|_{s=0} &= -i \Big \langle 
\big [ 
{\hat{\mathcal A}}_{k'}(t'),{\hat{\mathcal A}}^{\dag}_{k}(t) 
 \big ] 
 \Big \rangle , &t>t', 
\end{aligned} 
\label{eq:54a} 
\end{align}%
which is nothing but complex conjugation of (\ref{eq:53a}) with the replacement $t\leftrightarrow t'$. 

At the same time, Eq.\ (\ref{eq:50a}) resists any attempt to extend it to equal time arguments. Within the causal reqularisation, 
\begin{align} 
\begin{aligned} 
\frac{\delta \big \langle 
{\hat{\mathcal A}}'_{k}(t)
 \big \rangle 
}{\delta s_{k}(t)} = 
\frac{\delta \big \langle 
{\hat{\mathcal A}}^{\prime\dag}_{k}(t)
 \big \rangle 
}{\delta s^*_{k}(t)} = 0, 
\end{aligned} 
\label{eq:55a} 
\end{align}%
giving rise to a patently wrong result, 
\begin{align} 
\begin{aligned} 
\big \langle 
{\hat{\mathcal A}}^{\dag}_{k}(t){\hat{\mathcal A}}_{k}(t) 
 \big \rangle = \big \langle 
{\cal T}^W\!{\hat{\mathcal A}}^{\dag}_{k}(t){\hat{\mathcal A}}_{k}(t)
 \big \rangle . 
\end{aligned} 
\label{eq:56a} 
\end{align}%
Conversely, making the response correction in (\ref{eq:50a}) nonzero for $t=t'$ inevitably causes troubles with causality under regularisation. This shows, in particular, that the ``unwanted effects'' of regularisation (see section \ref{ch:SecCaus}) are unavoidable and cannot be eliminated by a better regularisation scheme. For this reason we regard it as ``good conduct'' to treat all quantum averages as generalised functions, making the question of their values at coinciding time arguments meaningless. 

\section{Conclusions}\label{sec:mais}

We have developed a generalisation of the symmetric ordering to multitime problems with nonlinear interactions. This includes generalisation of the symmetric (Weyl) ordering to time-symmetric ordering of Heisenberg operators, and of the renowned Wigner function to a path integral in phase-space. Among other results, continuity of time-symmetric operator products has been proven. A way of calculating time-normally-orderied operator products within the time-symmetric-ordering-based techniques has also been developed. 

\section*{Acknowledgments}

The authors are grateful to S.\ Stenholm, W.\ Schleich, M.\ Fleischhauer, and A.\ Polkovnikov for their 
comments on the manuscript. 
M.O.\ thanks the 
Institut f\"ur Quantenphysik at the Universit\"at Ulm for generous hospitality. 
L.P.\ is grateful to ARC Centre of Excellence for Quantum-Atom Optics at the University of Queensland for hospitality and for meeting the cost of his visit to Brisbane. 
This work was supported by the 
Program Atomoptik of the Landesstiftung Baden-W\"urttemberg and 
SFB/TR 21 ``Control of Quantum Correlations in Tailored Matter'' 
funded 
by the Deutsche Forschungsgemeinschaft (DFG), 
Australian Research Council (Grant ID: FT100100515) and a University of Queensland New Staff Grant. 
\appendix*
\section{Structural symmetric Wick theorem for a pair of fields}
In this appendix, we formulate an ultimate generalisation of the symmetric Wick theorem. It appears to cover all imaginable cases of interest and, if necessary, can be extended to fermionic fields. Firstly, let $\tau $ be a generalised time variable which belongs to some linearly ordered set with the succession symbol $\succ$. Of practical relevance are the time axis and the C-contour. Secondly, let $\hat X(\xi ,\tau )$, $\hat {\bar X}(\xi ,\tau )$ be a pair of free-field operators defined in some Fock space, (in this appendix $\hbar \neq 1$)
\begin{align} 
\begin{aligned} 
\hat X(\xi ,\tau) = \sqrt{\hbar }\sum_{\kappa }\big [ 
u_{\kappa }(\xi )\hat a_{\kappa }(\tau) + v_{\kappa }(\xi )\hat b_{\kappa }^{\dag}(\tau)
 \big ] , \\ 
\hat {\bar X}(\xi ,\tau) = \sqrt{\hbar }\sum_{\kappa }\big [ 
\bar u_{\kappa }(\xi )\hat a_{\kappa }^{\dag}(\tau) + \bar v_{\kappa }(\xi )\hat b_{\kappa }(\tau)
 \big ] . 
\end{aligned} 
\label{eq:60HX} 
\end{align}%
Here, $\xi $ symbolises arguments of field operators except time; $\xi $ can contain coordinate, momentum, spin indices, and so on. The index $\kappa $ enumerates the modes of which the Fock space is built. With extentions to relativity or solid state in mind, for each mode we define two pairs of creation and annihilation operators, 
\begin{align} 
\begin{aligned} 
 &\hat a_{\kappa }(\tau ) = \text{e}^{-i\varphi_{\kappa }(\tau )}\hat a_{\kappa }, &
 &\hat a_{\kappa }^{\dag}(\tau ) = \text{e}^{i\varphi_{\kappa }(\tau )}\hat a_{\kappa }^{\dag}, \\ 
 &\hat b_{\kappa }(\tau ) = \text{e}^{-i\varphi_{\kappa }(\tau )}\hat b_{\kappa }, &
 &\hat b_{\kappa }^{\dag}(\tau ) = \text{e}^{i\varphi_{\kappa }(\tau )}\hat b_{\kappa }^{\dag}. 
\end{aligned} 
\label{eq:61HY} 
\end{align}%
The phases $\varphi _{\kappa }(\tau )$ are c-number functions of `time;'' note that Eq.\ (\ref{eq:61HY}) employs one phase for ``particles and antiparticles'' in each mode. The stationary operators obey standard commutational relations, 
\begin{align} 
\begin{aligned} 
\big [ 
\hat a_{\kappa },\hat a_{\kappa '}^{\dag}
 \big ] = 
\big [ 
\hat b_{\kappa },\hat b_{\kappa '}^{\dag}
 \big ] = \delta_{\kappa \kappa '}, 
\end{aligned} 
\label{eq:62HZ} 
\end{align}%
otherwise mode operators pairwise commute. 
The c-number functions $u_{\kappa }(\xi )$, $v_{\kappa }(\xi )$, $\bar u_{\kappa }(\xi )$, and $\bar v_{\kappa }(\xi )$ are replaced in particular problems by suitable solutions of a single-body equation. In nonrelativistic problems, $v_{\kappa }(\xi )=\bar v_{\kappa }(\xi )=0$. For our purposes here all c-number functions in Eqs.\ (\ref{eq:60HX}) and (\ref{eq:61HY}) are regarded as arbitrary. 

Specification of the quantities occuring in Eqs.\ (\ref{eq:60HX})--(\ref{eq:62HZ}) belongs to a particular problem. It is implied that such specification should meet certain conditions of algebraic consistency. For example, summation in (\ref{eq:60HX}) should be consistent with the ``Kronecker symbol'' in (\ref{eq:62HZ}), (with $f_{\kappa }$ being an arbitrary function of the mode index)
\begin{align} 
\begin{aligned} 
\sum_{\kappa '} \delta_{\kappa \kappa '}f_{\kappa '} = f_{\kappa }. 
\end{aligned} 
\label{eq:74JN} 
\end{align}%
Similar consistencies should exist between integrations and corresponding delta-functions, 
\begin{align} 
\begin{aligned} 
 &\int d\tau' \delta (\tau ,\tau ') f(\tau ') = f(\tau ), &
 &\int d\xi' \delta (\xi ,\xi ') f(\xi ') = f(\xi ) . 
\end{aligned} 
\label{eq:75JP} 
\end{align}%
The delta-functions and Kronecker symbols are symmetric. These consistencies extend to functional differentiations by c-number functions of $\xi $ and $\tau $. So, for the c-number functions $a_{\kappa }(\tau ),\bar a_{\kappa }(\tau )$ occuring in Eqs.\ (\ref{eq:52HP}), (\ref{eq:54HR}), 
\begin{align} 
\begin{aligned} 
 &\frac{\delta a_{\kappa }(\tau )}{\delta a_{\kappa '}(\tau ')} = 
\frac{\delta \bar a_{\kappa }(\tau )}{\delta \bar a_{\kappa '}(\tau ')} = 
\delta (\tau ,\tau ')\delta_{\kappa \kappa '} , 
\\
 &\frac{\delta F(\bar a)}{\delta a_{\kappa }(\tau )} = 
\frac{\delta F(a)}{\delta \bar a_{\kappa }(\tau )} = 
0 , 
\end{aligned} 
\label{eq:57HU} 
\end{align}%
and similarly for the c-number pair $X(\xi ,\tau ),\bar X(\xi ,\tau )$ in (\ref{eq:63JA}), (\ref{eq:64JB}). Conditions (\ref{eq:74JN})--(\ref{eq:57HU}) make all relations below algebraically defined. 
 
Linear order of the generalised time axis allows one to define the step-function, 
\begin{align} 
\begin{aligned} 
\theta(\tau ,\tau ') = 
\begin{cases}
1, & \tau \succ \tau ', \\
0, & \tau \prec \tau ' , 
\end{cases}
\end{aligned} 
\label{eq:48HK} 
\end{align}%
and the ``time'' ordering, 
\begin{multline} 
\hspace{0.4\columnwidth}\hspace{-0.4\twocolumnwidth}
T\hat X(\xi ,\tau)\hat{\bar X}(\xi,\tau) = 
\theta(\tau ,\tau ')\hat X(\xi ,\tau)\hat {\bar X}(\xi ,\tau) 
\\ + 
\theta(\tau' ,\tau )\hat {\bar X}(\xi ,\tau)\hat X(\xi ,\tau), 
\hspace{0.4\columnwidth}\hspace{-0.4\twocolumnwidth}%
\label{eq:49HL} 
\end{multline}%
cf.\ Eq.\ (\ref{eq:Tp}). Similar definitions apply to a larger number of factors. The time-symmetric ordering is defined replacing ``larger than'' by ``succeed'' in Eq.\ (\ref{eq:78JS}): 
\begin{multline} 
\hspace{0.4\columnwidth}\hspace{-0.4\twocolumnwidth}
{\cal T}^W\!{\hat{\mathcal X}}_1(\tau_1){\hat{\mathcal X}}_2(\tau_2)\cdots{\hat{\mathcal X}}_n(\tau_n)
= \frac{1}{2^{N-1}} \\ \times
\big [ \cdots
\big [ 
\big [ 
{\hat{\mathcal X}}_1(\tau_1),{\hat{\mathcal X}}_2(\tau_2)
 \big ]_+ ,{\hat{\mathcal X}}_3(\tau_3)
 \big ]_+ ,\cdots,{\hat{\mathcal X}}_N(\tau_N)
 \big ]_+ , \\ 
\tau_1\succ \tau_2\succ\cdots\succ \tau_N . 
\hspace{0.4\columnwidth}\hspace{-0.4\twocolumnwidth}%
\label{eq:51HN} 
\end{multline}%
The proof of section \ref{ch:EqTWtoW} that for the free operators symmetric and time-symmetric orderings are the same thing generalises literally with $t\to\tau $ and $\omega_0t\to\varphi_{\kappa }(\tau )$. 

The {\em generalised structural symmetric Wick theorem\/} that we now wish to prove reads, 
\begin{align} 
\begin{aligned} 
T P[\hat X,\hat {\bar X}] = {\cal T}^W\!\bigg \{ 
\exp
\mathcal{Z} \bigg [ 
\frac{\delta }{\delta X} , 
\frac{\delta }{\delta \bar X} 
 \bigg ]
\\ \times P[X,\bar X] |_{X\to\hat X, \bar X\to\hat{\bar X}}
 \bigg \} , 
\end{aligned} 
\label{eq:63JA} 
\end{align}%
where $X(\xi ,\tau ),\bar X(\xi ,\tau )$ are a pair of c-number functions and, 
\begin{multline} 
\hspace{0.4\columnwidth}\hspace{-0.4\twocolumnwidth}
\mathcal{Z} \bigg [ 
\frac{\delta }{\delta X} , 
\frac{\delta }{\delta \bar X} 
 \bigg ] = -i\hbar \int d\tau d\tau ' d\xi d\xi ' \\ 
\times G^W(\xi ,\tau ;\xi ',\tau ')
\frac{\delta ^2}{\delta X(\xi ,\tau )\delta \bar X(\xi ',\tau ')} . 
\hspace{0.4\columnwidth}\hspace{-0.4\twocolumnwidth}%
\label{eq:64JB} 
\end{multline}%
Applying (\ref{eq:63JA}) to $T\hat X(\xi ,\tau )\hat{\bar X}(\xi ',\tau ')$ we find that $G^W(\xi ,\tau ;\xi ',\tau ')$ coincides with the symmetric contraction of the pair (\ref{eq:60HX}), 
\begin{multline} 
\hspace{0.4\columnwidth}\hspace{-0.4\twocolumnwidth}
-i\hbar G^W(\xi ,\tau ;\xi ',\tau ') \\ = 
T\hat X(\xi ,\tau )\hat{\bar X}(\xi ',\tau ')
- 
{\cal T}^W\!\hat X(\xi ,\tau )\hat{\bar X}(\xi ',\tau '). 
\hspace{0.4\columnwidth}\hspace{-0.4\twocolumnwidth}%
\label{eq:65JC} 
\end{multline}%

Since all operator reorderings occur mode-wise, it suffices to verify (\ref{eq:63JA}) for one mode; the general formula then follows by direct calculation. In the one-mode case, 
\begin{multline} 
\hspace{0.4\columnwidth}\hspace{-0.4\twocolumnwidth}
\begin{aligned}
 &%
\begin{aligned} &\hat X(\xi ,\tau ) \to\hat a_{\kappa }(\tau ), &
 &\hat {\bar X}(\xi ,\tau ) \to\hat a_{\kappa }^{\dag}(\tau ),\end{aligned} \\ 
 &G^W(\xi ,\tau ;\xi ',\tau ')\to G_{\kappa }^W(\tau ,\tau ') 
\end{aligned} 
\\ = i \big [ 
T\hat a_{\kappa }(\tau )\hat a_{\kappa }^{\dag}(\tau ')- {\cal T}^W\!\hat a_{\kappa }(\tau )\hat a_{\kappa }^{\dag}(\tau ')
 \big ] \\ = 
\frac{i}{2}\text{e}^{-i\varphi_{\kappa }(\tau )+i\varphi_{\kappa }(\tau ')} 
\big [ 
\theta(\tau ,\tau ') - \theta(\tau ',\tau )
 \big ] , 
\hspace{0.4\columnwidth}\hspace{-0.4\twocolumnwidth}%
\label{eq:77JR} 
\end{multline}%
and Eq.\ (\ref{eq:63JA}) becomes, 
\begin{multline} 
\hspace{0.4\columnwidth}\hspace{-0.4\twocolumnwidth}
T P[\hat a,\hat a^{\dag}] = {\cal T}^W\!\bigg \{ 
\exp 
\mathcal{Z}_{\kappa }\bigg [ 
\frac{\delta }{\delta a} , 
\frac{\delta }{\delta \bar a} 
 \bigg ] 
\\ \times P[a,\bar a] |_{a\to\hat a_{\kappa }, \bar a\to\hat a_{\kappa }^{\dag}}
 \bigg \} , 
\hspace{0.4\columnwidth}\hspace{-0.4\twocolumnwidth}%
\label{eq:52HP} 
\end{multline}%
where $a(\tau ),\bar a(\tau )$ are a pair of arbitrary c-number functions of ``time.'' The ``per mode'' reordering exponent $\mathcal{Z}_{\kappa }$ reads, 
\begin{align} 
\begin{aligned} 
\mathcal{Z}_{\kappa }\bigg [ 
\frac{\delta }{\delta a} , 
\frac{\delta }{\delta \bar a} 
 \bigg ] = -i\int d\tau d\tau ' G_{\kappa }^W(\tau ,\tau ')
\frac{\delta ^2}{\delta a(\tau )\delta \bar a(\tau ')} . 
\end{aligned} 
\label{eq:54HR} 
\end{align}%
Verification of Eq.\ (\ref{eq:52HP}) goes in two steps. Firstly, we note that the equivalence between the verbal and algebraic forms of Wick's theorem proper \cite{Hori} does not depend on details of the contraction and hence equally applies to the symmetric Wick theorem. Secondly, the proof of the symmetric Wick theorem for the time axis in section \ref{ch:SWDyson} may be literally adjusted to generalised time replacing time $t$ by $\tau $, the symbol $>$ by $\succ$, and the contraction $G^W_{++}(t-t')$ by $G^W(\tau ,\tau ')$. We therefore regard Eq.\ (\ref{eq:52HP}) proven.

To verify the general relation (\ref{eq:63JA}) we note that any functional depending on the mode operators may be brought to the symmetric form by applying Eq.\ (\ref{eq:52HP}) mode-wise: 
\begin{widetext} 
\begin{align} 
\begin{aligned} 
T P[\hat X,\hat {\bar X}] = {\cal T}^W\!\bigg(
\exp\bigg \{ \sum_{\kappa }\bigg(
\mathcal{Z}_{\kappa }\bigg [ 
\frac{\delta }{\delta a_{\kappa }} , 
\frac{\delta }{\delta \bar a_{\kappa }} 
 \bigg ] + 
\mathcal{Z}_{\kappa }\bigg [ 
\frac{\delta }{\delta b_{\kappa }} , 
\frac{\delta }{\delta \bar b_{\kappa }} 
 \bigg ]
 \bigg) 
 \bigg \} 
P[X,\bar X] |_{a\to\hat a}
 \bigg) . 
\end{aligned} 
\label{eq:66JD} 
\end{align}%
\end{widetext}%
In this formula, 
the c-number fields $X(\xi ,\tau )$, $\bar X(\xi ,\tau )$ are given by Eq.\ (\ref{eq:60HX}) without hats, 
\begin{align} 
\begin{aligned} 
X(\xi ,\tau) = \sqrt{\hbar }\sum_{\kappa }\big [ 
u_{\kappa }(\xi ) a_{\kappa }(\tau) + v_{\kappa }(\xi ) \bar b_{\kappa }(\tau)
 \big ] , \\ 
{\bar X}(\xi ,\tau) = \sqrt{\hbar }\sum_{\kappa }\big [ 
\bar u_{\kappa }(\xi ) \bar a_{\kappa }(\tau) + \bar v_{\kappa }(\xi ) b_{\kappa }(\tau)
 \big ] , 
\end{aligned} 
\label{eq:67JE} 
\end{align}%
where $a_{\kappa }(\tau )$, $\bar a_{\kappa }(\tau )$, $b_{\kappa }(\tau )$, and $\bar b_{\kappa }(\tau )$ are four independent c-number fields per mode, and $a\to\hat a$ is a shorthand notation for the substitution, 
\begin{align} 
\begin{aligned} 
 &a_{\kappa }(\tau )\to\hat a_{\kappa }(\tau ), &
 &\bar a_{\kappa }(\tau )\to\hat a_{\kappa }^{\dag}(\tau ), \\ 
 &b_{\kappa }(\tau )\to\hat b_{\kappa }(\tau ), &
 &\bar b_{\kappa }(\tau )\to\hat b_{\kappa }^{\dag}(\tau ). 
\end{aligned} 
\label{eq:70JJ} 
\end{align}%
The ``per mode'' reordering exponents are given by (\ref{eq:54HR}). Applying the chain rule to differentiations in (\ref{eq:66JD}) we get, 
\begin{align} 
\begin{aligned} 
\frac{\delta }{\delta a(\tau )} P[X,\bar X] = 
\sqrt{\hbar}\int d\xi\, u_{\kappa }(\xi)
\frac{\delta }{\delta X(\xi ,\tau )} P[X,\bar X], \\ 
\frac{\delta }{\delta \bar a(\tau )} P[X,\bar X] = 
\sqrt{\hbar}\int d\xi\, \bar u_{\kappa }(\xi)
\frac{\delta }{\delta \bar X(\xi ,\tau )} P[X,\bar X], \\ 
\frac{\delta }{\delta b(\tau )} P[X,\bar X] = 
\sqrt{\hbar}\int d\xi\, \bar v_{\kappa }(\xi)
\frac{\delta }{\delta \bar X(\xi ,\tau )} P[X,\bar X], \\ 
\frac{\delta }{\delta \bar b(\tau )} P[X,\bar X] = 
\sqrt{\hbar}\int d\xi\, v_{\kappa }(\xi)
\frac{\delta }{\delta X(\xi ,\tau )} P[X,\bar X] . 
\end{aligned} 
\label{eq:71JK} 
\end{align}%
Substituting these relations into Eq.\ (\ref{eq:66JD}) we recover Eq.\ (\ref{eq:63JA}) with 
\begin{multline} 
\hspace{0.4\columnwidth}\hspace{-0.4\twocolumnwidth}
G^W(\xi ,\tau ;\xi ',\tau ') = \sum_{\kappa }\big [ 
u_{\kappa }(\xi)\bar u_{\kappa }(\xi')G^W_{\kappa }(\tau ,\tau ') \\ + v_{\kappa }(\xi')\bar v_{\kappa }(\xi)G^W_{\kappa }(\tau' ,\tau )
 \big ] . 
\hspace{0.4\columnwidth}\hspace{-0.4\twocolumnwidth}%
\label{eq:72JL} 
\end{multline}%
It is readily verified that this definition of $G^W(\xi ,\tau ;\xi ',\tau ')$ agrees with (\ref{eq:65JC}). The only remaining artifact of the underlying mode expansion is then the substitution (\ref{eq:70JJ}). However, with everything expressed in terms of $X(\xi ,\tau),\bar X(\xi ,\tau)$, it may be replaced by 
\begin{align} 
\begin{aligned} 
 &X(\xi ,\tau)\to\hat X(\xi ,\tau), &
 &\bar X(\xi ,\tau)\to\hat {\bar X}(\xi ,\tau). 
\end{aligned} 
\label{eq:73JM} 
\end{align}%
This concludes the proof of Eq.\ (\ref{eq:63JA}). 

We conclude this appendix with a remark on scaling of fields and contractions in the classical limit $\hbar \to 0$. {\em Commutators of physical quantities must scale as\/} $\hbar $. In this sense, quantised mode amplitudes $\hat a_{\kappa }$ are not physical. The physical amplitude is $\sqrt{\hbar }\hat a_{\kappa }$, which in the limit $\hbar \to 0$ becomes the classical mode amplitude \cite{OPO}. For this reason the factor $\sqrt{\hbar }$ is explicitly present in the definitions of ``physical fields'' (\ref{eq:60HX}). The scaling $\propto \hbar $ is then removed from the contraction (\ref{eq:65JC}) by the factor $\hbar $ on the LHS. The result is that, as was demonstrated in Refs.\ \cite{API,APII}, Green functions (propagators) of bosonic fields are to a large extent classical quantities. In particular, they are all expressed by the {\em response transformation\/} in terms of the retarded Green function of the corresponding classical field. For more details see \cite{OPO,API,APII,APIII}.


\begin{thebibliography}{53}
\expandafter\ifx\csname natexlab\endcsname\relax\def\natexlab#1{#1}\fi
\expandafter\ifx\csname bibnamefont\endcsname\relax
  \def\bibnamefont#1{#1}\fi
\expandafter\ifx\csname bibfnamefont\endcsname\relax
  \def\bibfnamefont#1{#1}\fi
\expandafter\ifx\csname citenamefont\endcsname\relax
  \def\citenamefont#1{#1}\fi
\expandafter\ifx\csname url\endcsname\relax
  \def\url#1{\texttt{#1}}\fi
\expandafter\ifx\csname urlprefix\endcsname\relax\def\urlprefix{URL }\fi
\providecommand{\bibinfo}[2]{#2}
\providecommand{\eprint}[2][]{\url{#2}}

\bibitem[{\citenamefont{Berg et~al.}(2009)\citenamefont{Berg, Plimak,
  Polkovnikov, Olsen, Fleischhauer, and Schleich}}]{Bettina}
\bibinfo{author}{\bibfnamefont{B.}~\bibnamefont{Berg}},
  \bibinfo{author}{\bibfnamefont{L.~I.} \bibnamefont{Plimak}},
  \bibinfo{author}{\bibfnamefont{A.}~\bibnamefont{Polkovnikov}},
  \bibinfo{author}{\bibfnamefont{M.~K.} \bibnamefont{Olsen}},
  \bibinfo{author}{\bibfnamefont{M.}~\bibnamefont{Fleischhauer}},
  \bibnamefont{and} \bibinfo{author}{\bibfnamefont{W.~P.}
  \bibnamefont{Schleich}}, \bibinfo{journal}{Phys. Rev. A}
  \textbf{\bibinfo{volume}{80}}, \bibinfo{pages}{033624}
  (\bibinfo{year}{2009}).

\bibitem[{\citenamefont{Werner and Drummond}(1997)}]{Wminus}
\bibinfo{author}{\bibfnamefont{M.~J.} \bibnamefont{Werner}} \bibnamefont{and}
  \bibinfo{author}{\bibfnamefont{P.~D.} \bibnamefont{Drummond}},
  \bibinfo{journal}{J.\ Comput.\ Phys.} \textbf{\bibinfo{volume}{132}},
  \bibinfo{pages}{312} (\bibinfo{year}{1997}).

\bibitem[{\citenamefont{Wolf and Mandel}(1995)}]{MandelWolf}
\bibinfo{author}{\bibfnamefont{E.}~\bibnamefont{Wolf}} \bibnamefont{and}
  \bibinfo{author}{\bibfnamefont{L.}~\bibnamefont{Mandel}},
  \emph{\bibinfo{title}{Optical Coherence and Quantum Optics}}
  (\bibinfo{publisher}{Cambridge University Press}, \bibinfo{year}{1995}).

\bibitem[{\citenamefont{Vasil'ev}(1998)}]{VasF}
\bibinfo{author}{\bibfnamefont{A.~N.} \bibnamefont{Vasil'ev}},
  \emph{\bibinfo{title}{Functional methods in quantum field theory and
  statistical physics}} (\bibinfo{publisher}{Gordon and Breach},
  \bibinfo{year}{1998}).

\bibitem[{\citenamefont{Plimak et~al.}(2001{\natexlab{a}})\citenamefont{Plimak,
  Fleischhauer, Olsen, and Collett}}]{PreprintGenPW}
\bibinfo{author}{\bibfnamefont{L.~I.} \bibnamefont{Plimak}},
  \bibinfo{author}{\bibfnamefont{M.}~\bibnamefont{Fleischhauer}},
  \bibinfo{author}{\bibfnamefont{M.~K.} \bibnamefont{Olsen}}, \bibnamefont{and}
  \bibinfo{author}{\bibfnamefont{M.~J.} \bibnamefont{Collett}},
  \bibinfo{journal}{e-print arXiv:cond-mat/0102483}
  (\bibinfo{year}{2001}{\natexlab{a}}).

\bibitem[{\citenamefont{Plimak and Stenholm}(2012)}]{DirResp}
\bibinfo{author}{\bibfnamefont{L.~I.} \bibnamefont{Plimak}} \bibnamefont{and}
  \bibinfo{author}{\bibfnamefont{S.}~\bibnamefont{Stenholm}},
  \bibinfo{journal}{Ann.~Phys.~(N.Y.)} \textbf{\bibinfo{volume}{327}},
  \bibinfo{pages}{2691} (\bibinfo{year}{2012}).

\bibitem[{\citenamefont{Steel et~al.}(1998)\citenamefont{Steel, Olsen, Plimak,
  Drummond, Tan, Collett, Walls, and Graham}}]{WBEC1}
\bibinfo{author}{\bibfnamefont{M.~J.} \bibnamefont{Steel}},
  \bibinfo{author}{\bibfnamefont{M.~K.} \bibnamefont{Olsen}},
  \bibinfo{author}{\bibfnamefont{L.~I.} \bibnamefont{Plimak}},
  \bibinfo{author}{\bibfnamefont{P.~D.} \bibnamefont{Drummond}},
  \bibinfo{author}{\bibfnamefont{S.~M.} \bibnamefont{Tan}},
  \bibinfo{author}{\bibfnamefont{M.~J.} \bibnamefont{Collett}},
  \bibinfo{author}{\bibfnamefont{D.~F.} \bibnamefont{Walls}}, \bibnamefont{and}
  \bibinfo{author}{\bibfnamefont{R.}~\bibnamefont{Graham}},
  \bibinfo{journal}{Phys. Rev. A} \textbf{\bibinfo{volume}{58}},
  \bibinfo{pages}{4824} (\bibinfo{year}{1998}).

\bibitem[{\citenamefont{Olsen and Plimak}(2003)}]{MeLev}
\bibinfo{author}{\bibfnamefont{M.~K.} \bibnamefont{Olsen}} \bibnamefont{and}
  \bibinfo{author}{\bibfnamefont{L.~I.} \bibnamefont{Plimak}},
  \bibinfo{journal}{Phys. Rev. A} \textbf{\bibinfo{volume}{68}},
  \bibinfo{pages}{031603} (\bibinfo{year}{2003}).

\bibitem[{\citenamefont{Olsen}(2004)}]{WigstateMe}
\bibinfo{author}{\bibfnamefont{M.~K.} \bibnamefont{Olsen}},
  \bibinfo{journal}{Phys. Rev. A} \textbf{\bibinfo{volume}{69}},
  \bibinfo{pages}{013601} (\bibinfo{year}{2004}).

\bibitem[{\citenamefont{Olsen et~al.}(2004)\citenamefont{Olsen, Bradley, and
  Cavalcanti}}]{UFAL}
\bibinfo{author}{\bibfnamefont{M.~K.} \bibnamefont{Olsen}},
  \bibinfo{author}{\bibfnamefont{A.~S.} \bibnamefont{Bradley}},
  \bibnamefont{and} \bibinfo{author}{\bibfnamefont{S.~B.}
  \bibnamefont{Cavalcanti}}, \bibinfo{journal}{Phys. Rev. A}
  \textbf{\bibinfo{volume}{70}}, \bibinfo{pages}{033611}
  (\bibinfo{year}{2004}).

\bibitem[{\citenamefont{Olsen and Bradley}(2009)}]{OCstates}
\bibinfo{author}{\bibfnamefont{M.~K.} \bibnamefont{Olsen}} \bibnamefont{and}
  \bibinfo{author}{\bibfnamefont{A.~S.} \bibnamefont{Bradley}},
  \bibinfo{journal}{Opt.\ Comm.} \textbf{\bibinfo{volume}{282}},
  \bibinfo{pages}{3924} (\bibinfo{year}{2009}).

\bibitem[{\citenamefont{Johnsson and Hope}(2007)}]{JohnssonHope}
\bibinfo{author}{\bibfnamefont{M.~T.} \bibnamefont{Johnsson}} \bibnamefont{and}
  \bibinfo{author}{\bibfnamefont{J.~J.} \bibnamefont{Hope}},
  \bibinfo{journal}{Phys. Rev. A} \textbf{\bibinfo{volume}{75}},
  \bibinfo{pages}{043619} (\bibinfo{year}{2007}).

\bibitem[{\citenamefont{Jain et~al.}(2007)\citenamefont{Jain, Bradley, and
  Gardiner}}]{Pyush}
\bibinfo{author}{\bibfnamefont{P.}~\bibnamefont{Jain}},
  \bibinfo{author}{\bibfnamefont{A.~S.} \bibnamefont{Bradley}},
  \bibnamefont{and} \bibinfo{author}{\bibfnamefont{C.~W.}
  \bibnamefont{Gardiner}}, \bibinfo{journal}{Phys. Rev. A}
  \textbf{\bibinfo{volume}{76}}, \bibinfo{pages}{023617}
  (\bibinfo{year}{2007}).

\bibitem[{\citenamefont{{Ferris {\em et al.\/}}}(2008)}]{AJFerris}
\bibinfo{author}{\bibfnamefont{A.~J.} \bibnamefont{{Ferris {\em et al.\/}}}},
  \bibinfo{journal}{Phys. Rev. A} \textbf{\bibinfo{volume}{77}},
  \bibinfo{pages}{012712} (\bibinfo{year}{2008}).

\bibitem[{\citenamefont{Hoffmann et~al.}(2008)\citenamefont{Hoffmann, Corney,
  and Drummond}}]{seatlicker}
\bibinfo{author}{\bibfnamefont{S.~E.} \bibnamefont{Hoffmann}},
  \bibinfo{author}{\bibfnamefont{J.~F.} \bibnamefont{Corney}},
  \bibnamefont{and} \bibinfo{author}{\bibfnamefont{P.~D.}
  \bibnamefont{Drummond}}, \bibinfo{journal}{Phys. Rev. A}
  \textbf{\bibinfo{volume}{78}}, \bibinfo{pages}{013622}
  (\bibinfo{year}{2008}).

\bibitem[{\citenamefont{{Corney {\em et al.\/}}}(2008)}]{Joelfibre}
\bibinfo{author}{\bibfnamefont{J.~F.} \bibnamefont{{Corney {\em et al.\/}}}},
  \bibinfo{journal}{Phys. Rev. A} \textbf{\bibinfo{volume}{78}},
  \bibinfo{pages}{023831} (\bibinfo{year}{2008}).

\bibitem[{\citenamefont{Olsen and Davis}(2006)}]{MeMavis}
\bibinfo{author}{\bibfnamefont{M.~K.} \bibnamefont{Olsen}} \bibnamefont{and}
  \bibinfo{author}{\bibfnamefont{M.~J.} \bibnamefont{Davis}},
  \bibinfo{journal}{Phys. Rev. A} \textbf{\bibinfo{volume}{73}},
  \bibinfo{pages}{063618} (\bibinfo{year}{2006}).

\bibitem[{\citenamefont{Ferris et~al.}(2009)\citenamefont{Ferris, Olsen, and
  Davis}}]{AndyMeMavis}
\bibinfo{author}{\bibfnamefont{A.~J.} \bibnamefont{Ferris}},
  \bibinfo{author}{\bibfnamefont{M.~K.} \bibnamefont{Olsen}}, \bibnamefont{and}
  \bibinfo{author}{\bibfnamefont{M.~J.} \bibnamefont{Davis}},
  \bibinfo{journal}{Phys. Rev. A} \textbf{\bibinfo{volume}{79}},
  \bibinfo{pages}{043634} (\bibinfo{year}{2009}).

\bibitem[{\citenamefont{Shrestha et~al.}(2009)\citenamefont{Shrestha,
  Javanainen, and Ruostekoski}}]{Finns}
\bibinfo{author}{\bibfnamefont{U.}~\bibnamefont{Shrestha}},
  \bibinfo{author}{\bibfnamefont{J.}~\bibnamefont{Javanainen}},
  \bibnamefont{and}
  \bibinfo{author}{\bibfnamefont{J.}~\bibnamefont{Ruostekoski}},
  \bibinfo{journal}{Phys. Rev. A} \textbf{\bibinfo{volume}{79}},
  \bibinfo{pages}{043617} (\bibinfo{year}{2009}).

\bibitem[{\citenamefont{{Midgley {\em et al \/}}}(2009)}]{BECSarah}
\bibinfo{author}{\bibfnamefont{S.~L.~W.} \bibnamefont{{Midgley {\em et al
  \/}}}}, \bibinfo{journal}{Phys. Rev. A} \textbf{\bibinfo{volume}{79}},
  \bibinfo{pages}{053632} (\bibinfo{year}{2009}).

\bibitem[{\citenamefont{{Opanchuk {\em et
  al.\/}}}(2012{\natexlab{a}})}]{Hoffperv}
\bibinfo{author}{\bibfnamefont{B.}~\bibnamefont{{Opanchuk {\em et al.\/}}}},
  \bibinfo{journal}{Europhys. Lett.} \textbf{\bibinfo{volume}{97}},
  \bibinfo{pages}{5003} (\bibinfo{year}{2012}{\natexlab{a}}).

\bibitem[{\citenamefont{{Sau {\em et al.\/}}}(2009)}]{Sau}
\bibinfo{author}{\bibfnamefont{J.~D.} \bibnamefont{{Sau {\em et al.\/}}}},
  \bibinfo{journal}{Phys. Rev. A} \textbf{\bibinfo{volume}{80}},
  \bibinfo{pages}{023622} (\bibinfo{year}{2009}).

\bibitem[{\citenamefont{Mathey and Polkovnikov}(2009)}]{unbind}
\bibinfo{author}{\bibfnamefont{L.}~\bibnamefont{Mathey}} \bibnamefont{and}
  \bibinfo{author}{\bibfnamefont{A.}~\bibnamefont{Polkovnikov}},
  \bibinfo{journal}{Phys. Rev. A} \textbf{\bibinfo{volume}{80}},
  \bibinfo{pages}{041601} (\bibinfo{year}{2009}).

\bibitem[{\citenamefont{Mathey and Polkovnikov}(2010)}]{reverseKZ}
\bibinfo{author}{\bibfnamefont{L.}~\bibnamefont{Mathey}} \bibnamefont{and}
  \bibinfo{author}{\bibfnamefont{A.}~\bibnamefont{Polkovnikov}},
  \bibinfo{journal}{Phys. Rev. A} \textbf{\bibinfo{volume}{81}},
  \bibinfo{pages}{033605} (\bibinfo{year}{2010}).

\bibitem[{\citenamefont{Martin and Ruostekoski}(2010)}]{AndyMartin}
\bibinfo{author}{\bibfnamefont{A.~D.} \bibnamefont{Martin}} \bibnamefont{and}
  \bibinfo{author}{\bibfnamefont{J.}~\bibnamefont{Ruostekoski}},
  \bibinfo{journal}{Phys. Rev. Lett.} \textbf{\bibinfo{volume}{104}},
  \bibinfo{pages}{194102} (\bibinfo{year}{2010}).

\bibitem[{\citenamefont{Chianca and Olsen}(2011{\natexlab{a}})}]{CVC1}
\bibinfo{author}{\bibfnamefont{C.~V.} \bibnamefont{Chianca}} \bibnamefont{and}
  \bibinfo{author}{\bibfnamefont{M.~K.} \bibnamefont{Olsen}},
  \bibinfo{journal}{Phys. Rev. A} \textbf{\bibinfo{volume}{83}},
  \bibinfo{pages}{043607} (\bibinfo{year}{2011}{\natexlab{a}}).

\bibitem[{\citenamefont{Chianca and Olsen}(2011{\natexlab{b}})}]{CVC2}
\bibinfo{author}{\bibfnamefont{C.~V.} \bibnamefont{Chianca}} \bibnamefont{and}
  \bibinfo{author}{\bibfnamefont{M.~K.} \bibnamefont{Olsen}},
  \bibinfo{journal}{Phys. Rev. A} \textbf{\bibinfo{volume}{84}},
  \bibinfo{pages}{043636} (\bibinfo{year}{2011}{\natexlab{b}}).

\bibitem[{\citenamefont{{Opanchuk {\em et
  al.\/}}}(2012{\natexlab{b}})}]{2wellEPR}
\bibinfo{author}{\bibfnamefont{B.}~\bibnamefont{{Opanchuk {\em et al.\/}}}},
  \bibinfo{journal}{Phys. Rev. A} \textbf{\bibinfo{volume}{86}},
  \bibinfo{pages}{023625} (\bibinfo{year}{2012}{\natexlab{b}}).

\bibitem[{\citenamefont{Kubo}(1957)}]{KuboIrrevI}
\bibinfo{author}{\bibfnamefont{R.}~\bibnamefont{Kubo}}, \bibinfo{journal}{J.\
  Phys.\ Soc.\ Jap.} \textbf{\bibinfo{volume}{12}}, \bibinfo{pages}{570}
  (\bibinfo{year}{1957}).

\bibitem[{\citenamefont{Kubo et~al.}(1985)\citenamefont{Kubo, Toda, and
  Hashitsume}}]{KuboTodaHashitsumeII}
\bibinfo{author}{\bibfnamefont{R.}~\bibnamefont{Kubo}},
  \bibinfo{author}{\bibfnamefont{M.}~\bibnamefont{Toda}}, \bibnamefont{and}
  \bibinfo{author}{\bibfnamefont{N.}~\bibnamefont{Hashitsume}},
  \emph{\bibinfo{title}{Statistical Physics II: Nonequilibrium Statistical
  Mechanics}} (\bibinfo{publisher}{Springer}, \bibinfo{year}{1985}).

\bibitem[{\citenamefont{Plimak et~al.}(2003)\citenamefont{Plimak, Fleischhauer,
  Olsen, and Collett}}]{BWO}
\bibinfo{author}{\bibfnamefont{L.~I.} \bibnamefont{Plimak}},
  \bibinfo{author}{\bibfnamefont{M.}~\bibnamefont{Fleischhauer}},
  \bibinfo{author}{\bibfnamefont{M.~K.} \bibnamefont{Olsen}}, \bibnamefont{and}
  \bibinfo{author}{\bibfnamefont{M.~J.} \bibnamefont{Collett}},
  \bibinfo{journal}{Phys. Rev. A} \textbf{\bibinfo{volume}{67}},
  \bibinfo{pages}{013812} (\bibinfo{year}{2003}).

\bibitem[{\citenamefont{Plimak and Stenholm}(2008{\natexlab{a}})}]{API}
\bibinfo{author}{\bibfnamefont{L.~I.} \bibnamefont{Plimak}} \bibnamefont{and}
  \bibinfo{author}{\bibfnamefont{S.}~\bibnamefont{Stenholm}},
  \bibinfo{journal}{Ann.~Phys.~(N.Y.)} \textbf{\bibinfo{volume}{323}},
  \bibinfo{pages}{1963} (\bibinfo{year}{2008}{\natexlab{a}}).

\bibitem[{\citenamefont{Plimak and Stenholm}(2008{\natexlab{b}})}]{APII}
\bibinfo{author}{\bibfnamefont{L.~I.} \bibnamefont{Plimak}} \bibnamefont{and}
  \bibinfo{author}{\bibfnamefont{S.}~\bibnamefont{Stenholm}},
  \bibinfo{journal}{Ann.~Phys.~(N.Y.)} \textbf{\bibinfo{volume}{323}},
  \bibinfo{pages}{1989} (\bibinfo{year}{2008}{\natexlab{b}}).

\bibitem[{\citenamefont{Plimak and Stenholm}(2009)}]{APIII}
\bibinfo{author}{\bibfnamefont{L.~I.} \bibnamefont{Plimak}} \bibnamefont{and}
  \bibinfo{author}{\bibfnamefont{S.}~\bibnamefont{Stenholm}},
  \bibinfo{journal}{Ann.~Phys.~(N.Y.)} \textbf{\bibinfo{volume}{324}},
  \bibinfo{pages}{600} (\bibinfo{year}{2009}).

\bibitem[{\citenamefont{Konstantinov and Perel}(1960)}]{Perel}
\bibinfo{author}{\bibfnamefont{O.~V.} \bibnamefont{Konstantinov}}
  \bibnamefont{and} \bibinfo{author}{\bibfnamefont{V.~I.} \bibnamefont{Perel}},
  \bibinfo{journal}{Zh. Eksp. Theor. Phys.} \textbf{\bibinfo{volume}{39}},
  \bibinfo{pages}{197} (\bibinfo{year}{1960}) \bibinfo{note}{[Sov. Phys. JETP
  {\bf 12}, 142 (1961)]}.

\bibitem[{\citenamefont{Keldysh}(1964)}]{Keldysh}
\bibinfo{author}{\bibfnamefont{L.~V.} \bibnamefont{Keldysh}},
  \bibinfo{journal}{Zh. Eksp. Theor. Phys.} \textbf{\bibinfo{volume}{47}},
  \bibinfo{pages}{1515} (\bibinfo{year}{1964}) \bibinfo{note}{[Sov. Phys. JETP
  {\bf 20}, 1018 (1965)]}.

\bibitem[{\citenamefont{Wyld}(1961)}]{Wyld}
\bibinfo{author}{\bibfnamefont{H.~W.} \bibnamefont{Wyld}},
  \bibinfo{journal}{Ann.~Phys.~(N.Y.)} \textbf{\bibinfo{volume}{14}},
  \bibinfo{pages}{143} (\bibinfo{year}{1961}).

\bibitem[{\citenamefont{{Zinn-Justin}}(1989)}]{ZinnJ}
\bibinfo{author}{\bibfnamefont{J.}~\bibnamefont{{Zinn-Justin}}},
  \emph{\bibinfo{title}{Quantum Field Theory and Critical Phenomena}}
  (\bibinfo{publisher}{Oxford University Press}, \bibinfo{year}{1989}).

\bibitem[{\citenamefont{Vasil'ev}(2004)}]{VasR}
\bibinfo{author}{\bibfnamefont{A.~N.} \bibnamefont{Vasil'ev}},
  \emph{\bibinfo{title}{The Field Theoretic Renormalization Group in Critical
  Behavior Theory and Stochastic Dynamics}} (\bibinfo{publisher}{CRC Press},
  \bibinfo{year}{2004}).

\bibitem[{\citenamefont{Pawula}(1967)}]{Pawula}
\bibinfo{author}{\bibfnamefont{R.~F.} \bibnamefont{Pawula}},
  \bibinfo{journal}{Phys. Rev. A} \textbf{\bibinfo{volume}{162}},
  \bibinfo{pages}{186} (\bibinfo{year}{1967}).

\bibitem[{\citenamefont{Plimak et~al.}(2001{\natexlab{b}})\citenamefont{Plimak,
  Olsen, Fleischhauer, and Collett}}]{EPL01}
\bibinfo{author}{\bibfnamefont{L.~I.} \bibnamefont{Plimak}},
  \bibinfo{author}{\bibfnamefont{M.~K.} \bibnamefont{Olsen}},
  \bibinfo{author}{\bibfnamefont{M.}~\bibnamefont{Fleischhauer}},
  \bibnamefont{and} \bibinfo{author}{\bibfnamefont{M.~J.}
  \bibnamefont{Collett}}, \bibinfo{journal}{Europhys. Lett.}
  \textbf{\bibinfo{volume}{56}}, \bibinfo{pages}{372}
  (\bibinfo{year}{2001}{\natexlab{b}}).

\bibitem[{\citenamefont{Plimak}(1994)}]{Corresp}
\bibinfo{author}{\bibfnamefont{L.~I.} \bibnamefont{Plimak}},
  \bibinfo{journal}{Phys. Rev. A} \textbf{\bibinfo{volume}{50}},
  \bibinfo{pages}{2120} (\bibinfo{year}{1994}).

\bibitem[{\citenamefont{Schwinger}(1961)}]{SchwingerC}
\bibinfo{author}{\bibfnamefont{J.~S.} \bibnamefont{Schwinger}},
  \bibinfo{journal}{J. Math. Phys.} \textbf{\bibinfo{volume}{2}},
  \bibinfo{pages}{407} (\bibinfo{year}{1961}).

\bibitem[{\citenamefont{Kamenev and Levchenko}(2009)}]{KamenevLevchenko}
\bibinfo{author}{\bibfnamefont{A.}~\bibnamefont{Kamenev}} \bibnamefont{and}
  \bibinfo{author}{\bibfnamefont{A.}~\bibnamefont{Levchenko}},
  \bibinfo{journal}{Advances in Physics} \textbf{\bibinfo{volume}{58}},
  \bibinfo{pages}{197} (\bibinfo{year}{2009}), \bibinfo{note}{also available as
  e-print arXiv:0901.3586v3}.

\bibitem[{\citenamefont{Jaksch et~al.}(1998)\citenamefont{Jaksch, Bruder,
  Cirac, Gardiner, and Zoller}}]{Jaksch}
\bibinfo{author}{\bibfnamefont{D.}~\bibnamefont{Jaksch}},
  \bibinfo{author}{\bibfnamefont{C.}~\bibnamefont{Bruder}},
  \bibinfo{author}{\bibfnamefont{J.~I.} \bibnamefont{Cirac}},
  \bibinfo{author}{\bibfnamefont{C.~W.} \bibnamefont{Gardiner}},
  \bibnamefont{and} \bibinfo{author}{\bibfnamefont{P.}~\bibnamefont{Zoller}},
  \bibinfo{journal}{Phys. Rev. Lett.} \textbf{\bibinfo{volume}{81}},
  \bibinfo{pages}{3108} (\bibinfo{year}{1998}).

\bibitem[{\citenamefont{Plimak et~al.}(1998)\citenamefont{Plimak, Fleischhauer,
  and Walls}}]{EPL98}
\bibinfo{author}{\bibfnamefont{L.~I.} \bibnamefont{Plimak}},
  \bibinfo{author}{\bibfnamefont{M.}~\bibnamefont{Fleischhauer}},
  \bibnamefont{and} \bibinfo{author}{\bibfnamefont{D.~F.} \bibnamefont{Walls}},
  \bibinfo{journal}{Europhys. Lett.} \textbf{\bibinfo{volume}{43}},
  \bibinfo{pages}{641} (\bibinfo{year}{1998}).

\bibitem[{\citenamefont{Plimak et~al.}(1999)\citenamefont{Plimak, Collett,
  Walls, and Fleischhauer}}]{ProcPathIntVI}
\bibinfo{author}{\bibfnamefont{L.~I.} \bibnamefont{Plimak}},
  \bibinfo{author}{\bibfnamefont{M.~J.} \bibnamefont{Collett}},
  \bibinfo{author}{\bibfnamefont{D.~F.} \bibnamefont{Walls}}, \bibnamefont{and}
  \bibinfo{author}{\bibfnamefont{M.}~\bibnamefont{Fleischhauer}}, in
  \emph{\bibinfo{booktitle}{Proceedings of the Sixth International Conference
  on Path Integrals from peV to TeV}}, edited by
  \bibinfo{editor}{\bibnamefont{{R.\ Casalbuoni {\em et al.\/}}}}
  (\bibinfo{publisher}{World Scientific, London}, \bibinfo{year}{1999}), p.
  \bibinfo{pages}{241}.

\bibitem[{\citenamefont{Schweber}(2005)}]{Schweber}
\bibinfo{author}{\bibfnamefont{S.}~\bibnamefont{Schweber}},
  \emph{\bibinfo{title}{An Introduction to Relativistic Quantum Field Theory}}
  (\bibinfo{publisher}{Dover}, \bibinfo{year}{2005}).

\bibitem[{\citenamefont{Bogoliubov and Shirkov}(1980)}]{Bogol}
\bibinfo{author}{\bibfnamefont{N.~N.} \bibnamefont{Bogoliubov}}
  \bibnamefont{and} \bibinfo{author}{\bibfnamefont{D.~V.}
  \bibnamefont{Shirkov}}, \emph{\bibinfo{title}{Introduction to the theory of
  quantized fields}} (\bibinfo{publisher}{Wiley, N.Y.}, \bibinfo{year}{1980}).

\bibitem[{\citenamefont{C.W.Gardiner}(1991)}]{GardinerQN}
\bibinfo{author}{\bibnamefont{C.W.Gardiner}}, \emph{\bibinfo{title}{Quantum
  Noise}} (\bibinfo{publisher}{Springer-Verlag, Berlin}, \bibinfo{year}{1991}).

\bibitem[{\citenamefont{Olsen et~al.}(2000)\citenamefont{Olsen, Plimak,
  Collett, and Walls}}]{QNDBWO}
\bibinfo{author}{\bibfnamefont{M.~K.} \bibnamefont{Olsen}},
  \bibinfo{author}{\bibfnamefont{L.~I.} \bibnamefont{Plimak}},
  \bibinfo{author}{\bibfnamefont{M.~J.} \bibnamefont{Collett}},
  \bibnamefont{and} \bibinfo{author}{\bibfnamefont{D.~F.} \bibnamefont{Walls}},
  \bibinfo{journal}{Phys. Rev. A} \textbf{\bibinfo{volume}{62}},
  \bibinfo{pages}{023802} (\bibinfo{year}{2000}).

\bibitem[{\citenamefont{Hori}(1952)}]{Hori}
\bibinfo{author}{\bibfnamefont{T.}~\bibnamefont{Hori}}, \bibinfo{journal}{Prog.
  Theor. Phys.} \textbf{\bibinfo{volume}{7}}, \bibinfo{pages}{378}
  (\bibinfo{year}{1952}).

\bibitem[{\citenamefont{Plimak and Walls}(1994)}]{OPO}
\bibinfo{author}{\bibfnamefont{L.~I.} \bibnamefont{Plimak}} \bibnamefont{and}
  \bibinfo{author}{\bibfnamefont{D.~F.} \bibnamefont{Walls}},
  \bibinfo{journal}{Phys. Rev. A} \textbf{\bibinfo{volume}{50}},
  \bibinfo{pages}{2627} (\bibinfo{year}{1994}).

\end{thebibliography}
\end{document}